\documentclass[12pt]{article}
\usepackage{amsmath,amssymb}
\usepackage[english]{babel}
\usepackage[cp866]{inputenc}
\usepackage{hyperref}
\numberwithin{equation}{section}

\newcommand{\bq}{\begin{eqnarray}}
\newcommand{\eq}{\end{eqnarray}}
\newcommand{\be}{\begin{equation}}
\newcommand{\ee}{\end{equation}}

\newcommand{\bbq}{\begin{equation*}}
\newcommand{\eeq}{\end{equation*}}

\newcommand{\ra}{\rightarrow}

\newcommand{\la}{\Lambda_Q}
\newcommand{\mph}{\mu_{\Phi}}
\newcommand{\ov}{\overline}
\newcommand{\lym}{\Lambda_{YM}}
\newcommand{\lt}{\tilde\Lambda}

\newcommand{\qq}{{\ov Q}Q}

\newcommand{\Qo}{({\ov Q}Q)_1}
\newcommand{\Qt}{({\ov Q}Q)_2}
\newcommand{\qo}{({\ov q}q)_1}
\newcommand{\qt}{{(\ov q}q)_2}

\newcommand{\mtq}{\langle m_Q^{\rm tot}\rangle}

\newcommand{\bo}{{\rm b_o}}
\newcommand{\bd}{{\rm\ov b}_{\rm o}}
\newcommand{\nd}{{\ov N}_c}
\newcommand{\mo}{\mu_{\Phi,\rm o}}

\newcommand{\tQ}{\textsf{Q}}
\newcommand{\otQ}{\ov{\textsf{Q}}}
\newcommand{\tq}{\textsf{q}}
\newcommand{\otq}{\ov{\textsf{q}}}
\newcommand{\tS}{\textsf{S}}

\newcommand{\tQt}{(\ov{\textsf{Q}}\textsf{Q})_2}
\newcommand{\tQo}{(\ov{\textsf{Q}}\textsf{Q})_1}
\newcommand{\dqt}{(\ov{\textsf{q}}\textsf{q})_2}

\newcommand{\tp}{M}

\newcommand{\mx}{\mu_{\rm x}}

\newcommand{\lm}{\Lambda_2}
\newcommand{\mos}{\mu_o^{\rm str}}
\newcommand{\bb}{2N_c-N_F}
\newcommand{\no}{{\rm n}_1}
\newcommand{\nt}{{\rm n}_2}
\newcommand{\tm}{\widetilde m}
\newcommand{\wt}{\widetilde}
\newcommand{\wmu}{\widetilde{\mu}_{\rm x}}

\begin{document}

\begin{center}{\bf \large Mass spectra in ${\cal N}=1$ SQCD with additional colorless but 
flavored fields.II} \end{center}
\vspace{1cm}

\begin{center}\bf Victor L. Chernyak $^{\,a,\,b}$ \end{center}
\begin{center}(e-mail: v.l.chernyak@inp.nsk.su) \end{center}
\begin{center} $^a\,$ Novosibirsk State University,\\ 630090 Novosibirsk, Russia\end{center}

\begin{center} $^b\,$ Budker Institute of Nuclear Physics SB RAS \\
 630090 Novosibirsk, Russia\end{center}
\vspace{1cm}
\begin{center}{\bf Abstract} \end{center}
\vspace{1cm}

This paper continues our previous study of similar theories in \cite{ch5}. We also consider here the ${\cal N}=1$ SQCD-like theories with $SU(N_c)$ colors (and their Seiberg's dual with $SU(N_F-N_c)$ dual colors) and $N_F$ flavors of light quarks, and with $N_F^2$ additional colorless flavored fields $\Phi^j_i$, but now with $N_F$ in the range $2N_c<N_F<3N_c$. The multiplicities of various vacua and quark and gluino condensates in these vacua are found.

The mass spectra of the direct and Seiberg's dual theories in various vacua are calculated within the dynamical scenario which assumes that quarks in such ${\cal N}=1$ SQCD-like theories can be in two {\it standard} phases only. These are either the HQ (heavy quark) phase where they are confined or the Higgs phase. The word {\it standard} implies here also that, in such ${\cal N}=1$ theories without elementary colored adjoint scalars, no {\it additional} parametrically light solitons (e.g. magnetic monopoles or dyons) are formed at those scales where quarks decouple as heavy or are higgsed. Recall that this scenario satisfies all those tests which were used as checks of the Seiberg hypothesis about the equivalence of the direct and dual theories.

The mass spectra of these direct and Seiberg's dual theories calculated within this framework were found to be different, in general. These parametrical differences of mass spectra of direct and dual theories show, in particular, that all those tests, which were used as checks of the Seiberg hypothesis about the equivalence of the direct and dual theories, although necessary, may well be insufficient.

Besides, the mass spectrum of the dual theory with $SU(N_F-N_c)$ colors and $N_c+1<N_F<3N_c/2$  dual quark flavors was calculated.

And finally, considered is the direct ${\cal N}=2$ SQCD with $SU(N_c)$ colors and $N_c+1<N_F<3N_c/2$ flavors of light quarks with the mass term $m{\rm Tr}({\overline Q}Q)$ in the superpotential. ${\cal N}=2$ SUSY is softly broken down to ${\cal N}=1$ by the mass term $\mu_{\rm x}{\rm Tr} (X^2)$ of the colored adjoint scalar field $X$,\,\,$m\ll\mu_{\rm x}\ll\Lambda_2$. The mass spectrum of this theory is calculated in vacua with the unbroken non-trivial discrete $Z_{2N_c-N_F\geq 2}$ symmetry, and compared with those in two different Seiberg's dual theories. In addition, we compare our results for this theory with those in two related papers arXiv:1304.0822,\,arXiv:1403.6086 of M.Shifman and A.Yung, and present the extensive criticism of these papers.

\newpage

\tableofcontents

\section{Introduction}

\hspace*{5mm} This article continues \cite{ch5}. We calculate below in sections 3 - 7 the mass spectra in ${\cal N}=1$ SQCD - like theories which include, in addition to standard quarks and gluons, also $N_F^2$ colorless fion fields $\Phi_i^j$ with the large mass parameter $\mph\gg\la$ in the superpotential. The Lagrangian of the direct $\Phi$-theory at the scale $\mu=\la$ is taken in the form, see \cite{ch5,ch4}
\footnote{\,
The gluon exponents are always implied in the Kahler terms. Besides, here and everywhere below in the text we neglect for simplicity all RG-evolution effects if they are logarithmic only.
}
\bbq
K={\rm Tr}\,\Bigl (\Phi^\dagger \Phi\Bigr )+{\rm Tr}\Bigl (\,Q^\dagger Q+(Q\ra {\ov Q})\,\Bigr )\,,
\quad {\cal W}=-\frac{2\pi}{\alpha(\mu=\la)} S+{\cal W}_{\rm matter}\,,\quad {\cal W}_{\rm matter}={\cal W}_{\Phi}+{\cal W}_Q\,,
\eeq
\bq
{\cal W}_{\Phi}=\frac{\mph}{2}\Biggl [{\rm Tr}\,(\Phi^2)-\frac{1}{\nd}\Bigl ({\rm Tr}\,\Phi\Bigr )^2\Biggr ],\quad {\cal W}_Q={\rm Tr}\,{\ov Q}(m_Q-\Phi) Q\,,\quad \nd\equiv N_F-N_c\,.\label{(1.1)}
\eq
Here\,: the gauge group is $SU(N_c)$,\,  $\mph$ and $m_Q$  are the mass parameters, $S=-W^{A}_{\beta}W^{A,\,\beta}/32\pi^2$, where $W^A_{\beta}$ is the gauge field strength, $A=1...N_c^2-1,\, \beta=1,2$,\, $a(\mu)=N_c\alpha(\mu)/2\pi=N_c g^2(\mu)/8\pi^2$ is the gauge coupling with its scale factor $\la$.

Besides, in addition to the direct $\Phi$ - theory in \eqref{(1.1)}, we calculate also the mass spectra in Seiberg's dual variant \cite{S2,IS}. The Lagrangian of the dual $d\Phi$-theory at the scale $\mu=\la$ looks as
\bbq
K={\rm Tr}\,(\Phi^\dagger \Phi)+ {\rm Tr}\Bigl ( q^\dagger q + (q\ra\ov q)\, \Bigr )+{\rm Tr}\, \Bigl (\frac{M^{\dagger}M}{\la^2}\Bigr )\,,\quad {\cal W}=\, -\,\frac{2\pi}{\ov \alpha(\mu=\la)}\, {\wt S}+{\cal W}_{\Phi}+{\cal W}_q\,,
\eeq
\bq
{\cal W}_q={\rm Tr}\, M(m_Q-\Phi) -\frac{1}{\la}\,\rm {Tr} \Bigl ({\ov q}\,M\, q \Bigr )\,.\label{(1.2)}
\eq
Here\,:\, the number of dual colors is ${\ov N}_c=(N_F-N_c)$ and $M^i_j\ra ({\ov Q}_j Q^i)$ are the $N_F^2$  elementary mion fields, ${\ov a}(\mu)=\nd{\ov \alpha}(\mu)/2\pi=\nd{\ov g}^2(\mu)/8\pi^2$ is the dual gauge coupling (with its scale parameter $\Lambda_q=\la$),\,\,${\wt S}=- {\wt W}^{B}_{\beta}\, {\wt W}^{B,\,\beta}/32\pi^2$,\,\, ${\wt W}^B_{\beta}$ is the dual gluon field strength, $B=1...\nd^2-1$. The gluino condensates of the direct and dual theories are matched as well as $\langle M^i_j\rangle$ and $\langle{\ov Q}_j Q^i \rangle\equiv\sum_{a=1}^{N_c}\langle{\ov Q}^{\,a}_j Q^i_{a} \rangle$,
\bq
\langle{-\,\wt S}\rangle=\langle S\rangle\,,\quad \langle M^i_j\rangle
\equiv\langle M^i_j(\mu=\la)\rangle=\delta^i_j \langle M_i\rangle=
\langle{\ov Q}_j Q^i\rangle\equiv\langle{\ov Q}_j Q^i (\mu=\la)\rangle =\delta^i_j\langle(\qq)_i \rangle.
\label{(1.3)}
\eq

In sections 2-7 the hierarchies of parameters entering \eqref{(1.1)},\eqref{(1.2)} are $m_Q\ll\la\ll\mph$,\, $\mph$ is varied while $m_Q$ and $\la$ stay intact.

In those cases when the fields $\Phi$ are too heavy and dynamically irrelevant, they can be integrated out
and the Lagrangian of the dual theory takes the form
\bq
K={\rm Tr}\Bigl ( q^\dagger q + (q\ra\ov q)\,\Bigr )+{\rm Tr}\, \frac{M^{\dagger}M}{\la^2}\,,\label{(1.4)}
\eq
\bbq
{\cal W}_{\rm matter}={\cal W}_M-\frac{1}{\la}\,\rm {Tr} \Bigl ({\ov q}\,M\, q \Bigr )\,,\quad {\cal W}_M=m_Q \rm {Tr}\,M-\frac{1}{2\mph}\Biggl [{\rm Tr}\, (M^2)- \frac{1}{N_c}({\rm Tr}\, M)^2 \Biggr ].
\eeq

The mass spectra of the direct and dual theories will be calculated below within the dynamical scenario introduced in \cite{ch3}. Recall that this scenario assumes that, when such ${\cal N}=1$ SQCD-like theories are in the strong coupling regime $a(\mu)\gtrsim 1$, the quarks can be in two {\it standard} phases only. - These are either the HQ (heavy quark) phase where they are confined or the Higgs phase where (some components of) $\langle Q\rangle=\langle{\ov Q}\rangle\neq 0$. The word {\it standard} also implies here that, unlike e.g. the very special ${\cal N}=2$ SQCD theories with {\it colored adjoint} scalar superfields, no {\it additional} parametrically lighter particles (like magnetic monopoles or dyons) appear in these ${\cal N}=1$ SQCD-like theories without colored adjoint scalars at those scales when quarks are either higgsed or decouple as heavy.
\footnote{\,
In those cases when the coupling is logarithmically small this scenario is literally standard. Besides, it is worth noting that the appearance of additional parametrically light composite solitons will influence the 't Hooft triangles.\label{(f2)}
}

The mass spectra were calculated before in \cite{ch3} within this scenario in the standard ${\cal N}=1$ SQCD and its dual variant (i.e. both without the additional fion fields $\Phi$ in \eqref{(1.1)} and \eqref{(1.2)}), and in \cite{ch5} for the direct and dual $\Phi$ - theories \eqref{(1.1)},\eqref{(1.2)} in the case $3N_c/2<N_F<2N_c$. The purpose of this paper is to extent the results \cite{ch5} to the region $2N_c<N_F<3N_c$. This is the content of sections 2 - 6. As will be shown therein, many properties of both direct $\Phi$-theories and dual $d\Phi$-theories at $2N_c<N_F<3N_c$, i.e. multiplicities of vacua and mass spectra differ, in general, from those calculated in \cite{ch5} at $3N_c/2<N_F<2N_c$.

Besides, we calculate in section 7 the mass spectrum of the dual $d\Phi$-theory \eqref{(1.2)} for the case $N_c+1<N_F<3N_c/2$. This is of interest by itself, but is also useful for the next section 8.\\

In this last section 8 we consider ${\cal N}=2$ SQCD with $SU(N_c)$ colors, with the scale factor $\lm$ of the gauge coupling, and with $N_c+1<N_F<3N_c/2$ flavors of original electric quarks $Q^i, {\ov Q}_j$ with the mass term $m {\rm Tr}\,{\ov Q}Q$ in the superpotential. ${\cal N}=2$ is broken down to ${\cal N}=1$ by the mass term $\mx{\rm Tr}\,(X^2)$ of the colored adjoint scalar superfield $X$. The considered hierarchies of parameters look as:\,\,$m\ll\mx\ll\lm$. The purpose here was to calculate the mass spectra of this theory and its various Seiberg's dual variants in vacua with the non-trivial unbroken $Z_{2N_c-N_F\geq 2}$ symmetry at scales $\mu<\mx$. Besides, we discuss in section 8.3 the transition from $\mx\ll\lm$ to $\mx\gg\lm$.

Recall now that, as was argued in detail in section 7 of \cite{ch1}, outside the conformal window $3N_c/2<N_F<3N_c$, the standard UV free direct theory \eqref{(1.1)} (i.e. without fields $\Phi$) with $N_c<N_F<3N_c/2$ and $m_Q\ll\la$ enters {\it smoothly} at scales $\mu<\la$ into {\it the strongly coupled effectively massless perturbative regime} with the gauge coupling $a(\mu\ll\la)\sim (\la/\mu)^{\,\nu_Q>\,0}\gg 1$. In short, the arguments were as follows.\\
{\bf I)} It was proposed by Seiberg in \cite{S2} that the standard direct (electric) ${\cal N}=1\,\, SU(N_c)$ SQCD (i.e. \eqref{(1.1)} without fields $\Phi$) at $N_c+1<N_F<3N_c/2$ and with $m_Q=0$ is in the "confinement without spontaneous chiral symmetry breaking" (and without R-symmetry breaking) regime (\, $\langle{\ov Q}_j Q^i\rangle=\langle S\rangle=\langle{\ov q}^{\,j} q_i\rangle=\langle M^i_j\rangle=0\,$). And the standard massless dual (magnetic) $SU(N_F-N_c)$ theory \eqref{(1.2)} (i.e. without fields $\Phi$) was proposed as its low energy form at scales $\mu<\la$. This implies: a) the confinement of all original electric quarks and gluons with the string tension $\sqrt{\sigma}\sim\la$ (with $\la$ the only dimensional parameter), b) all colorless hadrons made from electric quarks and gluons have masses $\sim\la$, c) the formation of solitonic IR free $SU(N_F-N_c)$ theory with massless at $m_Q=0$ magnetic quarks and gluons (and $M^i_j$ fields). Besides, both direct and dual theories are considered in \cite{S2} as nonsingular nontrivial interacting theories at $m_Q=0$.

As is well known, the Seiberg duality passed a number non-trivial checks (mainly, the 't Hooft triangles and the behavior in the conformal regime) but up to now, unfortunately, no proof has been given that the direct and dual theories are (or are not) equivalent. The reason is that such a proof needs real understanding of and the control over the dynamics of both theories. Therefore, in the framework of gauge theories, the Seiberg proposal remains a hypothesis up to now. We quote here three papers only.  1)\, \cite{AR} (with N.Seiberg among the authors): "By now many examples of such dualities have been found, and a lot of evidence has been collected
for their validity. However, there is still no general understanding of the origin of these dualities, nor a prescription to find the dual for a given gauge theory". 2)\, \cite{G}:\, "The most established example of such a duality is the Seiberg duality in ${\cal N}= 1$ SQCD... The validity of this duality is still a conjecture but it have anyway remarkably passed numerous checks...". 3)\, \cite{A}:\, "All of these dualities do not have any rigorous derivation so far, but they pass many consistency checks...".

As for the string theory, we quote here two papers. 4)\, The review \cite{K} :\, "Specifically, we have seen using branes that the quantum moduli spaces of vacua and quantum chiral rings of the electric and magnetic SQCD theories coincide. This leaves open the question whether Seiberg's duality extends to an equivalence of the full infrared theory, since in general the chiral ring does not fully specify the infrared {\it conformal} field theory. It is {\it believed} that in gauge theory the answer is yes, and to prove it in brane theory will require an understanding of the smoothness of the transition when fivebranes cross. It is important to emphasize that the question cannot be addressed using any currently available tools". 5)\, The most detailed discussion of implicit dynamical assumptions and weak points in attempts to derive Seiberg's duality with a help of moving branes has been given in \cite{V}, whose authors discussed their results with N.Seiberg. In addition to arguments similar to \cite{K}, the authors \cite{V} emphasize that, within the approach with moving branes, {\it no reasons are seen for the equivalence of the direct and dual ${\cal N}=1$ theories outside the conformal window}.

{\bf III)} As was argued in details in section 7 of \cite{ch1}, the problem with the Seiberg's scenario from {\bf I)} is that it is impossible to write at the scale $\mu\sim\la$ the effective Lagrangian of massive hadrons with masses $\sim\la$ (made of confined electric quarks and gluons), which will be nonsingular at $m_Q=0$ and preserving both the R-symmetry and chiral symmetry $SU(N_F)_L\times SU(N_F)_R\,$. (Recall that these arguments in \cite{ch1} about the effective hadron Lagrangian at the scale $\sim\la$ did not use any assumptions about possible dynamical scenarios).

The variant with possibly massive (confined or not) electric quarks with dynamically induced masses $\sim\la$ has the same problems as those for hadrons. The nonsingular dynamically induced quark mass term in the superpotential at the scale $\mu=\la$ will look then as $\delta{\cal W}\sim\sum_{i,j=1}^{N_F}
\langle\phi^j_i\rangle\,{\ov Q}_j Q^i,\,\,\langle\phi^j_i\rangle\sim\delta_i^j\la$, with some composite field $\phi^j_i$. But, in any case, this $\langle\phi^j_i\rangle\sim\delta_i^j\la$ will break spontaneously the exact chiral symmetry $SU(N_F)_L\times SU(N_F)_R$ of the $m_Q=0$ theory. Besides, with the unbroken R-symmetry, this (solitonic colorless or colorful) field $\phi$ has the positive R-charge, $R_{\phi}=2N_c/N_F$. But, first, $\langle\phi^j_i\rangle\sim\delta_i^j\la$ will break spontaneously also the R-symmetry of the $m_Q\ll\la$ theory, while R-symmetry is considered in \cite{S2} as unbroken in both direct and dual theories. And second, the unbroken at $m_Q\ll\la$ R-symmetry requires that any chiral (elementary or composite) superfield $\Psi$ with the R-charge $r > 0\,$ behaves as $\langle\Psi(\mu=\la)\rangle\sim (m_Q)^{r N_F/2N_c}\ra 0$ at $m_Q\ra 0$, and this forbids $\langle\phi^j_i\rangle\sim\delta_i^j\la\,$.

Besides, as was shown in section 7 of \cite{ch1} and section 7 of \cite{ch3}, at $N_c+1<N_F<3N_c/2$ another scenario (which is the only possibility in the conformal window $3N_c/2<N_F<3N_c$) also does not lead the same mass spectra of the direct and dual theories. In this scenario, according to reasonings given above, the direct theory with $m_Q\ll\la$ enters smoothly at $\mu<\la$ into the strongly coupled effectively massless perturbative regime with $a(\mu\ll\la)\sim (\la/\mu)^{\,\nu_Q>\,0}\gg 1$. The IR free massless Seiberg's dual theory is considered as {\it independent theory} which presumably becomes equivalent to the direct one at $\mu<\la$. As was shown in \cite{ch1,ch3} the mass spectra of the direct and dual theories are not equivalent in this case.\\

We also compare our results in section 8 with those essentially different results of M.Shifman and A.Yung from related papers \cite{SY1,SY2}. This comparison also contains the detailed critique of results from  \cite{SY1,SY2}.

\section{\hspace*{-2mm} Quark condensates and multiplicity of vacua at $N_F>2N_c$}

\hspace*{5mm} For the reader convenience, we reproduce here first some useful formulas for the theories \eqref{(1.1)},\eqref{(1.2)}, see section 3 in \cite{ch4} or section 4 in \cite{ch5}.

The Konishi anomalies \cite{Konishi} for the $i$-th flavor look in the $\Phi$-theory \eqref{(1.1)} as
\bbq
\langle\Phi^j_{i}\rangle\langle\frac{\partial W_{\Phi}}{\partial \Phi^j_{i}}\rangle=0\,,\quad
\langle m_{Q,i}^{\rm tot}\rangle\langle(\qq)_i\rangle=\langle S\rangle\,, \quad \langle\Phi^j_{i}\rangle=\delta^j_i \langle\Phi_{i}\rangle, \quad \langle m_{Q,\,i}^{\rm tot}\rangle=m_Q-\langle\Phi_{i}\rangle\,,
\eeq
\bq
\langle\Phi_j^i\rangle=\frac{1}{\mph}\Biggl ( \langle{\ov Q}_j Q^i \rangle-\delta^i_j\frac{1}{N_c}{\rm Tr}\,\langle\qq\rangle\Biggr )\,,\quad \langle{\ov Q}_j Q^i \rangle\equiv\sum_{a=1}^{N_c}\langle{\ov Q}^{\,a}_j Q^i_{a}\rangle =\delta^i_j\langle(\qq)_i\rangle\,,\quad {\it i}=1\, ...\, N_F\,,\label{(2.1)}
\eq
and $\langle m_{Q,i}^{\rm tot}\rangle$ is the value of the quark running mass at the scale $\mu=\la$.

At all scales until the field $\Phi$ remains too heavy and non-dynamical, i.e. until its perturbative running mass $\mu_{\Phi}^{\rm pert}(\mu)>\mu$, it can be integrated out and the superpotential takes the form
\bq
W_Q=m_Q{\rm Tr}({\ov Q} Q)-\frac{1}{2\mph}\Biggl (\,\sum_{i,j}\,({\ov Q}_j Q^i)({\ov Q}_i Q^j)-\frac{1}{N_c}\Bigl({\rm Tr}\,{\ov Q} Q \Bigr)^2 \Biggr ).\label{(2.2)}
\eq

The values of the quark condensates for the $i$-th flavor, $\langle{\ov Q}_i Q^i\rangle$, in various vacua can be obtained from the effective superpotential $W^{\rm eff}_{\rm tot}(\Pi)$ depending only on quark bilinears $\Pi^i_j=({\ov Q}_j Q^i)$ \cite{ch4,ch5}. ( It is worth recalling that this is {\it not} a genuine low energy superpotential, \eqref{(2.3)} can be used {\it only} for finding the values of vacuum condensates, see \cite{ch5,ch4}. The genuine low energy superpotentials in each vacuum are given below in the text ).
\bq
W^{\rm eff}_{\rm tot}(\Pi)=W_Q - \nd S\,,\quad S=\Bigl (\frac{\det{\ov Q} Q}{\la^{\bo}}\Bigr )^{1/\nd} \equiv \lym^3\,,\quad \bo=3N_c-N_F\,,\quad \nd=N_F-N_c\,. \label{(2.3)}
\eq

For the vacua with the spontaneously broken flavor symmetry, $U(N_F)\ra U(n_1)\times U(n_2),\,\,1\leq n_1\leq N_F/2,\,\, n_2\geq N_F/2$, the most convenient way to find the quark condensates is to use \cite{ch4,ch5}
\bbq
\langle \Qo+\Qt-\frac{1}{N_c}{\rm Tr}\, ({\ov Q} Q)\,\rangle_{\rm br}=m_Q\mph,\quad
\langle S\rangle_{\rm br}=\Bigl (\frac{\det \langle{\ov Q} Q\rangle_{\rm br}}{\la^{\bo}}\Bigr )^{1/\nd}=\frac{\langle\Qo \rangle_{\rm br}\langle\Qt\rangle_{\rm br}}{\mph},
\eeq
\bq
\det \langle{\ov Q} Q\rangle_{\rm br}=\langle\Qo\rangle^{{\rm n}_1}_{\rm br}\,\langle\Qt\rangle^{{\rm n}_2}_{\rm br}\,,\quad \Qo\equiv\sum_{a=1}^{N_c}{\ov Q}^{\,a}_1 Q^1_{a}\,,\quad \Qt\equiv\sum_{a=1}^{N_c}{\ov Q}^{\,a}_2 Q^2_{a}\,,\label{(2.4)}
\eq
\bbq
\langle m^{\rm tot}_{Q,1}\rangle_{\rm br}=m_Q-\langle\Phi_1\rangle_{\rm br}=\frac{\langle\Qt\rangle_{\rm br}}{\mph},\quad \langle m^{\rm tot}_{Q,2}\rangle_{\rm br}=m_Q-\langle\Phi_2\rangle_{\rm br}=\frac{\langle\Qo\rangle_{\rm br}}{\mph}\,.
\eeq

The Konishi anomalies for the $i$-th flavor look in the dual $d\Phi$-theory \eqref{(1.2)} as
\bq
\langle M_i\rangle \langle {\ov q}^{\,i} q_i\rangle=\la\langle S\rangle\,,\quad \frac{\langle{\ov q}^{\,i} q_i\rangle}{\la}=\langle m_{Q,i}^{\rm tot}\rangle=m_Q-\frac{1}{\mph}\,\langle M_i-\frac{1}{N_c}{\rm Tr}\,M \rangle\,,\quad {\it i}=1\, ...\, N_F\,.\label{(2.5)}
\eq

In vacua with the broken flavor symmetry these can be rewritten as (remind that $\langle M^i_j\rangle=
\delta^i_j\langle M_i\rangle,\\ \langle M_1\rangle=\langle\Qo\rangle,\, \langle M_2\rangle=\langle\Qt\rangle$)
\bbq
\langle M_1+M_2-\frac{1}{N_c}{\rm Tr}\, M\rangle_{\rm br}=m_Q\mph,\quad\langle S\rangle_{\rm br}
=\Bigl (\frac{\det \langle M\rangle_{\rm br}}{\la^{\bo}}\Bigr )^{1/\nd}=\frac{1}{\mph}\langle M_1\rangle_{\rm br}\langle M_2\rangle_{\rm br}\,,
\,\,
\eeq
\bq
\frac{\langle\qo\rangle_{\rm br}}{\la}=\frac{\langle S\rangle_{\rm br}}{\langle M_{1}\rangle_{\rm br}}=\frac{\langle M_{2}\rangle_{\rm br}}{\mph}=\langle m^{\rm tot}_{Q,1}\rangle_{\rm br},\quad
\frac{\langle \qt\rangle_{\rm br}}{\la}=\frac{\langle S\rangle_{\rm br}}{\langle M_{2}\rangle_{\rm br}}=\frac{\langle M_{1}\rangle_{\rm br}}{\mph}=\langle m^{\rm tot}_{Q,2}\rangle_{\rm br}\,,\label{(2.6)}
\eq
\bbq
\frac{\langle\qt\rangle_{\rm br}}{\langle\qo\rangle_{\rm br}}=\frac{\langle\Qo\rangle_{\rm br}}{\langle\Qt \rangle_{\rm br}}\,,\quad \qo\equiv\sum_{b=1}^{\nd}{\ov q}_{b}^{\,1} q^{b}_1\,,\quad \qt\equiv\sum_{b=1}^{\nd} {\ov q}_{b}^{\,2} q^b_2\,,\quad \nd\equiv N_F-N_c\,.
\eeq

\subsection{Vacua with the unbroken flavor symmetry}

One obtains from \eqref{(2.3)} at $\mph\lessgtr \mo$ and with $\langle{\ov Q}_j Q^i\rangle=\delta^i_j\langle{\ov Q} Q\rangle\,,\,\, \langle{\ov Q} Q\rangle=\sum_{a=1}^{N_c}{\ov Q}^{\,a}_1 Q^1_{a}\,$. -

{\bf a)} There are only $\nd=(N_F-N_c)$ nearly classical S - vacua at $\mph\ll\mo$ with
\footnote{\,
Here and everywhere below\,: $A\approx B$ has to be understood as an equality neglecting smaller power corrections, and $A\ll B$ has to be understood as $|A|\ll |B|$. \label{(f4)}
}
\bq
\langle\qq\rangle_S\equiv\langle\qq(\mu=\la)\rangle_S\approx -\frac{N_c}{\nd}\, m_Q\mph\,,\quad \mo=\la\Bigl (\frac{m_Q}{\la}\Bigr )^{\frac{N_F-2N_c}{N_c}}\ll\la\,,\label{(2.7)}
\eq
\bbq
\langle\lym^{(S)}\rangle^3\equiv\langle S\rangle_S=\Bigl (\frac{\det\langle\qq\rangle_S}{\la^{\rm \bo}}\Bigr )^{1/\nd}\sim\la^3\Bigl (\frac{m\mph}{\la^2}\Bigr )^{N_F/\nd}\,.
\eeq

{\bf b)} There are $(N_F-2N_c)$ quantum L - vacua at $\mph\gg\mo$ with
\bq
\langle\qq\rangle_L\equiv\langle\qq(\mu=\la)\rangle_L\sim \la^2\Biggl (\frac{\mph}{\la}\Biggr )^{\frac{\nd}{N_F-2N_c}}\,,\label{(2.8)}
\eq
\bbq
\langle\lym^{(L)}\rangle^3\equiv\langle S\rangle_L=\Bigl (\frac{\det\langle\qq\rangle_L}{\la^{\rm \bo}}\Bigr )^{1/\nd}\sim\la^3\Bigl (\frac{\mph}{\la}\Bigr )^{\frac{N_F}{N_F-2N_c}}\,.
\eeq

{\bf c)} There are $N_c$ quantum QCD  vacua at $\mph\gg\mo$ with
\bq
\langle\qq\rangle_{QCD}=\langle\qq(\mu=\la)\rangle_{QCD}\approx\frac{\langle S\rangle_{\rm QCD}\equiv \langle
\lym^{(\rm QCD)}\rangle^3}{m_Q}\approx\frac{1}{m_Q}\Bigl (\la^{\bo}m_Q^{N_F}\Bigr)^{1/N_c}\,.\label{(2.9)}
\eq

The total number of vacua with the unbroken flavor symmetry is
\bq
N^{\rm tot}_{\rm unbr}=(N_F-2N_c)+N_c=\nd\,.\label{(2.10)}
\eq

\subsection{Vacua with the broken flavor symmetry, $U(N_F)\ra U(n_1)\times U(n_2)$}

{\bf a)} There are $(n_1-N_c){\ov C}^{\,\rm n_1}_{N_F}$ br1 -vacua
\footnote {\,
${\ov C}^{\, n_1}_{N_F}$ differ from the standard $C^{\, n_1}_{N_F}=N_F!/[n_1!\,n_2!]$ only for ${\ov C}^{\,n_1={\rm k}}_{N_F=2{\rm k}}=C^{\,n_1={\rm k}}_{N_F=2{\rm k}}/2$.

Besides, by convention, we ignore the continuous multiplicity of vacua due to the spontaneous flavor symmetry breaking. Another way, one can separate slightly all quark masses, so that all Nambu-Goldstone particles will acquire small masses $O(\delta m_Q)\ll {\ov m}_Q$.
}
at $N_c<n_1\leq [N_F/2]$ and $\mph\ll\mo$ with
\bq
\langle\Qo\rangle_{\rm br1}\approx \frac{N_c}{N_c-n_1} m_Q\mph,\quad  \langle\Qt \rangle_{\rm br1}\sim \la^2\Bigl (\frac{\mph}{\la}\Bigr )^{\frac{n_1}{n_1-N_c}}\Bigl (\frac{\la}{m_Q}\Bigr )^{\frac{n_2-N_c}{n_1-N_c}}\,,\label{(2.11)}
\eq
\bbq
\langle S\rangle_{\rm br1}=\frac{\langle\Qo\rangle_{\rm br1}\langle\Qt\rangle_{\rm br1}}{\mph}\sim
\la^3\Bigl (\frac{\mph}{\la}\Bigr )^{\frac{n_1}{n_1-N_c}}\Bigl (\frac{\la}{m_Q}\Bigr )^{\frac{n_2-n_1}{n_1-N_c}}\,,\quad\frac{\langle\Qt\rangle_{\rm br1}}{\langle\Qo\rangle
_{\rm br1}}\sim \Bigl (\frac{\mph}{\mo}\Bigr )^{\frac{N_c}{n_1-N_c}}\ll 1\,.
\eeq

{\bf b)} There are at $\mph\ll\mo$ $\,(n_2-N_c){\ov C}^{\,\rm n_1}_{N_F}$ br2-vacua at all values $1\leq n_1\leq [N_F/2]$ with
\bq
\langle\Qt\rangle_{\rm br2}\approx \frac{N_c}{N_c-n_2} m_Q\mph\,,\quad \langle\Qo\rangle_{\rm br2}\sim \la^2\Bigl (\frac{\mph}{\la}\Bigr )^{\frac{n_2}{n_2-N_c}}\Bigl (\frac{m_Q}{\la}\Bigr )^{\frac{N_c-n_1}{n_2-N_c}}\,,\label{(2.12)}
\eq
\bbq
\langle\lym^{(\rm br2)}\rangle^3\equiv\langle S\rangle_{\rm br2}\sim\la^3\Bigl (\frac{\mph}{\la}\Bigr )^{\frac{n_2}{n_2-N_c}}\Bigl (\frac{m_Q}{\la}\Bigr )^{\frac{n_2-n_1}{n_2-N_c}}\,,\quad
\frac{\langle\Qo\rangle_{\rm br2}}{\langle\Qt\rangle_{\rm br2}}\sim \Bigl (\frac{\mph}{\mo}\Bigr )^{\frac{N_c}{n_2-N_c}}\ll 1\,.
\eeq

On the whole, there are ($\,\theta(z)$ is the step function)
\bq
N^{\rm tot}_{\rm br}(n_1)=\Bigl [ (n_2-N_c)+\theta(n_1-N_c)(n_1-N_c)\Bigr ]{\ov C}^{\,\rm n_1}_{N_F},\,\,
N^{\rm tot}_{\rm br}=\sum_{n_1=1}^{[N_F/2]}N^{\rm tot}_{\rm br}(n_1)\,,\,\, \no+\nt=N_F \label{(2.13)}
\eq
vacua with the broken flavor symmetry at $\mph\ll\mo$.

\vspace{2mm}

{\bf c)} There are $(N_c-n_1)C^{\rm\,n_1}_{N_F}$ br1 -vacua at $1\le n_1<N_c$ and $\mph\gg\mo$ with
\bq
\langle\Qo\rangle_{\rm br1}\approx \frac{N_c}{N_c-n_1} m_Q\mph,\quad
\langle\Qt\rangle_{\rm br1}\sim \la^2\Bigl (\frac{\la}{\mph}\Bigr )^{\frac{n_1}{N_c-n_1}}\Bigl (\frac{m_Q}{\la}\Bigr )^{\frac{n_2-N_c}{N_c-n_1}}\,,\label{(2.14)}
\eq
\bbq
\langle\lym^{(\rm br1)}\rangle^3\equiv\langle S\rangle_{\rm br1}\sim\la^3\Bigl (\frac{\la}{\mph}\Bigr )^{\frac{n_1}{N_c-n_1}}\Bigl (\frac{m_Q}{\la}\Bigr )^{\frac{n_2-n_1}{N_c-n_1}}\,,\quad
\quad \frac{\langle\Qt\rangle_{\rm br1}}{\langle\Qo\rangle_{\rm br1}}\sim \Bigl (\frac{\mo}{\mph}\Bigr )^{\frac{N_c}{N_c-n_1}}\ll 1\,.
\eeq

{\bf d)} There are $(N_F-2N_c){\ov C}^{\rm\,n_1}_{N_F}$ Lt (i.e. L - type) vacua  at $n_1\neq N_c$ and $\mph\gg\mo$ with
\bq
(1-\frac{n_1}{N_c})\langle\Qo\rangle_{\rm Lt}\approx -(1-\frac{n_2}{N_c})\langle\Qt\rangle_{\rm Lt}\sim \la^2\Biggl (\frac{\mph}{\la}\Biggr )^{\frac{\nd}{N_F-2N_c}},\label{(2.15)}
\eq
i.e. as in the L - vacua in \eqref{(2.8)} but $\langle\Qo\rangle_{\rm Lt}\neq\langle\Qt\rangle_{\rm Lt}$ here.

{\bf e)} There are $(N_F-2N_c) C^{\rm\,n_1}_{N_F}$ special vacua at $n_1=N_c$ and $\mph\gg\mo$ with
\bq
\langle\Qt\rangle_{\rm spec}=\frac{N_c}{2N_c-N_F} m_Q\mph\,,\quad \langle\Qo\rangle_{\rm spec}= \la^2
\Bigl (\frac{\mph}{\la}\Bigr )^{\frac{\nd}{N_F-2N_c}}\,,\label{(2.16)}
\eq
\bbq
\langle\lym^{(\rm spec)}\rangle^3\equiv\langle S\rangle_{\rm spec}\sim m\la^2\Bigl (\frac{\mph}{\la}\Bigr )^{\frac{\nd}{N_F-2N_c}}\,,\quad \frac{\langle\Qt\rangle_{\rm spec}}{\langle\Qo\rangle_{\rm spec}}\sim \Bigl (\frac{\mo}{\mph}\Bigr )^{\frac{N_c}{N_F-2N_c}}\ll 1\,.
\eeq

As one can see from the above, similarly to \cite{ch5} with $N_c<N_F<2N_c$, all quark condensates become parametrically the same at $\mph\sim\mo$, see \eqref{(2.7)}. Clearly, just this region $\mph\sim\mo$ and not $\mph\sim \la$ is very special and most of the quark condensates change their parametric behavior and hierarchies at $\mph\lessgtr\mo$. For example, the br1 - vacua with $1\le n_1<N_c\,,\,\,\langle\Qo \rangle\sim m_Q\mph\gg\langle\Qt\rangle$ at $\mph\gg\mo$ evolve into br2 - vacua with $\langle\Qt\rangle\sim m_Q\mph\gg\langle\Qo\rangle$ at $\mph\ll\mo$, while the br1 - vacua with $n_1>N_c\,,\,\,\langle\Qo\rangle\sim m_Q\mph\gg\langle\Qt\rangle$ at $\mph\ll\mo$ evolve into the L-type  vacua with $\langle\Qo\rangle
\sim\langle\Qt\rangle\sim\la^2\, (\mph/\la)^{\nd/(N_F-2N_c)}$ at $\mph\gg\mo$, etc.

The exceptions are the special vacua with $n_1=N_c\,,\, n_2=\nd$\,. In these, the parametric behavior $\langle\Qt\rangle\sim m_Q\mph, \,\langle\Qo\rangle\sim \la^2\,(\mph/\la)^{\nd/(N_F-2N_c)}$ remains the same but the hierarchy is reversed at $\mph\lessgtr\mo\, :\, \langle\Qo\rangle/\langle\Qt\rangle\sim (\mph/\mo)^{N_c/(N_F-2N_c)}$.\\

On the whole, there are
\bq
N^{\rm tot}_{\rm br}(n_1)=\Bigl [ (N_F-2N_c)+\theta(N_c-n_1)(N_c-n_1)\Bigr ]{\ov C}^{\,\rm n_1}_{N_F},
\quad N^{\rm tot}_{\rm br}=\sum_{n_1=1}^{[N_F/2]}N^{\rm tot}_{\rm br}(n_1) \label{(2.17)}
\eq
vacua with the broken flavor symmetry at $\mph\gg\mo$. The total number of vacua is the same at $\mph\lessgtr\mo$\,, as it should be.

We point out finally that the multiplicities of vacua at $N_F>2N_c$ are not the analytic continuations of those at $N_c<N_F<2N_c$\,, see \cite{ch4,ch5}.

\section{Direct theory. Unbroken flavor symmetry.\\ $\hspace*{3cm} 2N_c<N_F<3N_c\,,\,\mph\gg\la$}

\hspace*{5mm} It is worth noting first that in all calculations below in the text we use the NSVZ $\beta$ - function \cite{NSVZ1, NSVZ2} ( remind also the footnote 1).

Because $\mo=\la(m_Q/\la)^{(N_F-2N_c)/N_c}\ll\la$ and $\mph\gg\la$, we are here in the region $\mph\gg\mo$. There are $N_c$ SQCD vacua (these were considered previously in \cite{ch3}, all fields $\Phi^j_i$ are too heavy and dynamically irrelevant in these vacua for the case considered) and $(N_F-2N_c)$ L - vacua \eqref{(2.8)} with
\bq
\frac{\langle\qq\rangle_L}{\la^2}\sim\Biggl (\frac{\mph}{\la}\Biggr )^{\frac{\nd}{N_F-2N_c}}\gg 1\,.
\eq
So, the gluon masses due to possible higgsing of some quark flavors are large (neglecting here and below in this section all logarithmic factors due to the RG-evolution), $\mu^{\rm pole}_{\rm gl}\sim \langle\qq\rangle_L^{1/2}\gg\la$. \label{(3.1)}

But the quark masses are even larger, see \eqref{(2.1)},
\bq
m_Q^{\rm pole}\sim\mtq_L\sim\langle\Phi\rangle_L\sim\frac{\langle\qq\rangle_L}{\mph}\sim
\la\Bigl (\frac{\mph}{\la} \Bigr )^{\frac{N_c}{N_F-2N_c}},\quad\frac{\mu^{\rm pole}_{\rm gl}}{m_Q^{\rm pole}}\sim \Bigl (\frac{\la}{\mph}\Bigr )^{\frac{3N_c-N_F}{2(N_F-2N_c)}}\ll 1\,.\label{(3.2)}
\eq
Therefore, all quarks are in the HQ (heavy quark) phase and are not higgsed but confined. After integrating out all quarks as heavy ones at scales $\mu<m^{\rm pole}_Q\sim \mtq$ in the weak coupling regime, there remain $N_F^2$ fions $\Phi$ and the $SU(N_c)$ SYM with the scale factor of its gauge coupling
\bq
\lym^{(L)}=\Bigl (\la^{\bo}\det m^{\rm tot}_Q \Bigr )^{1/3N_c},\quad \frac{\langle\lym^{(L)}\rangle=\langle S\rangle^{1/3}_L}{\la}\sim\Bigl (\frac{\mph}{\la} \Bigr )^{\frac{N_F}{3(N_F-2N_c)}}\gg \frac{\mph}{\la},\quad \bo=3N_c-N_F\,.\label{(3.3)}
\eq

After integrating out all gluons at the scale $\mu<\langle\lym^{(L)}\rangle$ through the Veneziano-Yankielowics (VY) procedure \cite{VY,TVY},  the Lagrangian of $N_F^2$ fions looks as, see \eqref{(1.1)},
\bq
K_{\Phi}\sim{\rm Tr}\,(\Phi^\dagger\Phi),\quad {\cal W}= N_c\Bigl (\la^{\bo}\det m^{\rm tot}_Q \Bigr )^{1/N_c}+{\cal W}_{\Phi}\,,\quad m^{\rm tot}_Q=m_Q-\Phi\,,\quad \langle m^{\rm tot}_Q\rangle_L\approx -\langle\Phi\rangle_L\,.\label{(3.4)}
\eq
From \eqref{(3.4)},\eqref{(3.2)} the fion masses are $\mu^{\rm pole}(\Phi)\sim\mph$.

On the whole, the mass spectrum looks in these $(N_F-2N_c)$ L - vacua as follows:\\
a) there is a large number of hadrons made of weakly coupled and weakly confined non-relativistic quarks
with masses $m^{\rm pole}_Q\sim\la(\mph/\la)^{N_c/(N_F-2N_c)}$\, (the tension of the confining string originating from the $SU(N_c)$ SYM is $\sqrt\sigma\sim\langle\lym^{(L)}\rangle\ll m^{\rm pole}_Q)$\,;\\
b) a large number of gluonia with the mass scale $\sim \langle\lym^{(L)}\rangle\sim\la (\mph/\la)^{N_F/3(N_F-2N_c)}$\,;\\
c) the lightest are $N^2_F$ fions with the masses $\mu^{\rm pole}(\Phi)\sim\mph\gg\la$.

The overall hierarchies look as
\bbq
\la\ll\mu^{\rm pole}(\Phi)\sim\mph\ll\langle\lym^{(L)}\rangle\ll m^{\rm pole}_Q\,.
\eeq

\section{Dual theory. Unbroken flavor symmetry.\\ $\hspace*{3cm} 2N_c<N_F<3N_c\,,\,\mph\gg\la$}

\hspace*{5mm} The dual theory in the UV region $\mu>\la$ and $2N_c<N_F<3N_c$ is taken as UV free. The largest mass in the case considered is that of dual quarks (neglecting here and below in this section all logarithmic factors due to the RG-evolution)
\bq
\mu^{\rm pole}_q\sim \frac{\langle M\rangle_L=\langle\qq\rangle_L}{\la}\sim\la\Biggl (\frac{\mph}{\la}\Biggr )^{\frac{\nd}{N_F-2N_c}}\gg\la,\label{(4.1)}
\eq
while the gluon masses due to possible higgsing of dual quarks are smaller
\bbq
{\ov\mu}^{\,2}_{\rm gl}\sim \langle{\ov q}q \rangle_L=\mtq\la\sim \langle\Phi\rangle\la\sim\frac{\la\langle\qq\rangle_L}{\mph}\sim\la^2\Biggl (\frac{\mph}{\la}\Biggr )^{\frac{N_c}{N_F-2N_c}}\,,
\eeq
\bq
\frac{{\ov\mu}_{\rm gl}}{\mu^{\rm pole}_q}\sim \Bigl (\frac{\la}{\mph}\Biggr )^{\frac{2N_F-3N_c}{2(N_F-2N_c)}}\ll 1\,.\label{(4.2)}
\eq

Therefore, the dual quarks are in the Hq (heavy quark) phase and are not higgsed but confined confined. After they are integrated out at scales $\mu<\mu^{\rm pole}_q$ in the weak coupling regime, there remain $N_F^2$ fions $\Phi$, $N_F^2$ mions $M$ and the $SU(\nd)$ SYM with the scale factor of its gauge coupling
\bq
\lym^{(L)}=\Bigl (\la^{\bd}\det\frac{M}{\la} \Bigr )^{1/3\nd},\quad \frac{\langle\lym^{(L)}\rangle}{\la} \sim\Bigl (\frac{\mph}{\la} \Bigr )^{\frac{N_F}{3(N_F-2N_c)}}\gg \frac{\mph}{\la},\quad \bd=3\nd-N_F>0\,.\label{(4.3)}
\eq

After integrating out all dual gluons at scales $\mu<\langle\lym^{(L)}\rangle$ through the Veneziano-Yankielowics (VY) procedure \cite{VY,TVY}, the Lagrangian looks as, see \eqref{(1.1)} for ${\cal W}_{\Phi}$,
\bq
K\sim {\rm Tr}\,\Bigl (\frac{M^\dagger M}{\la^2}+ \Phi^\dagger\Phi\,\Bigr )\,,\quad{\cal W}=-\nd
\Bigl (\la^{\bd}\det\frac{M}{\la}\Bigr )^{1/\nd}+{\rm Tr}\,(m_Q-\Phi)M+{\cal W}_{\Phi}\,.\label{(4.4)}
\eq

All fions $\Phi$ have masses $\mu^{\rm pole}(\Phi)\sim \mph$ and can be integrated out at scales $\mu<\mph$, resulting in the lower energy Lagrangian of mions
\bq
K_M\sim {\rm Tr}\,\frac{M^\dagger M}{\la^2}\,,\,\,\quad {\cal W}= -\nd\Bigl (\frac{\det M}{\la^{\bo}}\Bigr )^{1/\nd}+{\cal W}_{M}\,,\label{(4.5)}
\eq
\bbq
{\cal W}_M=m_Q{\rm Tr}\,M -\frac{1}{2\mph}\Biggl [{\rm Tr}\, (M^2)- \frac{1}{N_c}({\rm Tr}\, M)^2 \Biggr ].
\eeq
From \eqref{(4.5)} the mion masses are
\bq
\mu^{\rm pole}(M)\sim\frac{\la^2}{\mph}\ll\la \,.\label{(4.6)}
\eq

On the whole, the mass spectrum looks in these $(N_F-2N_c)$ dual L - vacua as follows:\\
a) there is a large number of hadrons made of the weakly coupled and weakly confined non-relativistic dual quarks with masses $\mu^{\rm pole}_q\sim\la(\mph/\la)^{\nd/(N_F-2N_c)}$ (the tension of the confining string originating from the $SU(\nd)$ dual SYM is $\sqrt\sigma\sim
\langle\lym^{(L)}\rangle\ll \mu^{\rm pole}_q)$\,;\\
b) a large number of gluonia with the mass scale $\sim \langle\lym^{(L)}\rangle\sim\la (\mph/\la)
^{N_F/3(N_F-2N_c)}$\,;\\
c) there are $N_F^2$ fions with masses $\mu^{\rm pole}(\Phi)\sim \mph$\,;\\
d) the lightest are $N^2_F$ mions with masses $\mu^{\rm pole}(M)\sim (\la^2/\mph)\ll\la$.

The overall hierarchies look here as:
\bbq
\mu^{\rm pole}(M)\ll\la\ll\mu^{\rm pole}(\Phi)\sim\mph\ll\langle\lym^{(L)}\rangle\ll \mu^{\rm pole}_q\,.
\eeq

Comparing the mass spectra in these L - vacua of the direct theory in section 3 and the dual one in this section it is seen that they are parametrically different.

We would like to emphasize also that quarks in both direct and dual theories are simultaneously weakly coupled in these L - vacua.

\section{Direct theory. Broken flavor symmetry.\\ $\hspace*{3cm} 2N_c<N_F<3N_c\,,\,\mph\gg\la$}

\subsection{Lt - vacua}

\hspace*{5mm} The main difference with the L - vacua in section 3 is that the flavor symmetry is broken spontaneously in these Lt - vacua, $\langle\Qo\rangle\neq\langle\Qt\rangle$ and, from \eqref{(3.4)}, the fions $\Phi_1^2$ and $\Phi_2^1$ are the Nambu-Goldstone particles here and are exactly massless.

\subsection{br1 - vacua}

\hspace*{5mm} In these vacua with $n_1<N_c$ and $\mo\ll\la\ll\mph\ll \la^2/m_Q$, the regime is conformal at scales $m^{\rm pole}_{Q,2}\ll\mu\ll\la$ (see below) and potentially most important masses  look here as follows.

The gluon masses due to possible higgsing of quarks are
\bq
\hspace*{-2mm}\Bigl (\mu^{\rm pole}_{\rm gl,1}\Bigr )^2\sim z_Q(\la,\mu^{\rm pole}_{\rm gl,1})\langle\Qo\rangle_{\rm br1}\sim\la^2\Bigl (\frac{m_Q\mph}{\la^2}\Bigr )^{2N_F/3\nd}\gg\mu^2_{\rm gl,2}\,,\,\, z_Q(\la,\mu^{\rm pole}_{\rm gl,1})\sim\Bigl (\frac{\mu^{\rm pole}_{\rm gl,1}}{\la}\Bigr )^{\bo/N_F},\label{(5.1)}
\eq
while the quark masses are
\bq
\quad \langle m^{\rm tot}_{Q,2}\rangle=\frac{\langle\Qo\rangle_{\rm br1}}{\mph}\sim m_Q\,,\quad {\tilde m}^{\rm pole}_{Q,2}\sim\frac{\langle m^{\rm tot}_{Q,2}\rangle}{z_Q(\la,{\tilde m}^{\rm pole}_{Q,2})}\sim\la\Bigl (\frac{m_Q}{\la}\Bigr )^{N_F/3N_c}\gg m^{\rm pole}_{Q,1}\,,\label{(5.2)}
\eq
\bq
\frac{{\tilde m}^{\rm pole}_{Q,2}}{\mu^{\rm pole}_{\rm gl,1}}\sim \Bigl (\frac{\mo}{\mph} \Bigr )^{N_F/3\nd}\ll 1\,.\label{(5.3)}
\eq

Therefore, the quarks $Q^1, {\ov Q}_1$ are higgsed and the overall phase is $Higgs_1-HQ_2$. If we take
$2n_1<\bo$, then ${\rm b}^\prime_{\rm o}=(\bo-2n_1)>0$ and the lower energy theory with $SU(N_c-n_1)$ colors and $n_2$ flavors will be in the conformal regime at scales $m^{\rm pole}_{Q,2}\sim\langle\lym^{(\rm br1)}\rangle\ll\mu\ll\mu^{\rm pole}_{\rm gl,1}$. Then all results for the mass spectra will be the same as in section 11.1 of \cite{ch5}.

\section{Dual theory. Broken flavor symmetry.\\ $\hspace*{3cm} 2N_c<N_F<3N_c\,,\,\mph\gg\la$}

\subsection{Lt - vacua}

\hspace*{5mm} The main difference with the L - vacua in section 4 is that the flavor symmetry is broken spontaneously in these dual Lt - vacua, $\langle M_1\rangle\neq\langle M_2\rangle$. The fions have masses $\mu^{\rm pole}(\Phi)\sim\mph\gg\la$ and are dynamically irrelevant at scales $\mu<\mph$. The low energy Lagrangian of mions is \eqref{(4.4)}, but the mion masses look now as
\bq
\mu^{\rm pole}(M_1^1)\sim \mu^{\rm pole}(M_2^2)\sim\frac{\la^2}{\mph}\ll\la,\quad \mu^{\rm pole}(M_1^2)=\mu^{\rm pole}(M_2^1)=0\,.\label{(6.1)}
\eq

Clearly, the parametric differences remain in mass spectra of the direct and dual theories in these Lt-vacua.

\subsection{br1 - vacua}

\hspace*{5mm} Not going into any details we give here the results only. The overall phase is $Hq_1-Hq_2$ (heavy quarks) and the massless Nambu-Goldstone particles here are the mions $M_1^2$ and $M_2^1$. The mass spectra are as in section 7.1 of \cite{ch5}, the only difference is that $Z_q\ra 1$ here because $\bd/N_F=O(1)$ now.

\section{Dual theory. $N_c+1<N_F<3N_c/2,\,\la\ll\mph\ll\mo$}

This dual theory \eqref{(1.2)} is in the IR free logarithmic regime in some range of scales $\mu_o<\mu<\la$ in all vacua. We neglect in this section all logarithmic factors due to the RG flow for simplicity.

\subsection{L  and Lt - vacua}

\hspace*{5mm} The condensates in L - vacua look as \cite{ch5}
\bq
\langle M\rangle_{\rm L}=\langle {\ov Q} Q\rangle_{\rm L}\sim\la^2\Bigl (\frac{\la}{\mph} \Bigr )^{\frac{\nd}{2N_c-N_F}}, \,\, \langle {\ov q} q\rangle_{\rm L}=\frac{\langle S\rangle_{\rm L}\la}{\langle M\rangle_{\rm L}}\sim\la^2\Bigl (\frac{\la}{\mph}\Bigr )^{\frac{N_c}{2N_c-N_F}}\,,\,\, S=\Bigl (\frac{\det M}{\la^{\bo}}\Bigr )^{1/\nd}\,,\label{(7.1)}
\eq
\bbq
\la\ll\mph\ll\mo=\la(\la/m_Q)^{(2N_c-N_F)/N_c}\,.
\eeq

Therefore, the pole masses $\mu^{\rm pole}_{q}$ of dual quarks and the possible gluon masses ${\ov\mu}_{\rm gl}$ due to their higgsing look as
\bq
\mu^{\rm pole}_q\sim \frac{\langle M\rangle_{\rm L}}{\la}\,,\quad {\ov\mu}_{\rm gl}^{\rm pole}\sim\langle{\ov q}q\rangle_{\rm L}^{1/2},\quad \frac{{\ov\mu}_{\rm gl}^{\rm pole}}{\mu^{\rm pole}_q}\sim\Bigl (\frac{\la}{\mph} \Bigr )^{\frac{3N_c-2N_F}{2(2N_c-N_F)}}\ll 1 \label{(7.2)}
\eq
and therefore the overall phase is $Hq$.

After integrating out all quarks as heavy ones at $\mu<\mu^{\rm pole}_q$ and then all $SU(\nd)$ gluons at $\mu<\lym^{(L)}\ll\mu^{\rm pole}_q$ through the VY-procedure \cite{VY,TVY}, the Lagrangian of $N_F^2$ mions looks as, see \eqref{(4.4)},\eqref{(4.5)}
\bq
K=\frac{M^\dagger M}{\la^2},\quad W=-\nd\Bigl (\frac{\det M}{\la^{\bo}}\Bigr )^{1/\nd}+W_M\,.\label{(7.3)}
\eq
From \eqref{(7.3)}, the masses of all $N_F^2$ mions are $\mu(M)\sim\la^2/\mph$.\\

On the whole, the mass spectrum in these L - vacua includes. -

1) A large number of hadrons made of weakly interacting non-relativistic and weakly confined dual quarks, the scale of their masses is $\mu^{\rm pole}_q\sim \langle M\rangle_{\rm L}/\la$ \eqref{(7.1)} (the tension of the confining string originating from $SU(\nd)$ SYM is much smaller, $\sqrt \sigma\sim\langle\lym^{(\rm L)}\rangle\ll\mu^{\rm pole}_q$).

2) A large number of gluonia made of $SU(\nd)$ gluons with their mass scale $\langle\lym^{(\rm L)}\rangle=\langle S\rangle_L^{1/3}\sim \la (\la/\mph)^{N_F/3(2N_c-N_F)}$.

3) $N_F^2$ mions with masses $\mu(M)\sim\la^2/\mph$.\\

The mass hierarchies look as $\mu(M)\ll\langle\lym^{(\rm L)}\rangle\ll\mu^{\rm pole}_{q}\ll\la$.\\

In comparison with these L - vacua, the only qualitative difference in Lt - vacua with the broken flavor symmetry is that the hybrid mions $M_{12}$ and $M_{21}$ are the Nambu-Goldstone particles there and are exactly massless.

\subsection{S - vacua}

\hspace*{5mm} The condensates look here as \cite{ch5}
\bq
\langle{\ov Q} Q\rangle_S=\langle M\rangle_S\sim m_Q\mph,\quad \langle {\ov q} q\rangle_S=\frac{\langle S\rangle_S\la}{\langle M\rangle}\sim\la^2\Bigl (\frac{m_Q\mph}{\la^2}\Bigr )^{N_c/\nd},\quad \langle S\rangle=\Bigl (\frac{\det \langle M\rangle_S}{\la^{\bo}}\Bigr )^{1/\nd},\label{(7.4)}
\eq
\bq
\mu^{\rm pole}_q\sim \frac{\langle M\rangle_S}{\la},\quad {\ov\mu}_{\rm gl}^{\,\rm pole}\sim\langle {\ov q}q
\rangle_S^{1/2}, \quad \frac{{\ov\mu}_{\rm gl}^{\,\rm pole}}{\mu^{\rm pole}_q}\sim\Bigl (\frac{m_Q\mph}{\la^2} \Bigr )^{(3N_c-2N_F)/2\nd}\ll 1 \label{(7.5)}
\eq
and so the overall phase is also $Hq$. Proceeding as before we obtain the Lagrangian \eqref{(7.3)} (but now in S - vacua). From this, the masses of all $N_F^2$ mions are $\mu^{\rm pole}(M)\sim\la^2/\mph$, while the scale of gluonia masses is $\langle\lym^{(\rm S)}\rangle=\langle S\rangle_S^{1/3}\sim\la (m_Q\mph/\la^2)^{N_F/3\nd}\ll\mu^{\rm pole}_q$.

As a result, the hierarchies of masses (except for $\mu(\Phi)\sim\mph\gg\la$) look as:\\
a) $\la\gg\mu^{\rm pole}(M)\gg\mu^{\rm pole}_q\gg\langle\lym^{(\rm S)}\rangle$ at $\la\ll\mph\ll{\mu_{\Phi}^\prime}=\la (\la/m_Q)^{1/2}\,$;\\
b) $\la\gg\mu^{\rm pole}_q\gg\mu^{\rm pole}(M)\gg\langle\lym^{(\rm S)}\rangle$ at ${\mu_{\Phi}^\prime}\ll\mph\ll {\mu^{\prime\prime}}_{\Phi}=\la (\la/m_Q)^{N_F/(4N_F-3N_c)}\,$;\\
c) $\la\gg\mu^{\rm pole}_q\gg\langle\lym^{(\rm S)}\rangle\gg\mu^{\rm pole}(M)$ at ${\mu^{\prime\prime}}_{\Phi}\ll\mph\ll\mo=\la (\la/m_Q)^{(2N_c-N_F)/N_c}\,$\,.

\subsection{br2 - vacua}

\hspace*{5mm} The condensates look in these vacua as \cite{ch4},
\bbq
\langle\Qt\rangle_{\rm br2}=\langle M_2\rangle_{\rm br2}\sim m_Q\mph\gg \langle\Qo\rangle_{\rm br2}=\langle M_1\rangle_{\rm br2}\sim \la^2\Bigl (\frac{\mph}{\la}\Bigr )^{\frac{n_2}{n_2-N_c}}\Bigl (\frac{m_Q}{\la}\Bigr )^{\frac{N_c-n_1}{n_2-N_c}},
\eeq
\bbq
\langle \qo\rangle_{\rm br2}=\frac{\langle S\rangle_{\rm br2}\la}{\langle M_1\rangle_{\rm br2}}=\frac
{\langle M_2\rangle_{\rm br2}\la}{\mph}\sim m_Q\la\gg \langle \qt\rangle_{\rm br2}=\frac
{\langle M_1\rangle_{\rm br2}\la}{\mph},\quad n_2>N_c\,,
\eeq
\bq
{\ov\mu}^{\,\rm pole}_{\rm gl,1}\sim (\langle \qo\rangle)^{1/2}\sim (m_Q\la)^{1/2}, \quad \mu^{\rm pole}_{q,2}\sim\frac{\langle M_2\rangle_{\rm br2}}{\la}\sim\frac{m_Q\mph}{\la}\gg\mu^{\rm pole}_{q,1},\quad \frac{\mu^{\rm pole}_{q,2}}{{\ov\mu}_{\,\rm gl,1}^{\,\rm pole}}\sim\Bigl (\frac{m_Q\mph^2}{\la^3}\Bigr )^{1/2}. \label{(7.6)}
\eq

{\bf A) The range} $\mathbf{\la\ll\mph\ll \la(\la/m_Q)^{1/2}}$
\vspace*{2mm}

The largest mass ${\ov\mu}_{\,\rm gl,1}^{\,\rm pole}$ have in this case gluons (and their superpartners) due to higgsing of dual quarks $q_1,{\ov q}^1$, $\,SU(\nd)\ra SU(\nd-n_1)$. The new scale factor of the gauge coupling is
\bq
\Bigl (\Lambda^\prime\Bigr )^{\bd-2n_1}=\la^{\bd}/\det N_1^1\,,\quad \mu^{\rm pole}_{q,2}\ll\langle\Lambda^\prime\rangle\ll\la,\quad \bd=3\nd-N_F <0\,,\quad \langle N_1^1\rangle=\langle {\ov q}^1 q_1\rangle \label{(7.7)}
\eq
and the overall phase is $Higgs_1-Hq_2$.

After integrating out massive gluons and higgsed quarks, the Lagrangian of the matter at the scale $\mu=\langle\Lambda^\prime\rangle$ looks as
\bq
K={\rm Tr}\,\Biggl (\frac{M^\dagger M}{\la^2}+2\sqrt{(N_1^1)^\dagger N_1^1}+K_{\rm hybrid}+\Bigl ( ({q^{\,\prime}_2})^\dagger\, q^{\,\prime}_2+({\ov q}^{\,\prime,\, 2})^\dagger\, {\ov q}^{\,\prime,\, 2}
\Bigl )\Biggr ),\label{(7.8)}
\eq
\bbq
K_{\rm hybrid}=N_2^1\frac{1}{\sqrt{(N^1_1) (N_1^1)^\dagger}}(N_2^1)^\dagger+(N_1^2)^\dagger\frac{1}
{\sqrt{(N_1^1)^\dagger N_1^1}}N_1^2\,,
\eeq
where $N_1^1=({\ov q}^1 q_1)$ are $n^2_1$ nion fields remained from the higgsed quarks $q_1, {\ov q}^1$, while the hybrids $N^1_2=\langle{\ov q}^{\,1}\rangle q_2$ and $N^2_1={\ov q}^{\,2}\langle q_1\rangle$ are in essence the quarks $q_2, {\ov q}^2$ with broken colors,
\bbq
{\cal W}={\cal W}_{M}-{\cal W}_{MN}-{\cal W}_{q},\quad {\cal W}_{MN}=\frac{1}{\la}{\rm Tr}\,\Bigl (M_1^1 N_1^1+(M_1^2 N_2^1+M_2^1 N_1^2)+M_2^2\, (N_2^1\frac{1}{N_1^1}N_1^2)\,\Bigr ),
\eeq
\bq
{\cal W}_M=m_Q {\rm Tr}\,M-\frac{1}{2\mph}\Bigl ({\rm Tr}\,(M^2)-\frac{1}{N_c}({\rm Tr}\,M)^2\Bigr ),\quad {\cal W}_q={\rm Tr}\, \Bigl ({\ov q}^{\,\prime,\, 2}\,\frac{M_{22}}{\la}\, q^{\,\prime}_2 \Bigr ),\label{(7.9)}
\eq
where $q^{\,\prime}_2$ and ${\ov q}^{\,\prime,\, 2}$ are quarks with unbroken colors.

Next, at scales $\mu<\mu^{\rm pole}_{q,2}\sim m_Q\mph/\la$, we integrate out remained $q^{\,\prime}_2,\, {\ov q}^{\,\prime,\, 2}$ quarks with unbroken colors as heavy ones and then $SU(\nd-n_1)$ gluons at $\mu<\langle\lym^{(\rm br2)}\rangle\ll\mu^{\rm pole}_{q,2}$. The Lagrangian of remaining mions and nions is
\bbq
K={\rm Tr}\,\Bigl (\frac{M^\dagger M}{\la^2}+2\sqrt{(N_1^1)^\dagger N_1^1}+K_{\rm hybrid}\Bigr ),
\eeq
\bq
{\cal W}=-(\nd-n_1)\Biggl (\la^{\bd}\frac{\det (M_2^2/\la)}{\det N_1^1}\Biggr )^{1/(\nd-n_1)} +{\cal W}_{M}-{\cal W}_{MN}\,.\label{(7.10)}
\eq

We obtain from \eqref{(7.10)} at $m_Q\mph^2/\la^3\ll 1$:\\
a) the mixing of mions and nions is parametrically small and masses of all $N_F^2$ mions are \\ $\hspace*{3cm} \mu^{\rm pole}(M)\sim\la^2/\mph$\,;\\
b) the masses of $N_1^1$ nions are much smaller,\\ $\hspace*{3cm}\mu^{\rm pole}(N_1^1)\sim m_Q\mph/\la,\quad \mu^{\rm pole}(N_1^1)/\mu^{\rm pole}(M)\sim m_Q\mph^2/\la^3\ll 1$\,;\\
c) the hybrid nions $N_1^2$ and $N_2^1$ are massless.\\

On the whole for this case the mass spectrum looks as follows.

1) The heaviest (among the masses $<\la$) are $N_F^2$ mions $M$ with masses $\mu^{\rm pole}(M)\sim \la^2/\mph\,$.

2) There are $n_1(2\nd-n_1)$ massive dual gluons (and their superpartners) with masses \\ ${\ov\mu}_{\rm gl,1}^{\,\rm pole}\sim\langle N_1\rangle^{1/2}\sim (m_Q\la)^{1/2}$.

3) There is a large number of hadrons made of weakly interacting non-relativistic and weakly confined quarks $q^{\,\prime}_2$ and ${\ov q}^{\,\prime,\, 2}$ with unbroken colors, the scale of their masses is $\mu^{\rm pole}_{q,2}\sim m_Q\mph/\la$ (the tension of confining string originating from the non-Abelian ${\cal N}=1\,\, SU(\nd-n_1$) SYM is much smaller, $\sqrt \sigma\sim\langle\lym^{(\rm br2)}\rangle\ll\mu^{\rm pole}_{q,2}$).

4) The masses of $n_1^2$ nions $N_1^1$ (dual pions) are also $\mu^{\rm pole}(N_1^1)\sim m_Q\mph/\la$.

5) There is a large number of $SU(\nd-n_1)$ SYM gluonia, the scale of their masses is \\ $\sim\langle\lym^{(\rm br2)}\rangle\sim (m_Q\langle M_1\rangle_{\rm br2})^{1/3}$, see \eqref{(7.6)}.

6) Finally, $2n_1 n_2$ Nambu-Goldstone hybrid nions $N_1^2, N_2^1$ are massless.\\

The overall hierarchy of nonzero masses looks as:
\bbq
\langle\lym^{(\rm br2)}\rangle\ll\mu^{\rm pole}(N_1^1)\sim\mu^{\rm pole}_{q,2}\ll{\ov\mu}_{\,\rm gl,1}^{\rm pole}\ll\mu(M)\ll\la\,.
\eeq

{\bf B) The range} $\mathbf{\la(\la/m_Q)^{1/2}\ll\mph\ll \mo=\la(\la/m_Q)^{(2N_c-N_F)/N_c}}$
\vspace*{2mm}

The largest mass have in this case $q_2, {\ov q}^2$ quarks, see \eqref{(7.6)}. After integrating them out at $\mu<\mu^{\rm pole}_{q,2}$ the new scale factor of the gauge coupling is
\bq
\langle\Lambda^{\prime\prime}\,\rangle^{3\nd-n_1}=\la^{\bd}\Bigl (\frac{m_Q\mph}{\la}\Bigr )^{n_2},
\quad\langle\Lambda^{\prime\prime}\,\rangle\ll \mu^{\rm pole}_{q,2}\,,\quad n_1<\nd\,.\label{(7.11)}
\eq

The ratio of the pole masses $\mu^{\rm pole}_{q,1}$ of $q_1, {\ov q}^1$ quarks to the possible gluon masses ${\ov\mu}_{\rm gl,1}^{\,\rm pole}$ due to their higgsing looks as
\bq
\mu^{\rm pole}_{q,1}\sim\frac{\langle M_1\rangle_{\rm br2}}{\la},\quad {\ov\mu}_{\rm gl,1}^{\,\rm pole}\sim (m_Q\la)^{1/2},\quad \frac{\mu^{\rm pole}_{q,1}}{{\ov\mu}_{\rm gl,1}^{\,\rm pole}}\sim \Bigl (\frac{\mph}{{\hat\mu}_{\Phi}}\Bigr )^{n_2/(\nd-n_1)}\,,\label{(7.12)}
\eq
\bbq
\frac{\langle\Lambda^{\prime\prime}\,\rangle}{{\ov\mu}^{\,\rm pole}_{\rm gl,1}}\sim\Bigl (\frac{\mph}{{\hat\mu}_
{\Phi}}\Bigr )^{n_2/2(3\nd-n_1)}\,, \quad {\hat\mu}_{\Phi}\sim\la\Bigl (\frac{\la}{m_Q} \Bigr )^{\frac{3N_c-N_F-n_1}{2n_2}>0}\,.
\eeq

Therefore, in the range $(\la^3/m_Q)^{1/2}\ll\mph\ll {\hat\mu}_{\Phi}$ the hierarchies look as: ${\ov\mu}_{\rm gl,1}^{\,\rm pole}\gg\langle\Lambda^{\prime\prime}\,\rangle\gg\mu^{\,\rm pole}_{q,1}$ and the quarks $q_1, {\ov q}^1$ are higgsed in the weak coupling region at $\mu={\ov\mu}^{\,\rm pole}_{\rm gl,1}\gg\langle\Lambda^{\prime\prime}\,\rangle$, the overall phase is also $Higgs_1-Hq_2$. But $\mu^{\rm pole}_{q,1}\gg\langle\Lambda^{\prime\prime}\,\rangle\gg{\ov\mu}^{\,\rm pole}_{\rm gl,1}$ at
${\hat\mu}_{\Phi}\ll\mph\ll\mo$, the quarks $q_1, {\ov q}^1$ are too heavy and not higgsed, the overall phase is $Hq_2-Hq_1$.\\

{\bf 1)}\quad $(\la^3/m_Q)^{1/2}\ll\mph\ll {\hat\mu}_{\Phi}$\\

After integrating out all heaviest quarks $q_2, {\ov q}^2$ at $\mu<\mu^{\,\rm pole}_{q,2}$, then higgsed quarks $q_1, {\ov q}^1$ and massive gluons at $\mu<{\ov\mu}^{\,\rm pole}_{\rm gl,1}\ll\mu^{\rm pole}_{q,2}$, and finally $SU(\nd-n_1)$ gluons at $\mu<\langle\lym^{(\rm br2)}\rangle\ll{\ov\mu}^{\,\rm pole}_{\rm gl,1}$, there remain only $N_F^2$ mions $M^i_j$ and $n_1^2$ nions $N_1^1$ (dual pions). The Lagrangian looks as, see \eqref{(4.5)}
\bq
K={\rm Tr}\,\Bigl (\,\frac{1}{\la^2}\,M^\dagger M+2\sqrt{(N_1^1)^\dagger N_1^1}\, \,\Bigr ),\label{(7.13)}
\eq
\bbq
{\cal W}=-(\nd-n_1)\Biggl (\frac{\la^{\bd}\det (M_2^2/\la)}{\det N_1^1} \Biggr )^{1/(\nd-n_1)}
+\frac{1}{\la}{\rm Tr}\,N_1^1\Bigl (M_1^2\frac{1}{M_2^2} M_2^1-M_1^1\Bigr )+{\cal W}_M\,.
\eeq

We obtain from \eqref{(7.13)}:
\bq
\mu^{\rm pole}(N_1^1)\sim\mu^{\rm pole}(M_1^1)\sim (m_Q\la)^{1/2}\,,\quad \mu^{\rm pole}(M_2^2)\sim\la^2/\mph\,,\quad \mu^{\rm pole}(M_1^2)=\mu^{\rm pole}( M^1_2)=0\,.\label{(7.14)}
\eq

{\bf 2)}\quad ${\hat\mu}_{\Phi}\ll\mph\ll\mo$\\

After integrating out all quarks $q_2, {\ov q}^2$ and $q_1, {\ov q}^1$ as heavy ones and then $SU(\nd)$ gluons at $\mu<\langle\lym^{(\rm br2)}\rangle\ll\mu^{\rm pole}_{q,1}\ll\mu^{\rm pole}_{q,2}$, the Lagrangian of $N_F^2$ mions $M$ looks as
\bq
K={\rm Tr}\,\Bigl (\frac{1}{\la^2}\,M^\dagger M\Bigr )\,, \quad {\cal W}=-\nd\Bigl (\frac{\det M}{\la^{\bo}}\Bigr )^{1/\nd}+{\cal W}_M\,.\label{(7.15)}
\eq

From \eqref{(7.15)}, see \eqref{(4.5)} and \eqref{(7.6)},
\bq
\mu^{\rm pole}(M_2^2)\sim\frac{\la^2}{\mph}\,,\quad \mu^{\rm pole}(M_1^1)\sim\frac{\langle M_2\rangle}{\langle M_1\rangle}\,\frac{\la^2}{\mph}\gg\mu^{\rm pole}(M_{22}),\quad \mu^{\rm pole} (M_{12})=\mu^{\rm pole}( M_{12})=0\,.\label{(7.16)}
\eq

\section{Broken $\mathbf{{\cal N}=2}$ SQCD and exercises with Seiberg's duality}

\hspace*{3mm}\, It was considered long ago in \cite{APS,CKM} and reconsidered e.g. recently in \cite{SY1,SY2} ${\cal N}=2$ SQCD direct (electric) theory broken down to ${\cal N}=1$ by the mass term $\mx{\rm Tr}(X^2)$ of the adjoint field $X$, with $SU(N_c)$ gauge group
\footnote{\,
unlike \cite{SY1,SY2}, we do not introduce the extra $U(1)$ gauge field because it is not needed really
}
and $N_c+1<N_F<3N_c/2$ flavors of direct fundamental quarks $Q^i_a\,, a=1\,...\,N_c,\, i=1\,...\,N_F$, and antifundamental quarks $\ov Q^{\,a}_j$. The matter Lagrangian at the scale $\mu\gg\lm$ looks as ( the gluon exponents are implied here and everywhere below in Kahler terms, $\lm$ is the scale factor of the gauge coupling)
\bbq
K=\frac{2}{g^2(\mu)}{\rm Tr}\,(X^\dagger X)+{\rm Tr}\,(Q^\dagger Q+{\ov Q}^{\,\dagger} {\ov Q}),\,\,
\,\, X=T^A X^A,\,\, {\rm Tr}\,(T^{A_1} T^{A_2})=\frac{1}{2}\delta^{A_1 A_2}, \,\, A=1\,...\,N_c^2-1,
\eeq
\bq
{\cal W}_{\rm matter}=\mx{\rm Tr}\,(X^2) +{\rm Tr}\,\Bigl (m\,{\ov Q}\, Q-\sqrt{2}\,{\ov Q} X Q \Bigr )\,.\label{(8.1)}
\eq
We consider in sections 8.1 and 8.2 only the case with hierarchies $0<m\ll\mx\ll\lm$, while $\mx\gg\lm$ in section 8.3\,.

\numberwithin{equation}{subsection}

\subsection{Broken $\mathbf {U(N_F)\ra U(n_1)\times U(n_2)}$ and unbroken $\mathbf{Z}_{\mathbf{2N_c-N_F\geq 2}}$}

\hspace*{3mm} There are br2 - vacua in this theory at $1\leq n_1<\nd=(N_F-N_c),\, N_F=\no+\nt,\,\,\mx\ll\mu_{\rm x,o}$, with the multiplicity $(\nd-n_1)C_{N_F}^{\,\rm n_1}$ (these are the vacua of the baryonic branch in the language of \cite{APS, CKM}, or zero vacua in the language of \cite{SY1,SY2}), with the condensates of direct quarks $Q^i_a, {\ov Q}_j^{\,a}\,\, (\,\la^{3N_c-N_F}=\lm^{2N_c-N_F}\mx^{N_c},\,\,N_c+1<N_F<3N_c/2$\,)\,, see e.g. (4.9)
in \cite{ch5} and \eqref{(2.2)}-\eqref{(2.4)},
\footnote{\,
Here and everywhere below\,: $A\approx B$ has to be understood as an equality neglecting smaller power corrections, and $A\ll B$ has to be understood as $|A|\ll |B|$.
}
\bbq
\langle({\ov Q}Q)_2\rangle\approx\mx m_1,\,\, \langle({\ov Q}Q)_1\rangle\approx\mx m_1\Bigl (\frac{m_1}{\lm}\Bigr)^{\frac{\bb}{{\rm n}_2-N_c}},\,\,
\frac{\langle ({\ov Q}Q)_1\rangle}{\langle({\ov Q}Q)_2\rangle}\sim\Bigl(\frac{m}{\lm}\Bigr)^
{\frac{2N_c-N_F}{{\rm n}_2-N_c}}\ll 1\,, \,\,m_1=\frac{N_c}{N_c-n_2}\,m\,,\,
\eeq
\bq
\frac{\mu_{\rm x,o}}{\mx}=\frac{\la}{\mx}\Bigl (\frac{\la}{m}\Bigr )^{\frac{2 N_c-N_F}{N_c}}=\Bigl (\frac{\lm}{m}\Bigr )^{\frac{2N_c-N_F}{N_c}}\gg 1,\,\, \langle S\rangle=\frac{\langle\Qo\rangle\langle\Qt\rangle}{\mx}\approx\mx m_1^2\Bigl (\frac{m_1}{\lm}\Bigr)^{\frac{\bb}{{\rm n}_2-N_c}},\label{(8.1.1)}
\eq
\bbq
\quad\mx\langle\,{\rm Tr}\,(\sqrt{2}\,X^{\rm adj}_{SU(N_c)})^2\rangle=
(2N_c-N_F)\langle S\rangle+m\langle{\rm Tr\,}{\ov Q} Q\rangle\approx m (\nt\,\mx m_1)\,,
\eeq
\bbq
\langle{\ov Q}_j Q^i\,\rangle\equiv\langle\,\sum_{a=1}^{N_c}{\ov Q}_j^{\,a} Q^i_a\,\rangle=\delta^i_j\langle({\ov Q}Q)_1\rangle,\, i,j=1\,...\,n_1\,;\,\, \langle\,\sum_{a=1}^{N_c}{\ov Q}_j^{\,a} Q^i_a\,\rangle=\delta^i_j\langle({\ov Q}Q)_2\rangle,\, i,j=n_1+1\,...\,N_F\,.
\eeq

Only leading terms are given in \eqref{(8.1.1)}, all nonleading parametrically small power corrections are neglected here and everywhere below.

At $m\ll\mx\ll\lm$, because the field $X$ is higgsed in these br2-vacua as $\langle X^A\rangle_{\rm br2}\sim\lm,\,\, A=\nd+1,\,...,\,N_c$\,, the lower energy gauge group at scales $\mu=\lm/(\rm several)$ is $SU(\nd)\times U(1)^{2N_c-N_F}$ \cite{APS}.

This can be understood as follows. The perturbative NSVZ $\beta$-function of the massless ${\cal N}=2$ SQCD is exactly one loop \cite{NSVZ1,NSVZ2}. Therefore, at $(2N_c-N_F)>0$, the coupling $g^2(\mu)$ is well defined at $\mu\gg\lm$, has a pole at $\mu=\lm$ and becomes negative at $\mu<\lm$. To avoid this unphysical behavior, the field $X$ is necessarily higgsed breaking the $SU(N_c)$ group, with (at least some) components $\langle X^A\rangle\sim\lm$. (It is seen from \eqref{(8.1.1)} that even if any quarks are higgsed, this gives at best the masses $\sim (\mx m)^{1/2}\ll\lm$. Therefore, all particles will remain effectively massless at the scale $\mu\sim\lm$ if all $\langle X^A\rangle\ll\lm$, and this will not help). It is worth to remind also that in the considered ${\cal N}=2$ SQCD with $m\ll\mx\ll\lm$, if there will be no some components $\langle X^A\rangle\sim\lm$, i.e. all $\langle X\rangle\ll\lm$, the nonperturbative instanton contributions will be also non operative at the scale $\lm$ without both the corresponding fermion masses $\sim\lm$ and the infrared cut off $\,\, \rho\lesssim\, 1/\lm$ supplied by some $\langle X^A\rangle\sim\lm$, and the problem with $g^2(\mu<\lm)<0$ will survive. Moreover, for the same reasons, if there is the non-Abelian subgroup at the lower scale $\mu<\lm$, it has to be IR free (or at least conformal).

Besides, there is the residual non-trivial discrete R-symmetry $Z_{2N_c-N_F\geq 2}$ in the theory \eqref{(8.1)} \cite{APS}. The charges of fields and parameters in the superpotential of \eqref{(8.1)} under $Z_{2N_c-N_F}=\exp\{ i\pi/(2N_c-N_F)\}$ transformation are:\, $q_{\lambda}=q_{\theta}=1,\,\,q_X=q_{\rm m}=2,\,\, q_Q=q_{\ov Q}=q_{\lm}=0,\,\, q_{\mx}=-2$. It is broken spontaneously at $m\ll\lm$ (resulting in the multiplicity factor $2N_c-N_F$) in L vacua with $\no=0$ and the unbroken global $U(N_F)$, in Lt vacua with $1\leq\no<N_F/2$, and in special vacua with $\no=\nd$ , while it is unbroken in br2 vacua with $\nt>N_c$ and in S vacua with the unbroken $U(N_F)$, see section 4 in \cite{ch5} and \eqref{(8.1.1)}. From all this it follows that in br2 and S vacua the adjoint field $X$ is higgsed necessarily at the scale $\mu\sim\lm$ as $SU(N_c)\ra SU(\nd)\times U^{(1)}(1)\times U^{2N_c-N_F-1}(1)$, with the largest $X^{\rm adj}_{SU(\bb)}$ part of the form \cite{APS}
\bbq
\langle\sqrt{2} X^{\rm adj}_{SU(\bb)}\rangle\sim\lm\, {\rm diag}\,(\,\underbrace{0,..., 0}_{\nd}\,;\, \underbrace{\omega^0,...,\,\omega^{2N_c-N_F-1}}_{2N_c-N_F}\,\,)\,,
\eeq
where $\omega=\exp\{2\pi i/(2N_c-N_F)\}$ is a $(2N_c-N_F)$-th root of unity. All other components of $\langle X\rangle$ in the remained $SU(\nd)\times U^{(1)}(1)$ part are either $\sim m$, or much smaller $\sim m f(z),\,\,f(z\ra 0)\ra 0,\,\, z=(m/\lm)^{2N_c-N_F}\ll~1$,\, or simply zero, see Introduction and section 2 in \cite{ch7.3} for detailed discussions and results). In particular, the $U^{(1)}(1)$ part of $\langle X\rangle$ looks as \cite{ch7.3}
\bq
\langle\sqrt{2}X_1\rangle=\langle a_1\rangle\, {\rm diag}\,(\,\underbrace{1}_{\nd}\,,\,\underbrace{c_1}
_{\bb}),\quad c_1=-\frac{\nd}{\bb}\,,\quad \langle a_1\rangle=-\frac{\bb}{\nd}\, m\,.\label{(8.1.2)}
\eq

At $m\ll\lm$, for vacua of the baryonic branch with the lower energy ${\cal N}=2\,\,\, SU(\nd=N_F-N_c)\times U^{2N_c-N_F}(1)$ gauge group at $\mu<\lm$ \cite{APS}, it was first proposed naturally by M.Shifman and A.Yung in \cite{SY1} that all light charged particles of ${\cal N}=2\,\, SU(\nd)$ are the original pure electric particles, as they have not received masses $\sim\lm$ directly from higgsed $\langle X^{\rm adj}_{SU(N_c)}\rangle$ ( this is forbidden in these vacua by the unbroken non-trivial $Z_{\bb\geq 2}$ symmetry \cite{APS}).

But they changed later their mind to the opposite and, extrapolating freely by analogy their previous results for $r=N_c=3,\, N_F=5,\, \no=2$ very special $U(N_c=3)$ vacua in \cite{SY3} (see also \cite{SY4}) to all br2 and very special $U(N_c)$ vacua with $\no=\nd,\,\nt=N_c,\,\langle S\rangle=0$, claimed in their subsequent paper \cite{SY2} that this $SU(\nd)$ group is not pure electric but dyonic, i.e. all its charged particles have nonzero both electric and magnetic charges. This means that at $m\ll\lm$ all original pure electrically charged particles of $SU(\nd)$ have received large masses $\sim\lm$, but now not directly from $\langle X^{\rm adj}_{SU(N_c)}\rangle\sim\lm$ (because this is forbidden by the non-trivial unbroken $Z_{\bb\geq 2}$ symmetry), but from some mysterious "outside" sources and decoupled at $\mu<\lm$, while the same gauge group $SU(\nd)$ of light composite dyonic solitons was formed. Unfortunately, it was overlooked in \cite{SY2} that, in distinction with their example with $U(N_c=3),\,\,N_F=5,\,\,\no=2$ in \cite{SY3} with the trivial $Z_{2N_c-N_F=1}$ symmetry giving no restrictions on the form of $\langle X\rangle$, the appearance in the superpotential at $\mu\sim\lm$ of e.g. mass terms of original $SU(\nd)$ quarks like $\sim\lm{\rm Tr}_{\,\nd}
({\otQ}\tQ)$ is forbidden in vacua with the non-trivial unbroken $Z_{2N_c-N_F\geq 2}$ symmetry, independently of whether these terms originated directly from $\langle X\rangle$ or from unrecognized "outside". Therefore, we proceed below in this paper considering all original charged particles of $SU(\nd)$ as they were, i.e. pure electrical light particles with masses $\ll\lm$ (see Introduction in \cite{ch7.3} for additional more detailed discussions of this point).

Besides, due to the strong coupling, i.e. $a(\mu\sim\lm)=N_c g^2(\mu\sim\lm)/8\pi^2\sim 1$, perturbative loops and non-perturbative instanton contributions, and very special properties of enhanced ${\cal N}=2$ SUSY, $\bb$ light composite dyons ${\ov D}_{\rm n}, D_{\rm n}$ (this number $\bb$ is required by the unbroken $Z_{\bb}\geq 2$ discrete symmetry which acts interchanging them with each other) are formed in the $U^{(1)}(1)\times U^{\bb-1}(1)$ sector (see section 2.1 in \cite{ch7.3} for detailed discussions of these dyons). These dyons are massless at $\mx\ra 0$ and are higgsed as $\langle D_{\rm n}\rangle=\langle {\ov D}_{\rm n}\rangle\sim (\mx\lm)^{1/2}$ at $\mx\neq 0$. They have, in particular, $SU(\bb)$ adjoint magnetic charges and nonzero electric $SU(N_c)$ baryon and $U^{(1)}(1)$ charges, this last is a source of the ${\cal N}=2\,\, U^{(1)}(1)$ photon multiplet (with its scalar superpartner $a_1$, see \eqref{(8.1.2)}\,).

Therefore, all these dyons are mutually local with respect to all charged original electric particles in the $SU(\nd)$ sector, and higgsing of these dyons does not result in confinement of these particles. On the other hand, e.g. all original pure electrically $SU(\bb)$ charged particles with largest masses $\sim\lm$ become confined at $\mx\neq 0$ and form a large number of hadrons with masses $\sim\lm$. But this confinement is weak, in the sense that the tension of the confining string is much smaller than particle masses, $\sigma^{1/2}\sim (\mx\lm)^{1/2}\ll\lm$.

At $\mx\ll\lm$, because all these $\bb$ dyons are higgsed, all $\bb$ Abelian photon mutiplets $U^{\bb}(1)$  acquire masses $\sim (\mx\lm)^{1/2}$ and, together with dyons, form $\bb$ long ${\cal N}=2$ photon multiplets. There are no massless particles in this sector at $\mx\neq 0$. But because the quarks $\tQ^i_a, {\otQ}^{\,a}_j$ from $SU(\nd)$ are also coupled with $a_1$ as $\delta{\cal W}_{a_1}=(m-a_1){\rm Tr}(\otQ \tQ)$, see \eqref{(8.1.2)}, this gives the additional contribution to their masses, $m\ra \tm= m-\langle a_1\rangle=m N_c/\nd$. Besides, $\mx\ra \wmu=-\mx$ in the adjoint sector of $SU(\nd)$, see section 2.1 in \cite{ch7.3} for all additional details.

\subsubsection{The first pair of Seiberg's dual theories}

The lower energy ${\cal N}=2$ theory in the $SU(\nd)$ sector at scales $\mu\ll (\mx\lm)^{1/2}$ contains the light adjoint fields $\textsf{x}^B,\, B=1\,...\,\nd^{\,2}-1$  (scalar partners of remained light $SU(\nd)$\, electric gluons) and $N_F$ flavors of remained light (i.e. with masses $\ll\lm$) original electric quarks $Q^i_b$ and ${\ov Q}^{\,b}_i,\, i=1\,...\,N_F,\, b=1\,...\,\nd$. These light quarks will be denoted below as ${\textsf{Q}}^i_b,\,\otQ^{\,b}_j,\, b=1\,...\,\nd,\,$ to distinguish them and their condensates $\langle\otQ_j\tQ^i\rangle$ {\it summed over $\nd$ unbroken colors} from condensates $\langle {\ov Q}_j Q^i\rangle$ of $Q^i_a,\, {\ov Q}_j^{\,a}$ in (8.1.1) {summed over all $N_c$ colors}.

This theory is IR free in the range of scales $m\ll\mu\ll\lm$ at $N_c<N_F<2N_c/3$, and the scale factor of the $SU(\nd)$ gauge coupling at scales $\mx\ll\mu\ll\lm$ is $\Lambda_{SU(\nd)}= - \Lambda_2$, see (2.1.15) in \cite{ch7.3}.

Therefore the Lagrangian of these light original electric fields with masses $\lesssim\mx\ll\lm$ (see below) looks at the scale $\mu=(\rm several)\mx$ as
\bbq
K=\frac{2}{g^2(\mu)}{\rm Tr}\,(\textsf{x}^2)+{\rm Tr}\,(\tQ^\dagger\tQ + {\otQ}^\dagger\otQ)\,,\,\, \textsf{x}=T^B \textsf{x}^B\,,\,\, B=1,\,...,\,\nd^{\,2}-1\,,\quad {\rm Tr}\, (T^{B_1} T^{B_2})=\frac{1}{2}\,\delta^{B_1 B_2}\,,
\eeq
\bq
W_{\rm matter}=\wmu\,{\rm Tr}\,({\textsf{x}}^{\,2} )+{\rm Tr}\,\Bigl (\tm\,\otQ \tQ-\otQ \sqrt{2}\,\textsf{x}\, \tQ \Bigr )\,,\,\,  \tm=\frac{N_c}{\nd}\,m\,,\quad \wmu=-\mx\,, \quad m\ll\mx\ll\lm\,.\label{(8.1.3)}
\eq

After integrating out in \eqref{(8.1.3)} all fields $\textsf{x}^B$ as heavy ones at $\mu<\mu^{\rm pole}(\rm x)={\it g}^2 (\mu=\mu^{\rm pole}(\rm x))\mx$, the new scale factor of the $SU(\nd)$ gauge coupling is
\bq
{\lt}^{\bd}=\Lambda_{SU(\nd)}^{\rm {\ov b}_2}\,\wmu^{\,\nd}\,,\quad\Lambda_{SU(\nd)}=
-\lm\,,\quad \wmu=-\mx\,,\quad \lt\gg\Lambda_2\,,\label{(8.1.4)}
\eq
\bbq
 \rm{\ov b}_2=(2\nd-N_F)<0\,,\quad \bd=(3\nd-N_F)<0\,,
\eeq
and the Lagrangian at the scale $\mu=\mx/(\rm several)$ has the form (all logarithmic factors due to the ${\cal N}=1$ RG evolution at $\mu<\mx$ are ignored here and everywhere below in the text for simplicity)
\bq
K={\rm Tr}\,(\tQ^\dagger \tQ+{\otQ}^\dagger\, {\otQ}),\quad W_{\rm matter}=\tm{\,\rm Tr}\, (\,\otQ \tQ\,)+
\frac{1}{2\mx}\Bigl (\,{\rm Tr\,}(\otQ \tQ)^2 -\frac{1}{\nd}({\rm Tr\,}\otQ\tQ)^2\,\Bigr )\,.\label{(8.1.5)}
\eq

The above br2 -vacua \eqref{(8.1.1)} of the theory \eqref{(8.1)} with $N_c$ colors, $N_c+1<N_F<3N_c/2$ and with $\langle({\ov Q}Q)_2\rangle_{\rm br2}\sim\mx m\gg\langle ({\ov Q}Q)_1\rangle_{\rm br2},\, 1\leq n_1<\nd\,,\, \mx\ll\mu_{\rm x,o}$, will correspond now to the br1 -vacua of the theory \eqref{(8.1.5)} with $\nd$ colors, $N_F>3\nd\,,\, \mx\gg {\tilde\mu}_{\rm x,o}$ and with (\, see \eqref{(2.4)},\eqref{(2.14)},\eqref{(8.1.4)}, and \eqref{(8.1.6)},\eqref{(8.1.8)} below),
\be
\langle\tQo\rangle\sim \mx m\gg\langle\tQt\rangle\,,\quad {\tilde\mu}_{\rm x,o}=\lt(\frac{\lt}{m})^{(2\nd-N_F)/\nd}\,,\quad \frac{{\tilde\mu}_{\rm x,o}}{\mx}\sim (\frac{m}{\lm})^{(2N_c-N_F)/\nd}\ll 1\,,\label{(8.1.6)}
\ee
\bbq
\langle\,\sum_{b=1}^{\nd}{\otQ}^{\,b}_j\tQ^i_b\,\rangle=\delta^i_j\langle\tQo\rangle ,\,\,i,j=1\,...\,\no\,,\,\, \langle\,\sum_{b=1}^{\nd}{\otQ}^{\,b}_j\tQ^i_b\,\rangle =\delta^i_j\langle\tQt\rangle,\,\,i,j=n_1+1\,...\,N_F\,.
\eeq
\hspace*{3mm} {\bf A)}\, To calculate the spectrum of masses smaller than $\mx$ in the theory \eqref{(8.1.5)} it will be convenient to introduce $N^2_F$ additional colorless but flavored fion fields $\Phi^j_i$ (these will be dynamically irrelevant finally and unobservable as real particles at energies $\mu<\mx$, see below). Therefore, the direct ${\cal N}=1$ $\Phi$ - theory is defined as follows. It has the same quark and $SU(\nd)$ gluon fields as in \eqref{(8.1.5)}, the scale factor of the $SU(\nd)$ gauge coupling is $\lt\gg\lm$ \eqref{(8.1.4)}, and there are $N_F^2$ fion fields $\Phi^j_i$ in addition. It is IR free in the range of scales $m\ll\mu\ll\lt$, and we will deal with its br1 - vacua with the multiplicity $(\nd-n_1)C_{N_F}^{\,\rm n_1}$ at $\mx\gg {\tilde\mu}_{\rm x,o}$, see \eqref{(8.1.7)} and \eqref{(8.1.8)} below. Its Lagrangian at the scale $\mu=\lt$ looks as
\bq
K={\rm Tr}\,(\Phi^\dagger\Phi)+{\rm Tr}\,({\tQ}^\dagger \tQ+{\otQ}^\dagger {\otQ})\,,\quad W_{\rm matter}=
{\cal W}_{\Phi}+{\rm Tr}\,\Bigl (\,{\otQ}(\tm- \Phi) \tQ \,\Bigr )\,,\label{(8.1.7)}
\eq
\bbq
{\cal W}_{\Phi}=\frac{\wmu}{2}\,\Bigl (\, {\rm Tr}\, (\Phi^2)-\frac{1}{N_c}(\, {\rm Tr}\,\Phi)^2\Bigr )\,,\quad \tm=\frac{N_c}{\nd}\,m\,,\quad \wmu=-\mx\,.
\eeq

In what follows the parameter $\mx$ can be varied in the range $m\ll\mx\ll\Lambda_2$ while $m$ and $\Lambda_2$ will stay intact. Then, with the change of notations: $N_c\ra \nd,\, \mph\ra \wmu,\, \la\ra\lt,\, m_Q\ra \tm$, we can use the results from \eqref{(2.4)},\eqref{(2.14)} for the quark condensates, $1\leq
n_1<\nd,\, n_2>N_c$,
\bq
\langle\tQo\rangle= [\,\frac{\nd\,\wmu \tm}{\nd-\no}=\wmu m_2=\mx m_1\,]+\frac{N_c-\no}
{\nd-\no}\langle\tQt\rangle\approx\mx m_1\approx\langle\Qt\rangle\,,\label{(8.1.8)}
\eq
\bbq
\langle\tQt\rangle\approx\wmu m_2\Bigl (\frac{m_1}{\lm}\Bigr)^{\frac{2N_c-N_F}{\nt-N_c}},\quad
\langle\lym^{\rm (br1)}\rangle^3\equiv\langle \tS\rangle=\frac{\langle\tQo\rangle\langle\tQt\rangle}
{\wmu}\approx\wmu m^2_2\Bigl (\frac{m_1}{\lm}\Bigr)^{\frac{2N_c-N_F}{\nt-N_c}}\,,
\eeq
\bbq
\frac{\langle\tQt\rangle}{\langle\tQo\rangle}\sim\Bigl (\frac{m}{\Lambda_2}\Bigr )^{\frac{2N_c-N_F}{{\rm n}_2-N_c}}\ll 1\,,\quad m_2=\frac{N_c}{\nt-N_c}\,m=-m_1\,,\quad 1\leq n_1<\nd\,,\quad n_2>N_c\,.
\eeq

We obtain from \eqref{(8.1.7)},\eqref{(8.1.8)} for the potentially most important quark masses and gluon masses due to possible higgsing of quarks (remind that all logarithmic effects of the ${\cal N}=1$ RG evolution are ignored for simplicity)\,:
\bq
\mu_{\rm gl,2}\ll\mu_{\rm gl,1}\sim\langle\tQo\rangle_{\rm br1}^{1/2}\sim (m \mx)^{1/2}\ll\Lambda_2\,,\label{(8.1.9)}
\eq
\bbq
\Bigl (m_{\tQ,1}^{\rm pole}\Bigr )_{\rm br1}\ll \Bigl (m_{\tQ,2}^{\rm pole}\Bigr )_{\rm br1}\sim\langle m_{\tQ,2}^{\rm tot}\rangle_{\rm br1}=\langle \tm -\Phi_2\rangle_{\rm br1} =\frac{\langle\tQo\rangle_{\rm br1}}{\wmu}\sim m\ll\mu_{\rm gl,1}\,,
\eeq
so that the overall phase is ${\rm Higgs}_1-{\rm H}\tQ_2$ (${\rm H}\tQ$=heavy quark). Proceeding now as in previous sections, we can calculate the mass spectrum. Not going into any details (see \cite{ch7}\,) we give here the results only (with logarithmic accuracy).\\

a) The mixing of $\no^2$ pions $\Pi_{1}^{1^\prime}\ra (\otQ_1 \tQ^{1^\prime})$ (originating from higgsing of $\tQ^1,\,\otQ_1$ quarks) and $2\no\nt$ hybrids $\Pi_{1}^{2},\,\Pi_{2}^{1}$ (these are in essence the quarks $\tQ^2,\,\otQ_2$ with broken colors) with fions $\Phi$ is parametrically small and neglected. The masses of all $N_F^2$ fions are the largest ones, $\mu^{\rm pole}(\Phi)\sim\mx$. Therefore, they are dynamically irrelevant and unobservable as real particles at scales $\mu<\mx$ we are interested in.

b) There are $n_1 (2\nd-n_1)$ massive electric gluons (and their ${\cal N}=1$ fermion superpartners) with masses $\mu_{\rm gl,1}\sim (m \mx)^{1/2}$\,.

c) There is a large number of hadrons made of the weakly interacting and weakly confined quarks $\,\otQ_2^{\,\prime},\,(\tQ^2)^{\,\prime}$ with $\nd-\no$ unbroken colors and with masses $m_{\tQ,2}^{\rm pole}\sim m$ (the tension of the confining string originating from the unbroken non-Abelian group $SU(\nd-n_1)$ is much smaller, $\sqrt{\sigma}\sim\langle\lym^{(\rm br1)}\rangle=\langle \tS\rangle^{1/3}_{(\rm br1)}\ll m$, see \eqref{(8.1.8)}\,).

d) The masses of $n_1^2$ pions $\Pi_{1}^{1}$ are $\mu^{\rm pole}(\Pi_{1}^{1})\sim m$ (the main contribution to $\mu^{\rm pole}(\Pi_{1}^{1})$ originates from the term $\sim (\Pi_{1}^{1})^2/\mx$ in the superpotential appearing after integrating out heavier fions $\Phi^1_{1}$).

e) There is a large number of strongly coupled gluonia made of $SU(\nd-n_1)$ gluons with the mass scale $\langle\lym^{(\rm br1)}\rangle=\langle \tS\rangle_{\rm br1}^{1/3}\sim [\,m\langle\tQt\rangle_{\rm br1}]^{1/3}$, see \eqref{(8.1.8)}.
\footnote{\,
The case ${\rm n}_1=\nd-1$ is different. The whole $SU(\nd)$ group is higgsed at the scale $\sim (m\mx)^{1/2}$, there is no confinement and no gluonia with masses $\sim \langle\lym^{(\rm br1)}\rangle$ from the unbroken non-Abelian ${\cal N}=1\,\, SU(\nd-\no)$ SYM with the scale factor $\langle\lym^{(\rm br1)}\rangle\ll (m\mx)^{1/2}$ of its gauge coupling, originated after the quarks $\,\otQ_2^{\,\prime},\,(\tQ^2)^{\,\prime}$ with $\nd-\no$ colors and masses $m_{\tQ,2}^{\rm pole}\sim m$ decoupled as heavy at scales $\mu<m$. The non-perturbative contribution to the low-energy superpotential originates at ${\rm n}_1=\nd-1$ directly from the instanton operating in this case at the higher scale $\sim\mu_{\rm gl,1}\sim (m\mx)^{1/2}$ \cite{ADS}.
}

f) $2n_1 n_2$ hybrid pions $\Pi_{1}^{2}\ra ({\otQ}_1 \tQ^2),\, \Pi_{2}^{1}\ra ({\otQ}_2 \tQ^1)$ are the Nambu-Goldstone particles of spontaneously broken global flavor symmetry $U(N_F)\ra U(\no)\times U(\nt)$ and are massless (in essence, these are quarks $\tQ^2,\,{\otQ}_2$ with broken $\no$ colors and $\nt$ flavors).

The overall hierarchies of nonzero masses look as
\bbq
\langle\lym^{(\rm br1)}\rangle\ll\mu^{\rm pole}(\Pi_{1}^{1})\sim m_{\tQ,2}^{\rm pole}\sim m \ll\mu_{\rm gl,1}\sim (m \mx)^{1/2} \ll\mu^{\rm pole}(\Phi)\sim\mx\ll\Lambda_2\ll\lt\,.
\eeq
\vspace*{2mm}

{\bf B)}\, Now consider the ${\cal N}=1\,\, d\Phi$ - theory which is the literal Seiberg dual to the direct $\Phi$ - theory \eqref{(8.1.7)}. According to the Seiberg rules \cite{S2,ISS}, this is the theory with $SU(N_c=N_F-\nd)$ dual colors, the scale factor of the dual gauge coupling is $\Lambda_\tq=C_o\lt,\, C_o=O(1)$, with $N_c+1<N_F<3N_c/2$ flavors of quarks ${\tq}_{ i}$ and ${\otq}^{\, j}$ (dual to $\tQ^{ i}, \otQ_{ j}$),\, $N_F^2$ fions $\Phi^j_i$ and with additional $N_F^2$ elementary mions $\tp^{i}_{j}\ra (\otQ_j\tQ^i)$.

According to \cite{ch1} (see section 7 therein) it will be in the strong coupling regime with the dual $SU(N_c)$ gauge coupling ${\ov a}(\mu\sim\lt)\sim 1$ and ${\ov a}(\mu\ll\lt)\sim (|\lt|/\mu)^{\,\nu^{(+)}\,>\,0}\gg 1$. The dual Lagrangian at the scale $\mu=|\lt|$ is
\bbq
K={\rm Tr}\,(\Phi^\dagger\Phi)+\frac{1}{\lt^2}{\rm Tr}\,(\tp^\dagger \tp)+{\rm Tr}\,({\tq}^\dagger \tq+
{\otq}^\dagger {\otq}),
\eeq
\bq
W_{\rm matter}={\cal W}_{\Phi}+ {\rm Tr}\,\,(\tm-\Phi)\tp\,-\frac{1}{\lt}{\rm Tr}\,(\,{\otq}\, \tp \tq \,)\,,
\quad {\cal W}_{\Phi}=\frac{\wmu}{2}\Bigl (\,{\rm Tr}\,(\Phi^2)-\frac{1}{N_c}({\rm Tr}\,\Phi)^2\Bigr ).\label{(8.1.10)}
\eq

Proceeding as in \cite{ch3,ch7} and not going into details we give below mainly the results only (see \cite{ch7} for all details). The potentially important masses of dual quarks and gluons (due to possible higgsing of quarks) look here as follows, see \eqref{(8.1.8)} and Appendix,
\bq
\mu_{\tq,1}\equiv\mu_{\tq,1}(\mu=|\lt|)=\frac{\langle\tp_1\rangle_{\rm br1}=\langle\tQo\rangle_{\rm br1}}{\lt}\,,\quad \mu_{\tq,1}^{\rm pole}=\frac{\mu_{\tq,1}}{z^{(+)}_\tq(\lt,\mu_{\tq,1}^{\rm pole})}\sim\Lambda_2\Bigl (\frac{m}{\Lambda_2}\Bigr )^{\nd/N_c}\,,\label{(8.1.11)}
\eq
\bbq
{\ov\mu}^{\,2}_{\rm gl,1}\ll {\ov\mu}^{\,2}_{\rm gl,2}\sim \rho^{(+)}\rho^{(-)}\Bigl[\langle\dqt\rangle_{\rm br1}=\frac{\langle\tQo\rangle_{\rm br1}\lt}{\wmu}\sim m\lt\Bigr ],\quad\quad {\ov\mu}_
{\rm gl,2}\sim m\ll\mu_{\tq,1}^{\rm pole}\,,
\eeq
\bbq
\rho^{(+)}=\Bigl [{\ov a}^{\,(+)}(\mu_{\tq,1}^{\rm pole})=\Bigl (\frac{|\lt|}{\mu_{\tq,1}^{\rm pole}}\Bigr )^{\nu^{(+)}}\Bigr ]\Bigl [z^{(+)}_{\tq}(\lt,\mu_{\tq,1}^{\rm pole})=\Bigl (\frac{\mu_{\tq,1}^{\rm pole}}{|\lt|} \Bigr )^{\gamma_\tq^{(+)}}\Bigr ]=\frac{\mu_{\tq,1}^{\rm pole}}{|\lt|}\,,
\eeq
\bbq
\rho^{(-)}=\Bigl (\frac{\mu_{\tq,1}^{\rm pole}}{{\ov\mu}_{\rm gl,2}}\Bigr )^{\nu^{(-)}}\,\Bigl [z^{(-)}_{\tq}(\mu_{\tq,1}^{\rm pole},{\ov\mu}_{\rm gl,2})=\Bigl (\frac{{\ov\mu}_{\rm gl,2}}{\mu_{\tq,1}^{\rm pole}} \Bigr )^{\gamma_\tq^{(-)}}\Bigr ]=\frac{{\ov\mu}_{\rm gl,2}}{\mu_{\tq,1}^{\rm pole}}\,,\quad \rho^{(+)}\rho^{(-)}=\frac{{\ov\mu}_{\rm gl,2}}{|\lt|}\,,
\eeq
\bbq
\mu_{\tq,2}^{\rm pole}\sim\frac{\langle\tp_2\rangle_{\rm br1}=\langle\tQt\rangle_{\rm br1}}{\lt}\frac{1}
{z^{(+)}_\tq(\lt,\mu_{\tq,1}^{\rm pole})z^{(-)}_\tq(\mu_{\tq,1}^{\rm pole}, \mu_{\tq,2}^{\rm pole})}\sim\, (\rm several)\,{\ov\mu}_{\rm gl,2}\ll\mu_{\tq,1}^{\rm pole}\,\,,
\eeq
where the anomalous dimensions are (see section 7 in \cite{ch1} and Appendix)
\bq
\gamma_\tq^{(+)}=\frac{2N_c-N_F}{N_F-N_c}>\,1\,,\quad \gamma_{\tp}^{(+)}=-2\gamma_\tq^{(+)}<\, -2\,,
\quad \nu^{(+)}=\frac{3N_c-2N_F}{N_F-N_c}>0\,,\label{(8.1.12)}
\eq
\bbq
\gamma_\tq^{(-)}=\frac{2N_c-n_2}{n_2-N_c}>\gamma_\tq^{(+)}\,,\quad\gamma_{\tp}^{(-)}=-2\gamma_\tq^{(-)}
<\gamma_{\tp}^{(+)},,\quad \nu^{(-)}=\frac{3N_c-2 n_2}{n_2-N_c}>\nu^{(+)}\,,\quad \gamma_{\Phi}=0\,.
\eeq

Therefore, the overall phase in these $(\nd-n_1)C_{N_F}^{\,\rm n_1}$ br1 - vacua  of the dual theory \eqref{(8.1.10)} is ${\rm H}\tq_1-{\rm H}\tq_2$ (heavy unhiggsed but confined quarks) and the mass spectrum looks as follows.

a) The mixing of all $N_F^2$ mions $M$ and $N_F^2$ fions $\Phi$ is small at scales $\mu<\mx$ and neglected. The masses of all $N_F^2$ fions, $\mu^{\rm pole}(\Phi)\sim\mx$, will be the largest ones provided that $\mx\gg\mu_{\tq,1}^{\rm pole}$ \eqref{(8.1.11)} which is always possible to adjust. So, all fions are dynamically irrelevant at scales $\mu<\mx$ and all other masses will be smaller than $\mx$. Therefore, all $N_F^2$ fion fields $\Phi$ can be integrated out as heavy ones at the scale $\mu<\mx$ we are interested in, and the Lagrangian at the scale $\mu=\mx$ will be
\bq
K=z^{(+)}_{M}(\lt,\mx)\,{\rm Tr}\,\Bigl (\frac{\tp^\dagger \tp}{\lt^2} \Bigr)+z^{(+)}_{\tq}(\lt,\mx)\,{\rm Tr}\,({\tq}^\dagger \tq+{\otq}^\dagger {\otq})\,,\label{(8.1.13)}
\eq
\bbq
z^{(+)}_{M}(\lt,\mx)=\Bigl (\frac{|\lt|}{\mx}\Bigr )^{2 \gamma^{(+)}_{\tq}}\gg 1\,,\quad z^{(+)}_{\tq}(\lt,\mx)=\Bigl (\frac{\mx}{|\lt|}\Bigr )^{\gamma^{(+)}_{\tq}}\ll 1\,,
\eeq
\bbq
{\cal W}_{\rm matter}={\widetilde{\cal W}}_{\tp}-\frac{1}{\lt}
{\rm Tr}\,(\,{\otq}\, \tp \tq \,)\,,\quad {\widetilde{\cal W}}_{\tp}=\tm\,{\rm Tr}\,\tp+\frac{1}{2\mx}\Bigl (\,{\rm Tr}\,(\tp^2)-\frac{1}{\nd}({\rm Tr}\,\tp)^2\Bigr ).
\eeq

b)  There is a large number of hadrons made of weakly confined $\tq_1, {\otq}^1$ quarks (and hybrids made of $\tq_1$ and ${\otq}^2$ quarks), the quark masses are $\mu_{\tq,1}^{\rm pole}\sim\Lambda_2(m/\Lambda_2)^{\nd/N_c}$, see \eqref{(8.1.11)}.

c) Further, integrating out in \eqref{(8.1.13)} ${\ov q}^1, q_1$ quarks as heavy ones at $\mu<\mu_{\tq,1}^{\rm pole}$, the matter Lagrangian at $\mu=\mu_{\tq,1}^{\rm pole}$ looks as
\bbq
K=z^{(+)}_{M}(\lt,\mu_{\tq,1}^{\rm pole})\,{\rm Tr}\,\Bigl (\frac{\tp^{\dagger} \tp}{\lt^2} \Bigr)+z^{(+)}_{\tq}(\lt,\mu_{\tq,1}^{\rm pole})\,{\rm Tr}\,(\tq^\dagger_2 \tq_2),\,\,
\eeq
\bq
{\cal W}_{\rm matter}={\widetilde{\cal W}}_{\tp}-{\cal W}_{\tq}\,,\quad {\cal W}_{\tq}={\rm Tr}\,\Biggl [\otq^{\,2}\frac{\widehat M^{\,2}_{\,2}}{\lt}\,\tq_2\Biggr ]\,,\quad
{\widehat M^{\,2}_{\,2}}=M^2_2-M^1_2\frac{1}{M^1_1}M^2_1\,.\label{(8.1.14)}
\eq

Evolving further down in energy and integrating out ${\ov q}^2, q_2$ quarks as heavy ones at $\mu<\mu_{\tq,2}^{\rm pole}\ll\mu_{\tq,1}^{\rm pole}$, the scale factor of remained $SU(N_c)$ SYM is determined from the matching, see \eqref{(8.1.11)},\eqref{(8.1.12)},\eqref{(8.1.8)},
\bbq
{\ov a}(\mu=\mu_{\tq,2}^{\rm pole})=\Bigl (\frac{|\lt|}{\mu_{\tq,1}^{\rm pole}}\Bigr )^{\nu^{(+)}}\Bigl (\frac{\mu_{\tq,1}^{\rm pole}}{\mu_{\tq,2}^{\rm pole}}\Bigr )^{\nu^{(-)}}={\ov a}_{\,YM}(\mu=\mu_{\tq,2}^{\rm pole})=\Bigl (\frac{\mu_{\tq,2}^{\rm pole}}{\lambda_{YM}}\Bigr )^3\,\,\ra\,\, \lambda_{YM}=\langle\lym^{\rm (br1)}\rangle\,,
\eeq
as it should be.

There is a large number of hadrons made of weakly confined $\tq_2, {\otq}^{\,2}$ quarks, the quark masses are $\mu_{\tq,2}^{\rm pole}\sim m\ll\mu_{\tq,1}^{\rm pole}$ (the tension of the confining string originating from $SU(N_c)$ SYM with the scale factor $\langle\lym^{(\rm br1)}\rangle$ \eqref{(8.1.8)} of its gauge coupling is much smaller, $\sqrt{\sigma}\sim\langle\lym^{(\rm br1)}\rangle\ll m)$.

d) And finally, integrating out all $SU(N_c)$ gluons at $\mu<\langle\lym^{(\rm br1)}\rangle\ll
\mu_{\tq,2}^{\rm pole}\sim m$ through the VY-procedure \cite{VY,TVY}, the lower energy Lagrangian of mions $M$ looks as, see \eqref{(8.1.12)}
\bq
K_{M}=\frac{z_{M}^{(+)}}{|\lt|^2}\,\,{\rm Tr}\,\Biggl [\,({M}_1^1)^\dagger {M}_1^1+\Bigl ( ({M}_2^1)^\dagger {M}_2^1+({M}_1^2)^\dagger {M}_1^2\Bigr )+z_{M}^{(-)}\,({M}_2^2)^\dagger {M}_2^2 \Biggr ]\,, \label{(8.1.15)}
\eq
\bbq
z_{M}^{(+)}=z_{M}^{(+)}(\lt,\mu_{\tq,1}^{\rm pole})=\Bigl (\frac{|\lt|}{\mu_{\tq,1}^{\rm pole}}\Bigr)^
{2\gamma_\tq^{(+)}}\gg 1\,,\quad z_{M}^{(-)}=z_{M}^{(-)}(\mu_{\tq,1}^{\rm pole},\mu_{\tq,2}^{\rm pole})=\Bigl(\frac{\mu_{\tq,1}^{\rm pole}}{\mu_{\tq,2}^{\rm pole}}\Bigr )^{2\gamma_\tq^{(-)}}\gg 1\,,
\eeq
\bbq
{\cal W}={\widetilde{\cal W}}_{\tp}+{\cal W}_{\rm non-pert},\quad {\cal W}_{\rm non-pert}=\,-\,N_c\Bigl
(\lt^{\rm 3N_c-N_F}\det \frac{M^1_1}{\lt}\det\frac{\widehat M^{\,2}_{\,2}}{\lt}\Bigr )^{1/N_c}\,.
\eeq

There is a large number of gluonia made of $SU(N_c)$ dual gluons with the mass scale $\langle\lym^{(\rm br1)}\rangle=\langle \tS\rangle_{\rm br1}^{1/3}\sim [\,m\langle\tp_2\rangle_{\rm br1}=m\langle\tQt\rangle_{\rm br1}\,]^{1/3}\ll m$, see \eqref{(8.1.8)}.\\

e) From \eqref{(8.1.15)}, the masses of $n_1^2$ mions $\tp_1^1$ and $n_2^2$ mions $\tp_2^2$ are
\bq
\mu^{\rm pole}({M}_1^1)\sim\frac{\lt^2}{z^{(+)}_{M}(\lt,\mu_{\tq,1}^{\rm pole})\,\mx}\sim
\mx\Bigl (\frac{m}{\lm} \Bigr )^{\frac{2(2N_c-N_F)}{N_c}}\,,\quad
z^{(\pm)}_{M}(\mu_1,\mu_2)=\Bigl (\frac{\mu_1}{\mu_2}\Bigr )^{2\gamma_\tq^{(\pm)}}\,,\label{(8.1.16)}
\eq
\bbq
\frac{\mu^{\rm pole}({M}_1^1)}{\mu_{q,1}^{\rm pole}}\ll\frac{\mu^{\rm pole}({M}_1^1)}{\mu_{q,2}^{\rm pole}}\sim\frac{\mx}{\lm}\Bigl (\frac{m}{\lm}\Bigr )^{\frac{3N_c-2N_F}{N_c}}\ll 1\,,
\eeq
\bq
\mu^{\rm pole}({M}_2^2)\sim\frac{\langle \tS\rangle_{\rm br1} \lt^2}{\langle{M}_2 \rangle_{\rm br1}^2}\,\frac{1}{z^{(+)}_{M}(\lt,\mu_{\tq,1}^{\rm pole})z^{(-)}_{M}(\mu_{\tq,1}^{\rm pole},\mu_{\tq,2}^{\rm pole})}\sim\frac{\langle\lym^{(\rm br1)}\rangle^3}{m^2}\sim\mx\Bigl (\frac{m}{\lm}\Bigr )^{\frac{2N_c-N_F}{n_2-N_c}}\,,\label{(8.1.17)}
\eq
\bbq
\frac{\mu^{\rm pole}({M}_2^2)}{\langle\lym^{(\rm br1)}\rangle}\sim\frac{\langle\lym^{(\rm br1)}\rangle^2}{m^2}\ll 1\,.
\eeq
In \eqref{(8.1.16)},\eqref{(8.1.17)}, the main contribution to $\mu^{\rm pole}({M}_1^1)$ originates from the term $\widetilde{\cal W}_{\tp}\sim ({M})^2/\mx$ \eqref{(8.1.13)} in \eqref{(8.1.15)}\,), while $\mu^{\rm pole}({M}_2^2)$ is dominated by the contribution from ${\cal W}_{\rm non-pert}$.

f) $2n_1 n_2$ hybrid mions $\tp_1^2,\, \tp_2^1$ are the Nambu-Goldstone particles and are massless.

The overall hierarchies of nonzero masses look as
\bbq
\mu (\tp_2^2)\ll \langle\lym^{(\rm br1)}\rangle\ll \mu_{\tq,2}^{\rm pole}\sim m\ll\mu_{\tq,1}^{\rm pole}\ll\mu(\Phi)\sim\mx\ll\Lambda_2\ll\lt\,,\quad \mu (\tp_2^2)\ll\mu (\tp_1^1)\ll
\mu_{\tq,2}^{\rm pole}\,.
\eeq

Comparing the mass spectra of the direct $\Phi$ \eqref{(8.1.7)} and dual $d\Phi$ \eqref{(8.1.10)} theories it is seen that they are parametrically different.\\

\subsubsection{The second pair of Seiberg's dual theories}

Therefore, the next question is as follows. - Is it possible at all to start from the direct ${\cal N}=1$ theory with the original direct (electric) quark fields $Q^i_a, {\ov Q}^{\,a}_j$ with $N_c$ colors, i.e. those in \eqref{(8.1)}, and to adjust the parameters so that its Seiberg's dual variant with $\nd$ colors and dual quarks $q_i, {\ov q}^j$ will coincide at scales $\mu<\mx$ with the theory \eqref{(8.1.5)}\,? As will be shown below, the answer is "nearly yes" (the meaning of "nearly" will become clear below).\\

{\bf C})\,\, The Lagrangian of this desired direct ${\cal N}=1$ theory with $SU(N_c)$ colors, with some scale factor $\la$ of the gauge coupling, with $N_c+1<N_F<3 N_c/2$ quark flavors, and with $N_F^2$ additional colorless fion fields $\Phi^j_i$ looks at the scale $\la$ as, see \eqref{(8.1.10)} for ${\cal W}_{\Phi}$,
\bq
K={\rm Tr}\,(\Phi^\dagger\Phi)+{\rm Tr}\,({Q}^\dagger Q+{\ov Q}^\dagger {\ov Q}), \label{(8.1.18)}
\eq
\bbq
\quad W_{\rm matter}={\cal W}_{\Phi}+{\rm Tr}\,\Bigl (\,{\ov Q}(m_Q - \Phi) Q \,\Bigr )\,,\quad
{\cal W}_{\Phi}=\frac{\mph}{2}\Bigl (\,{\rm Tr}\,(\Phi^2)-\frac{1}{\nd}\,({\rm Tr}\,\Phi)^2\,\Bigr )\,,
\eeq
and the hierarchies of parameters in \eqref{(8.1.18)} are $m_Q\ll\la\ll\mph$.

{\bf D})\,\, The corresponding Seiberg's theory dual to \eqref{(8.1.18)} has the gauge group $SU(\nd)$, the quarks $q_i, {\ov q}^j$ dual to $Q^i,\, {\ov Q}_j$ and $N_F^2$ additional elementary mions $M^i_j\ra ({\ov Q}_j Q^i)$, the scale factor of the dual gauge coupling is $\Lambda_{q}\sim\la$ and the dual Lagrangian looks at the scale $\la$ as
\bbq
K={\rm Tr}\,(\Phi^\dagger\Phi)+\frac{1}{\mu_2^2}{\rm Tr}\,(M^\dagger M)+{\rm Tr}\,({q}^\dagger q+{\ov q} ^\dagger {\ov q})\,,\quad \mu_1=\mu_2=\la\,,
\eeq
\bq
W_{\rm matter}={\cal W}_{\Phi} + {\rm Tr}\,(m_Q-\Phi) M\,-\frac{1}{\mu_1}{\rm Tr}\,(\,{\ov q}
\, M q \,)\,,\quad \langle M^i_j\rangle=\langle{\ov Q}_j Q^i\rangle\,. \label{(8.1.19)}
\eq

Returning to \eqref{(8.1.18)}, we take now for our special purpose the specific values $\la=\lt$ and  $\mph\sim \lt^2/\mx\gg\lt\gg\lm$,\,\, $m_Q\sim m\mx/\lt\ll m\,,\, 0\,<\,m\ll\mx\ll\lm\ll\tilde\Lambda$, see \eqref{(8.1.4)}. With this choice, the masses of $N_F^2$ fion fields $\Phi$ in \eqref{(8.1.19)} will be very large, $\mu^{\rm pole}(\Phi)\sim\mph\gg\lt\gg\lm$, so that they all can be integrated out once and forever and the dual Lagrangian at the scale $\mu=|\lt|$ can be rewritten as
\bq
K=\frac{1}{|\lt|^2}{\rm Tr}\,(M^\dagger M)+{\rm Tr}\,({q }^\dagger q+{\ov q}^\dagger {\ov q}), \label{(8.1.20)}
\eq
\bbq
W_{\rm matter}={\cal W}_{M}-\frac{1}{\lt}{\rm Tr}\,(\,{\ov q}\, M q \,)\,,\quad {\cal W}_{M}=m_Q{\rm Tr}\,M\,-\frac{1}{2\mph}\Bigl (\,{\rm Tr}\,(M^2)-\frac{1}{N_c}\,({\rm Tr}\,M)^2\,\Bigr )\,.
\eeq

Besides, see \eqref{(8.1.1)},\eqref{(8.1.4)},
\bq
\mo=\la\Bigl (\frac{\la}{m_Q}\Bigr )^{\frac{2N_c-N_F}{N_c}}=\tilde\Lambda\Bigl (\frac{\tilde
\Lambda}{m_Q}\Bigr )^{\frac{2N_c-N_F}{N_c}}\sim\mph\Bigl (\frac{\lm}{m}\Bigr )^{\frac{2N_c-N_F}{N_c}}\,,\quad \frac{\mph}{\mo}\sim\Bigl (\frac{m}\lm\Bigr )^{\frac{2N_c-N_F}{N_c}}\ll 1\,,\label{(8.1.21)}
\eq
so that the $\rm br1$-vacua of the theory \eqref{(8.1.5)} with $\mx\gg {\tilde\mu}_{\rm x,o}$, see \eqref{(8.1.6)},\eqref{(8.1.8)}, correspond to the $\rm br2$-vacua of \eqref{(8.1.18)},\eqref{(8.1.20)} with $\mph/\mo\ll 1$, and for this dual pair of theories the condensates of quarks and mions $M$ at the scale $\la=\lt$ look in these br2-vacua with the multiplicity $(\nd-n_1)C_{N_F}^{{\,\rm n}_1}$ as, see section 4 in \cite{ch5} and \eqref{(7.6)},
\bbq
\langle M_2\rangle_{\rm br2}=\langle{(\ov Q}Q)_2\rangle_{\rm br2}\sim m_Q\mph,\quad
\langle M_1\rangle_{\rm br2}=\langle({\ov Q}Q)_1\rangle_{\rm br2}\sim\lt^2\Bigl (\frac{\mph}{\lt}
\Bigr )^{\frac{n_2}{n_2-N_c}}\Bigl (\frac{m_Q}{\lt} \Bigr )^{\frac{N_c-n_1}{n_2-N_c}}\,,
\eeq
\bq
\langle({\ov q}q)_1\rangle_{\rm br2}=\frac{\langle M_2\rangle_{\rm br2}\lt}{\mph}\sim m_Q\lt\sim m\mx\,,\,\, \langle({\ov q}q)_2\rangle_{\rm br2}=\frac{\langle M_1\rangle_{\rm br2}\lt}{\mph}\sim\mx
m\Bigl (\frac{m}{\lm} \Bigr )^{\frac{\bb}{n_2-N_c}}\,,\label{(8.1.22)}
\eq
\bbq
\langle\lym^{(\rm br2)}\rangle^3=\langle S\rangle_{\rm br2}=\frac{\langle({\ov Q}Q)_1\rangle_{\rm br2}\langle ({\ov Q}Q)_2\rangle_{\rm br2}}{\mph}\sim\mx m^2\Bigl (\frac{m}{\lm}\Bigr )^{\frac{\bb}{n_2-N_c}}\,,
\eeq
\bbq
\frac{\langle({\ov q}q)_2\rangle_{\rm br2}}{\langle({\ov q}q)_1\rangle_{\rm br2}}\sim\Bigl (\frac{m}{\lm}\Bigr )^{\frac{2N_c-N_F}{n_2-N_c}}\ll 1\,,\quad \quad 1\leq n_1<\nd\,, \quad n_2>N_c\,.
\eeq
The dual theory \eqref{(8.1.20)} is in the IR free regime in the range of scales $m\ll\mu\ll\lt$, with only logarithmic RG evolution (which is neglected). The masses of $N_F^2$ mions $M$, $\mu^{\rm pole}(M)\sim\lt^2/\mph\sim\mx$ will be the largest ones among the masses smaller than $\lm$ and they all can be integrated out at scales $\mu<\mx\ll\lm$ we are interested in. As a result, the Langangian \eqref{(8.1.20)} at the scale $\mu=\mx$ will be
\bbq
\hspace*{-0.7cm}K={\rm Tr}\,({q }^\dagger q+{\ov q}^\dagger {\ov q})\,,\quad\quad
W_{\rm matter}=\frac{N_c}{\nd}\,\frac{m_Q\mph}{\lt}\,{\rm Tr}\, ({\ov q} q\,)+\frac{\mph}{2\lt^2}\,\Bigl [\,{\rm Tr}\,({\ov q} q)^2-\frac{1}{\nd}({\rm Tr}\,{\ov q} q)^2\,\Bigr ]=
\eeq
\bq
=\tm\,{\rm Tr}\, ({\ov q} q\,)+\frac{1}{2\mx}\,\Bigl [\,{\rm Tr}\,({\ov q} q)^2 -\frac{1}{\nd}({\rm Tr}\,{\ov q} q)^2\,\Bigr ]\,,\quad \tm=\frac{N_c}{\nd}\, m\,.\label{(8.1.23)}
\eq

Therefore, with the special choice of parameters in \eqref{(8.1.18)}, see \eqref{(8.1.4)},
\bq
\la=\lt\,,\quad m_Q= m\, \frac{\lt}{\mph}\,, \quad \mph=\frac{\lt^2}{\mx} \label{(8.1.24)}
\eq
the Lagrangian \eqref{(8.1.23)} will have the same form as \eqref{(8.1.5)}, with the same mass parameters $\tm$ and $\mx$, with the same gauge group $SU(\nd)$ and the same scale factor $\lt$ of its gauge coupling, and  all hierarchies will be as needed
\bq
m_Q\ll m\ll\mx\ll\Lambda_2\ll\lt\ll\mph\,.\label{(8.1.25)}
\eq
And the spectrum of masses smaller than $\mx$ will be the same in this dual theory \eqref{(8.1.20)},\eqref{(8.1.23)} with the quark fields $q, \ov q$ as those described above in the direct theory \eqref{(8.1.5)} with the direct quark fields $\tQ, {\otQ}$, both theories weakly coupled and with $\nd$ colors.

Does it mean that these two theories, \eqref{(8.1.5)} and \eqref{(8.1.23)}, will be completely equivalent at scales $\mu<\mu_{\rm x}$? The answer is negative. The reason is as follows. Let the direct (electric) quarks $Q^i_a$ in \eqref{(8.1)},\eqref{(8.1.18)} to have the positive baryon charge $B=1$ and to be in the {\it fundamental} representation $"N_F"$ of the $SU(N_F)_L$ flavor group, so that the baryon field with the baryon charge $N_c$ looks as $\sim Q^{N_c}$. The same baryon field looks then as $\sim q^{\nd}$ in terms of dual quarks, so that the quarks $q_i$ in \eqref{(8.1.23)} have the positive baryon charge $N_c/\nd$ and are in the {\it antifundamental} representation $"\ov N_F"$ of  $SU(N_F)_L$ (up to $q\leftrightarrow {\ov q}$ which is simply the change of notations). But the quarks $\tQ^i_b$ in \eqref{(8.1.5)} are in the same {\it fundamental} representation $"N_F"$ of the $SU(N_F)_L$ flavor group as the quarks $Q^i_{a}$ in \eqref{(8.1)} because electric quarks $\tQ$ and $Q$ differ only by a number of colors.

Now, in the interval of scales $(m\mx)^{1/2}\ll\mu\ll\mx$, all  terms in the superpotentials  \eqref{(8.1.5)} and \eqref{(8.1.23)} are dynamically irrelevant and all quark and gluon fields are effectively  massless in both theories, so that the non-Abelian flavor symmetry is enhanced and is $SU(N_F)_L\times SU(N_F)_R$\,. Therefore, the quarks $\tQ^i$ from \eqref{(8.1.5)} give the positive contribution, $"+\nd"$, to the flavor  't Hooft triangle $SU^{\,3}(N_F)_L$, while the contribution of quarks $q_i$ from \eqref{(8.1.23)} to this triangle is negative, $"-\,\nd"$. Clearly, this is a manifestation of the fact that the quarks $\tQ^i$ and $q_i$ behave differently under $SU(N_F)_L$ transformations. This conclusion differs from those in \cite{SY1} where it was claimed that the electric quarks $\tQ$ and Seiberg's dual quarks $q$ are the same.\\

Not going into details we only note here that, {\it with the choice \eqref{(8.1.24)} and with the correspondence $Q^i\leftrightarrow\tq_i,\, \Phi^j_i\leftrightarrow{M}^i_j$, the spectrum of masses smaller than $\mx$ in the direct theory \eqref{(8.1.18)} will be the same as in the dual theory \eqref{(8.1.10)},
\eqref{(8.1.13)}}, both theories with $N_c$ colors and strongly coupled (see \cite{ch7} for all details).

But the $SU^{\,3}(N_F)_L$ triangles will be different also.\\
1)\, In the range of scales $\mu^{\rm pole}_{\tq,1}\sim\lm(m/\lm)^{\nd/N_c}\ll\mu\ll\mx$\,. - In the dual theory \eqref{(8.1.13)}, the quarks $\tq_i$ contribute  "$-N_c$" and mions ${M}^i_j$ contribute "$+N_F$", i.e. "$+\nd$" on the whole. In the direct theory \eqref{(8.1.18)}, the quarks $Q^i$ contribute "$+N_c$" while the fions $\Phi^j_i$ contribute "$-N_F$", i.e. "$-\nd$" on the whole.\\
2)\, In the range of scales $\mu^{\rm pole}({M}^1_1)\sim\mu^{\rm pole}({\Phi^1_1})<\mu< \mu^{\rm pole}_{\tq,2}\sim m\,$. - In the dual theory \eqref{(8.1.15)} the mions ${M}^i_j$ contribute "$+N_F$", while in the direct theory \eqref{(8.1.18)} the fions $\Phi^j_i$ contribute "$-N_F$".\\

We can compare also the flavor triangles for the first Seiberg's dual pair of theories: the weakly coupled \eqref{(8.1.5)} with $\nd$ colors and strongly coupled \eqref{(8.1.13)} with $N_c$ colors.\\
{\bf I)} For $SU^{\,3}(N_F)_L$ triangles in the range of scales $\mu_{\tq,1}^{\rm pole}<\mu<\mx$, see \eqref{(8.1.11)},\eqref{(8.1.13)}. In the direct theory \eqref{(8.1.5)} the quarks $\textsf{Q}^i$ contribute "$+\nd$". In the dual theory \eqref{(8.1.13)} the quarks $\textsf{q}_i$ contribute "$-N_c$" and mions ${M}^i_j$ contribute "$+N_F$", i.e. also "$+\nd$" on the whole.\\
{\bf II)} As an example, at lower energies.\\
1)  For $SU^{\,3}(n_2)_L$ triangles. In the direct theory \eqref{(8.1.5)} in the range of scales $0<\mu<m^{\rm pole}_{Q,2}\sim m$ the massless hybrid pions $\Pi_1^2\ra (\otQ_1\tQ^2)$ contribute "$+n_1$". In the dual theory \eqref{(8.1.15)}: a) in the range of scales $\mu^{\rm pole}({M}^2_2)<\mu<\mu^{\rm pole}_{\textsf{q},2}\sim m$ the mions ${M}^2_j$ contribute "$+N_F$",\, b) in the range of scales $0<\mu<\mu^{\rm pole}({M}^2_2)\ll m$ the massless hybrid mions ${M}^2_1$ contribute "$+n_1$".\\
2) For $SU^{\,3}(n_1)_L$ triangles. In the direct theory \eqref{(8.1.5)} in the range of scales $0<\mu<m^{\rm pole}_{Q,2}\sim m$ the massless hybrid pions $\Pi^1_2\ra (\otQ_2\tQ^1)$ contribute "$+\nt$". In the dual theory \eqref{(8.1.15)}: a) in the range of scales $\mu^{\rm pole}({M}^1_1)<\mu<\mu^{\rm pole}_{\textsf{q},2}\sim m$ the mions ${M}^1_j$ contribute "$+N_F$",\, b) in the range of scales $0<\mu<\mu^{\rm pole}({M}^1_1)\ll m$ the massless hybrid mions ${M}^1_2$ contribute "$+n_2$".\\

Now, the same for the second Seiberg's dual pair: the weakly coupled \eqref{(8.1.23)} with $\nd$ colors and strongly coupled \eqref{(8.1.18)} with $N_c$ colors.\\
{\bf I)} For $SU^{\,3}(N_F)_L$ triangles in the range of scales $m_{Q,1}^{\rm pole}\sim\mu_{\tq,1}^{\rm pole}<\mu<\mx$\,, see \eqref{(8.1.11)}. - In the dual theory \eqref{(8.1.23)} the dual quarks $q_i$ contribute "$-\nd$". In the direct theory \eqref{(8.1.18)} the quarks $Q^i$ contribute "$N_c$" and the fions $\Phi^j_i$ contribute "$-N_F$", i.e. also "$-\nd$" on the whole. \\
{\bf II)} At lower energies. -\\
1)  For $SU^{\,3}(n_2)_L$ triangles. In the direct theory \eqref{(8.1.18)}: a) in the range of scales $\mu^{\rm pole}(\Phi^2_2)<\mu<m^{\rm pole}_{Q,2}\sim m$ the fions $\Phi^j_2$ contribute "$-N_F$",\, b) in the range of scales $0<\mu<\mu^{\rm pole}(\Phi^2_2)\ll m$ the massless hybrid fions $\Phi^1_2$ contribute "$-\no$". In the dual theory \eqref{(8.1.23)}: in the range of scales $0<\mu<\mu^{\rm pole}_{\textsf{q},2}\sim m$ the massless dual hybrid pions ${\tilde\Pi}^1_2\ra ({\ov q}^{\,1} q_2)$ contribute "$-\no$".\\
2) For $SU^{\,3}(n_1)_L$ triangles. In the direct theory \eqref{(8.1.18)}: a) in the range of scales $\mu^{\rm pole}(\Phi^1_1)<\mu<m^{\rm pole}_{Q,2}\sim m$ the fions $\Phi_1^j$ contribute "$-N_F$",\, b) in the range of scales $0<\mu<\mu^{\rm pole}(\Phi^1_1)\ll m$ the massless hybrid fions $\Phi_1^2$ contribute "$-\nt$". In the dual theory \eqref{(8.1.23)}: in the range of scales $0<\mu<\mu^{\rm pole}_{\textsf{q},2}\sim m$ the massless dual hybrid pions ${\tilde\Pi}^2_1\ra ({\ov q}^{\,2} q_1)$ contribute "$-\nt$".\\

\subsection{Unbroken $\mathbf {U(N_F)}$ and unbroken $\mathbf{Z}_{\mathbf{2N_c-N_F}}$}

The vacua with these properties in the parent electric theory \eqref{(8.1)} with $N_c$ colors and at $\mx\ll\mu_{\rm x,o}$, see \eqref{(8.1.1)}, are the S vacua with the multiplicity $\nd$, see section 4 in \cite{ch5}, $\langle{\ov Q}_j Q^i\rangle_S=\delta^i_j\langle\qq\rangle_S$\,,
\bq
\langle\qq\rangle_S\approx -\frac{N_c}{\nd}\,\mx m\,,\quad \langle S\rangle_S=\Bigl (\frac{\det\langle\qq\rangle_S}{\la^{\rm\bo}}\Bigr )^{1/\nd}\sim\mx m^2\Bigl (\frac{m}{\lm}\Bigr )^{(2N_c-N_F)/\nd}\,.\label{(8.2.1)}
\eq

It is interesting that the corresponding vacua with these properties in the theories \eqref{(8.1.3)},\eqref{(8.1.5)} with $\nd$ electric colors, with quarks $\tQ,\,\otQ$ and at $\mx\gg{\tilde\mu}_{{\rm x},o}$\,, see \eqref{(8.1.5)},\eqref{(8.1.6)}, are the ${\cal N}=1$ SQCD vacua with the multiplicity $\nd$ (\, see \eqref{(2.9)},\eqref{(8.1.4)} with replacements $N_c\ra\nd\,,\, \la\ra\lt\,,\, \mph\ra \wmu\,,\, m_Q\ra \tm$, and \eqref{(8.1.3)},\,\,  $\bd=3\nd-N_F<0$ \,)
\bbq
\langle S\rangle_{QCD}\equiv\langle\lym^{(QCD)}\rangle^3\sim \mx m^2\Bigl (\frac{m}{\lm}\Bigr )^{(2N_c-N_F)/\nd}\,,\quad \Bigl (\frac{\langle\lym^{(QCD)}\rangle}{m}\Bigr )^3\sim\frac{\mx}{\lm}\Bigl (\frac{m}{\lm}\Bigr )^{(3N_c-2N_F)/\nd}\ll 1\,,
\eeq
\bq
\langle{\ov\tQ}\tQ\rangle_{QCD}\approx\frac{\langle S\rangle_{QCD}}{\tm}\approx \frac{1}{\tm}\Bigl (\lt^{\bd}\det \tm \Bigr )^{1/\nd}\sim\mx m\Bigl (\frac{m}{\lm}\Bigr)^{(2N_c-N_F)/\nd},\quad \tm=\frac{N_c}{\nd}\, m\,, \label{(8.2.2)}
\eq
\bbq
\frac{\langle\Phi\rangle_{QCD}}{m}\sim\frac{\langle{\ov\tQ}\tQ\rangle_{QCD}}{m\mx}\sim\Bigl (\frac{m}{\lm}\Bigr )^{(2N_c-N_F)/\nd}\ll 1\,,\quad
\langle{\ov\tQ}_j{\tQ}^i\rangle_{QCD}=\delta^i_j \langle {\ov\tQ}\tQ\rangle_{QCD}\,,\,\,
\langle\Phi^j_i\rangle_{QCD}=\delta^j_i\langle\Phi\rangle_{QCD}\,.
\eeq

The direct theory \eqref{(8.1.5)} in these SQCD vacua is IR free in the range of scales $m\ll\mu\ll\lt$ and the RG-evolution is only logarithmic (and ignored). Therefore, the potentially competing masses look as
\bq
m^{\rm pole}_{\tQ}\sim m^{\rm tot}_{\tQ}=\langle m-\Phi\rangle_{QCD}=\frac{\langle S\rangle_{QCD}}{\langle{\ov\tQ}\tQ\rangle_{QCD}}\sim m\,,\label{(8.2.3.)}
\eq
\bbq
\mu^2_{\rm gl}\sim\langle{\ov\tQ}\tQ\rangle_{QCD}\,,\quad
\Bigl (\frac{\mu_{\rm gl}}{m^{\rm pole}_{\tQ}}\Bigr )^2\sim\Bigl (\frac{m}{\lt} \Bigr)^{(3N_c-2N_F)/\nd}\sim
\frac{\mx}{\lm}\Bigl (\frac{m}{\lm}\Bigr )^{(3N_c-2N_F)/\nd}\ll 1
\eeq
and so the overall phase is ${\rm H}\tQ$ (heavy quark).

The mass spectrum looks as follows.\\
1) The masses of all $N^2_F$ fions $\Phi$ are $\sim \mx$ and they are dynamically irrelevant at scales $\mu<\mx$ we are interested in.\\
2) There is a large number of hadrons made of weakly coupled and weakly confined non-relativistic direct quarks $\tQ,\, \ov\tQ$ with masses $m_{\tQ}^{\rm pole}\sim m$ (the tension of the confining string originating from $SU(\nd)$ SYM is $\sqrt\sigma\sim\langle\lym^{(QCD)}\rangle\ll m$).\\
3) There is a large number of gluonia made of $SU(\nd)$ gluons, the scale of their masses is $\sim\langle\lym^{(QCD)}\rangle=\langle S\rangle^{1/3}_{QCD}$, see \eqref{(8.2.2)}.\\

The mass spectrum of the strongly coupled $d\Phi$-theory \eqref{(8.1.10)} which is the literal Seiberg's dual to \eqref{(8.1.7)} looks as follows in these SQCD vacua with the multiplicity $\nd$, see \cite{ch7}. \\
a)\, All $N^2_F$ fields $\Phi$ also decouple as heavy ones at $\mu<\mu_{\rm x}$ and the Lagrangian at $\mu=\mu_{\rm x}$ is \eqref{(8.1.13)}.\\
b)\,\, The overall phase is also ${\rm H}\tq$ (heavy quark). There is a large number of hadrons made of weakly confined dual quarks $\tq, \otq$ (the tension of the confining string originating from $SU(N_c)$ SYM is $\sqrt\sigma\sim\langle\lym^{(QCD)}\rangle\ll m$), the quark pole mass is ($\gamma_{\tq}=\gamma^
{(+)}_{\tq},\,\,\gamma_{M}=\gamma^{(+)}_{M}$, see \eqref{(8.1.12)},\eqref{(8.1.13)},\eqref{(8.1.4)}, and \eqref{(8.2.2)}\,)
\bq
\mu_{\tq}\equiv\mu_{\tq}(\mu=|\lt|)=\frac{\langle{M\rangle_{QCD}}}{\lt}=
\frac{\langle{\otQ}\tQ\rangle_{QCD}}{\lt}\,,\quad \mu_{\tq}^{\rm pole}=\frac{\mu_{\tq}}{z^{(+)}_{\tq}
(\lt,\mu_{\tq}^{\rm pole})}\sim\,m\ll \mu_{\rm x}\,.\label{(8.2.4)}
\eq
c) There is a large number of gluonia made of $SU(N_c)$ gluons with the mass scale $\langle\lym^{(QCD)}
\rangle=\langle S\rangle_{QCD}^{1/3}\ll\mu_{\tq}^{\rm pole}$\,, see \eqref{(8.2.2)}.\\
d) The Lagrangian of $N_F^2$ mions ${M}^i_j$ at the scale $\mu<\langle\lym^{(QCD)}\rangle$ looks as, see \eqref{(8.1.4)},\eqref{(8.2.4)},
\bbq
K_{M}=z_{M}(\lt,\mu_{\tq}^{\rm pole})\,{\rm Tr}\,\Bigl (\,\frac{{M}^\dagger {M}}{\lt^2}\Bigr )\,,\quad z_{M}(\lt,\mu_{\tq}^{\rm pole})=\Bigl (\frac{|\lt|}{\mu_{\tq}^{\rm pole}}\Bigr )^{2(2N_c-N_F)/\nd}\gg 1\,,
\eeq
\bq
{\cal W}={\cal W}_{\tp}+ m\,{\rm Tr}\,\tp\,+{\cal W}_{\rm non-pert},\quad{\cal W}_{\rm non-pert}
=\,-\,N_c\Bigl (\lt^{\rm 3N_c-N_F}\det \frac{M}{\lt}\Bigr )^{1/N_c}\,.\label{(8.2.5)}
\eq
The main contribution to the masses of $N_F^2$ mions M originates from ${\cal W}_{\rm non-pert}$ in \eqref{(8.2.5)} and is
\bq
\hspace*{-5mm}\mu^{\rm pole}(M)=\frac{1}{z_{M}(\lt,\mu_{\tq}^{\rm pole})}\frac{\langle S\rangle_{QCD}\lt^2}{\langle M\rangle^2_{QCD}}\sim\frac{\langle\lym^{(QCD)}\rangle^3}{m^2}\sim\mu_{\rm x}\Bigl (\frac{m}{\Lambda_2}\Bigr )^{(2N_c-N_F)/\nd}\,,\label{(8.2.6)}
\eq
\bbq
\Biggl [\frac{\mu^{\rm pole}(M)}{\langle\lym^{(QCD)}\rangle}\Biggr ]^{3/2}\sim \frac{\mx}{\lm}\,\Bigl (\frac{m}{\lm}\Bigr )^{(3N_c-2N_F)/\nd}\ll 1\,.
\eeq

The overall hierarchy of masses looks as
\bbq
\quad \mu^{\rm pole}(M)\ll \langle\lym^{(QCD)}\rangle\ll \mu_{\tq}^{\rm pole}\sim m\ll\mu_{\rm x}\ll\lm\,.
\eeq

Therefore, the spectra of mass smaller than $\mu_{\rm x}$ are parametrically different in the direct theory \eqref{(8.1.7)} and in its Seiberg's dual \eqref{(8.1.10)},\eqref{(8.1.13)} in these SQCD vacua.\\

Consider now the strongly coupled direct theory \eqref{(8.1.18)} with $N_c$ colors and direct quarks $Q,\, \ov Q$ and its weakly coupled Seiberg's dual \eqref{(8.1.19)},\eqref{(8.1.20)} with $\nd$ colors and dual quarks $q,\, \ov q$. For this case, see \eqref{(8.1.24)},
\bq
\mo=\la\Bigl (\frac{\la}{m_Q}\Bigr )^{\frac{2N_c-N_F}{N_c}}=\lt\Bigl (\frac{\lt}{m_Q}\Bigr )^{\frac{2N_c-N_F}{N_c}}\sim\mph\Bigl (\frac{\lm}{m}\Bigr )^{\frac{2N_c-N_F}{N_c}}\gg\mph\,,\label{(8.2.7)}
\eq
and so for this theory \eqref{(8.1.18)} (and therefore for \eqref{(8.1.20)} also) the corresponding vacua with the unbroken $U(N_F)$ and $Z_{2N_c-N_F}$ symmetries are the S-vacua with the multiplicity $\nd$\,, see section 4 in \cite{ch5}. In these vacua ($\,\langle{\ov Q}Q\rangle_{S}\equiv\langle{\ov Q}Q(\mu=\lt)\rangle_{S}\,$)\,,
\bbq
\hspace*{-2mm}\langle{\ov Q}Q\rangle_S\approx -\frac{N_c}{\nd}\,m_Q\mph\sim \mx m\Bigl (\frac{\lm}{\mx}\Bigr )^{\frac{(2N_c-N_F)}{3N_c-2N_F}},\,\,\langle S\rangle_{S}=\Bigl (\frac{\det\langle{\ov Q}Q\rangle_{S}}{\lt^{3N_c-N_F}}\Bigr )^{1/\nd}\sim\mx m^2\Bigl (\frac{m}{\lm}\Bigr )^{(2N_c-N_F)/\nd},
\eeq
\bq
\langle M\rangle_{S}\equiv\langle M(\mu=\lt)\rangle_{S}=\langle{\ov Q}Q\rangle_{S}\,,\quad\langle{\ov q}q\rangle_{S}\equiv\langle{\ov q}q(\mu=\lt)\rangle_{S}=\frac{\langle S\rangle_{S}\lt}{\langle M\rangle_{S}}\sim\mx m\Bigl (\frac{m}{\lm}\Bigr )^{(2N_c-N_F)/\nd}\,,\label{(8.2.8)}
\eq
\bbq
\langle M^i_j \rangle_{S}=\delta^i_j\langle M\rangle_{S}=\langle{\ov Q}_j Q^i\rangle_{S}=\delta^i_j\langle{\ov Q}Q\rangle_{S}\,.
\eeq

The weakly coupled dual theory \eqref{(8.1.20)} with $\nd$ dual colors is in the IR free regime at scales $m\ll\mu\ll\la=\lt$ in these S-vacua and the RG-evolution is only logarithmic (and ignored). From \eqref{(8.2.8)},\eqref{(8.1.24)} the potentially important masses look as
\bq
\mu^{\rm pole}_q\sim \mu_q=\frac{\langle M\rangle_{S}}{\lt}\sim m\,,\quad {\ov\mu_{\rm gl}}\sim\langle{\ov q}q\rangle^{1/2}_{S}\,,\quad \Bigl (\frac{\ov\mu_{\rm gl}}{\mu^{\rm pole}_q}\Bigr )^{2}\sim\frac{\mx}{\lm}\Bigl (\frac{m}{\lm}\Bigr )^{(3N_c-2N_F)/\nd}\ll 1\,,\label{(8.2.9)}
\eq
so that the overall phase is $Hq$ (heavy quarks). The masses of all $N_F^2$ mions M are the largest ones, $\mu(M)\sim \lt^2/\mph=\mx$\,, they are dynamically irrelevant at scales $\mu<\mx$ and can be integrated out resulting in \eqref{(8.1.23)}. As one could expect with the choice \eqref{(8.1.24)}, the spectrum of masses smaller than $\mx$ in these $\nd\,\,$ S - vacua of the dual theory \eqref{(8.1.23)} with the dual quarks $q,\,\ov q$ is the same as those in the $\nd\,\,$ SQCD - vacua of the direct theory \eqref{(8.1.7)} with direct quarks $\tQ,\, \otQ$, both theories weakly coupled and with $\nd$ colors. But, clearly, the values of 't Hooft triangles $SU^{\,3}(N_F)_L$ in the range of scales $m\ll\mu\ll\mx$ where all quarks and gluons in both theories \eqref{(8.1.7)} and \eqref{(8.1.23)} are effectively massless are also different in these vacua with unbroken symmetries.\\

Not going into any details we only note here that, with the choice \eqref{(8.1.24)} and with the correspondence $Q^i\leftrightarrow\tq_i,\,\, \Phi^j_i\leftrightarrow{M}^i_j$\,, the spectrum of masses smaller than $\mx$ is also the same in the pair \eqref{(8.1.13)} and \eqref{(8.1.18)} of strongly coupled theories with $N_c$ colors in the considered vacua with the unbroken $U(N_F)$ and $Z_{2N_c-N_F}$ symmetries, see section 3.1 in \cite{ch7}. But the $SU^{\,3}(N_F)_L$ triangles are also different.

\numberwithin{equation}{subsection}
\subsection{Going to large $\mx\gg\lm$}

{\bf A)}\, We give here first a short justification of the form \eqref{(8.1.19)} (the same concerns also the form of \eqref{(8.1.10)}\,).

In his original paper \cite{S2} Seiberg considered the {\it standard} ${\cal N}=1$ SQCD with $SU(N_c)$ colors and e.g. $N_c+1<N_F<3N_c/2$ flavors as the direct theory, and \eqref{(8.1.19)} with $SU(\nd)$ colors as its dual one, both without fields $\Phi$, see \eqref{(8.1.18)},\eqref{(8.1.19)}. Moreover, the starting point was this standard SQCD with {\it massless quarks}, $m_Q\ra 0$, and some fixed scale factor of the gauge coupling in the direct theory, say $\la$.

We would like now to emphasize two points:\\
a) both direct and dual theories are conventionally implied to be nonsingular and {\it non-trivial interacting theories} at scales $\mu\lesssim\la$ even at $m_Q=0,\,\, \la={\rm const}$;\\
b) the dual theory can be considered either as the assumed lower energy form of the direct theory with $N_c+1<N_F<3N_c/2$ at scales $\mu<\la$, or (as is the only possibility in the conformal window $3N_c/2<N<3N_c$) as independent theory which becomes supposedly equivalent to the direct one at $\mu<\la$.

In any case (as in the conformal window), $\la$ is then {\it the only nonzero dimensional parameter} in both these theories (without fields $\Phi$)\, at $\,m_Q=0$. Then the scale factor $\Lambda_q$ of the dual gauge coupling has to be $\Lambda_q=C_q\la,\, C_q=O(1)$. For the same reasons, it should be $\mu_1\sim\mu_2\sim\la$.

Besides, it was obtained in \cite{IS} (for not especially chosen values of $N_F, N_c$): $\,\la^{3N_c-N_F}\Lambda_q^{3\nd-N_F}=C_o\mu^{N_F}_1,\,\,C_o=O(1)$, so that $\mu_1\sim\Lambda_q\sim\la$, see also \cite{ISS} and sections 2-4 in \cite{ch3}. And the only acceptable value of $r=\mu_2/\mu_1$ at $m_Q=0$ is $r\sim 1$.

Moreover, {\it these scale factors $\Lambda_q\sim\la$ and $\mu_{1,2}\sim\la$ remain the same after the additional colorless fields $\Phi^j_i$ are added}. Therefore, the dual Lagrangian should be of the form \eqref{(8.1.19)}.\\

Consider also the original theory \eqref{(8.1)} with the gauge group $U(N_c)$ (to comply with \cite{SY2}),\,$0<m\ll\la,\,\,N_c+1<N_F<3N_c/2$, and large $\mx\gg\lm$. All $N_c^2$ fields $X^{adj}_{U(N_c)}$ decouple as heavy at scales $\mu<\mx^{\rm pole}/(\rm several)= {\it g}^2\mx/(\rm several)$ in the weak coupling region and can be integrating out. The Lagrangian at the scale $\mu=(\rm several)\la\ll\mx,\,\, \la^{3N_c-N_F}\sim\lm^{2N_c-N_F}\mx^{N_c}$, looks then as (ignoring logarithmic factors for simplicity)
\bq
K={\rm Tr}\,(Q^{\dagger} Q+Q\ra {\ov Q}),\quad {\cal W}_{\rm matter}=m{\rm Tr}\,(\qq)-\frac{1}{2\mx}{\rm Tr}\,(\qq)^2\,.\label{(8.3.1)}
\eq

The form of Seiberg's theory with $\nd=N_F-N_c$ colors dual to \eqref{(8.3.1)}, as it is given and used in \cite{SY2}, looks as
\bbq
K=\frac{1}{\kappa^2}{\rm Tr}\,(M^\dagger M)+{\rm Tr}\,\Bigl [\,\frac{\mx}{\kappa}({h}^\dagger h+{\ov h}^\dagger {\ov h})=({q}^\dagger q+{\ov q}^\dagger {\ov q})\,\Bigr ]\,,
\eeq
\bq
\quad {\cal W}_{\rm matter}={\rm Tr}\,\Bigl (\,m M -\frac{1}{2\mx}\, M^{2}\Bigr )-\frac{1}{\mx}{\rm Tr}\,({\ov q}{ M} q)\,, \quad \langle M^i_j\rangle=\langle{\ov Q_j} Q^i\rangle \,.\label{(8.3.2)}
\eq

It is worth reminding that Lagrangian forms \eqref{(8.3.1)} and \eqref{(8.3.2)} {\it are the same for all vacua} at $\mx\gg\la$. Consider the SQCD vacua in the limit of sufficiently large $\mx$, with fixed both $\la$ and $0<m\ll\la$. In these vacua (neglecting small power corrections $\sim \langle{\ov Q} Q\rangle_{\rm QCD}/m\mx\ll 1$)\,:\, $\langle{\ov Q_j} Q^i\rangle=\delta^i_j\langle {\ov Q} Q\rangle_{\rm QCD}=\delta^i_j\langle M\rangle_{\rm QCD},\,\,i,j=1...N_F,\,\, \langle {\ov Q} Q\rangle_{\rm QCD}\approx\langle S\rangle_{\rm QCD}/m\approx [(\la^{3N_c-N_F}m^{N_F})^{1/N_c}]/m\sim "\rm const"$ in this limit. Therefore, the last term ${\rm Tr\,}({\ov Q}Q)^2/\mx\sim\la^4/\mx$ in \eqref{(8.3.1)} is parametrically small and irrelevant, and the standard ${\cal N}=1$ SQCD Lagrangian emerges. If $(\kappa^2/\mx)\ra 0$ at large $\mx$, the extra term $(M^2/\mx)$ in \eqref{(8.3.2)} will be also irrelevant and can be safely neglected (as it should be in  standard Seiberg's dual SQCD theory). The problem is with the last term ${\rm Tr\,}({\ov q} M q)/\mx$ in \eqref{(8.3.2)}. On the one hand, if we take by hand $\langle {\ov q}^{\,i} q_i\rangle_{QCD}\sim{\rm const}$, then this last term will disappear also, but the term $\sim {\rm Tr\,}({\ov q} M q)$ has to survive necessarily in Seiberg's dual theory. On the other hand, from the Konishi anomaly: $\,\langle {\ov q}^{\,i} q_i\rangle_{QCD}\langle M^i_i\rangle_{QCD}=\mx\langle{-\tilde S}\rangle_{QCD}, \,\, i=1...N_F$ (no summation over "i"), where $\langle{\tilde S}\rangle_{QCD}$ is the dual gluino condensate (the observable quantity). For the direct and dual theories to be equivalent, at least potentially, the necessary (but clearly not sufficient) condition is the corresponding relation $\langle{ - \tilde S}\rangle_{QCD}=\langle S\rangle_{QCD}$. But then $\langle {\ov q}^{\,i} q_i\rangle_{QCD}=\mx\langle{ - \tilde S}\rangle_{QCD}/\langle M^i_i\rangle_{QCD}=\mx\langle S\rangle_{QCD}/\langle{\ov Q_i} Q^i\rangle_{QCD}\approx \mx m$, and it will be parametrically large at sufficiently large $\mx$, $\,\langle {\ov q}^{\,i} q_i\rangle_{QCD}\gg\la^2$. (The same follows from $\langle\partial{\cal W}_{\rm matter}/\partial M^i_j\rangle=0$ in \eqref{(8.3.2)} at $(\langle M^i_i\rangle_{\rm QCD}=\langle{\ov Q_i} Q^i\rangle_{\rm QCD})/m\mx\ll 1$\,). This looks clearly unrealistic for SQCD vacua and, besides, this very large condensate, $\langle {\ov q}^i q_i\rangle_{QCD}\sim m\mx\gg \la^2,\,\,i=1...N_F$, implies that these dual quarks with $\nd$ colors and $N_F$ flavors will be higgsed. The rank condition at $N_F>N_c>\nd$ will result then in the spontaneous breaking of (at least) the global flavor symmetry in the dual theory, but this is not the case for the direct theory \eqref{(8.3.1)}. Therefore, in any case and {\it in all vacua}, the form of the last term in ${\cal W}_{\rm matter}$ in \eqref{(8.3.2)} is not right at large $\mx$ for the standard Seiberg theory dual to \eqref{(8.3.1)}.

As described above, instead of \eqref{(8.3.2)}, the appropriate dual form of \eqref{(8.3.1)} is
\bq
K=\frac{1}{\Lambda^2_Q}{\rm Tr}\,(M^\dagger M)+{\rm Tr}\,({q}^\dagger q+{\ov q}^\dagger {\ov q}),\quad
\quad W_{\rm matter}={\rm Tr}\,\Bigl (\,m M -\frac{1}{2\mx}\, M^2\Bigr )-\frac{1}{\la}{\rm Tr}\,({\ov q}M q),\label{(8.3.3)} \eq
and it behaves correctly at large $\mx$ in SQCD vacua with fixed $\la$ and small $m$.\\

{\bf B)}\, Consider now attempts to increase $\mx$ from $\mx\ll\lm$ to $\mx\gg\lm$ in the $SU(\nd)$ theory \eqref{(8.1.3)} {\it by itself}.

It is worth to emphasize that this theory \eqref{(8.1.3)} as it is, with $\nd$ colors,\, $N_F>2\nd,\, \mx\ll\lm$ and, in any case, $\langle\textsf{x}\rangle\lesssim m\ll\mx\ll\lm$, is a well defined IR free theory at scales $\mu<\lm/(\rm several)$, but it is not well defined {\it by itself} at larger scales $\mu>\lm$. The reason is that it is the effectively massless unbroken ${\cal N}=2$ SQCD already at scales $\mx\ll\mu<\lm$, with the pure one-loop NSVZ $\beta$-function \cite{NSVZ1,NSVZ2} and $\rm{\ov b}_2=(2\nd-N_F)<0$. And if nothing will happen when $\mu$ crosses $\lm$, its gauge coupling $g^2(\mu>\lm)$ will be negative, this is clearly unphysical (see also the text below \eqref{(8.1.1)} for more details).  Therefore, {\it as it is}, this theory \eqref{(8.1.3)} {\it needs necessarily the appropriate UV completion at scales $\mu>\lm$} to be a definite meaningful theory in this region. For this reason, if one plans to continue it from $\mx\ll\lm$ to $\mx\gg\lm$, {\it before doing this continuation one has to add first some appropriate UV completion at scales} $\mu\gtrsim\lm\,$. As the evident example, the theory \eqref{(8.1)} is the possible UV completion of \eqref{(8.1.3)} at $\mx>\lm$ (but together with the whole Abelian $U^{(1)}(1)\times U^{\bb-1}(1)$ sector).\\

Following \cite{APS}, the theory \eqref{(8.1.3)} was continued {\it by itself} in \cite{SY2} from $\mx\ll\lm$ into the region $\mx\gg\lm$ and, moreover, the relation \eqref{(8.1.4)} between $\tilde\Lambda$ and $\lm$ valid {\it only} at scales $m\ll\mu\ll\mx\ll\lm$ was used in \cite{SY2} also at $\mx\gg\lm$. It was ignored that, before continuing \eqref{(8.1.3)} to $\mx\gg\lm$, it certainly needs the appropriate UV completion at scales $\mu>\lm$ to avoid $g^2(\mu>\lm)<0$. And many its properties at $\mx\gg\lm$ (see below), in particular the value of the scale factor $\Lambda^\prime$ of its gauge coupling, depend essentially on properties of this UV-completion and {\it can not be determined without specifying its details}.

This qualitatively differs ${\cal N}=1$ theories from ${\cal N}=2$ ones as the non-trivial NSVZ $\beta$-functions of (effectively) massless IR free ${\cal N}=1$ SQCD theories \cite{NSVZ1,NSVZ2} allow a smooth continuation without the UV completion, see section 7 in \cite{ch1}. E.g., the standard ${\cal N}=1\,\, SU(\nd=N_F-N_c)$ SQCD with $N_c+1<N_F<3N_c/2$ flavors of light quarks can be smoothly continued from the weakly coupled region at $\mu\ll\la$ to the strongly coupled one at $\mu\gg\la$ without the UV completion by extra fields at $\mu\gtrsim\la$, with its gauge coupling ${\ov a}(\mu)=\nd{\ov g}^2(\mu)/8\pi^2 > 0$ at both sides $\mu\gtrless\la$, and with ${\ov a}(\mu\gg\la)\approx (\mu/\la)^{(2 N_F-3\nd)/(N_F-\nd)}\gg 1$ . Or vice versa, nothing changes with the field content and the mass spectrum of the effectively massless UV free $SU(N_c)$ with $N_c+1<N_F<3N_c/2$ of light quarks, and its gauge coupling $a(\mu)$ behaves smoothly at the transition from the weakly coupled region at $\mu\gg\la$ to the strongly coupled one at $\mu\ll\la$, with $a(\mu) > 0$ at both sides $\mu\gtrless\la$, and with $a(\mu\ll\la)=N_c g^2(\mu)/8\pi^2\approx (\la/\mu)^{(3 N_c-2 N_F)/(N_F-N_c)}\gg 1$, see section 7 in \cite{ch1}. But effectively massless IR free ${\cal N}=2$ SQCD theories with $\langle\textsf{x}\rangle\lesssim m\ll\mx\ll\lm$, like \eqref{(8.1.3)}, {\it with the pure one loop $\beta$-functions} \cite{NSVZ1,NSVZ2} do not allow such smooth continuation from $\mx\ll\lm$ to $\mx\gg\lm$ and require necessarily the appropriate UV completion at $\mu\gtrsim\lm$.

Because this is an important point, we will dwell in short on some details. As a simplest example, consider the theory \eqref{(8.1)} with $m\ll\mx\ll\lm$. The purpose is to continue it to $\mx\gg\lm$ and to find the scale factor $\la$ of the gauge coupling at scales $\mu\ll\mx^{\rm pole}$ where the heavy adjoint field $X$ decouples. This can be done as follows (see e.g. \cite{ch1,ch3} about the use of this procedure for ${\cal N}=1$ SQCD with unequal mass quarks).\\
1) The first step is to define the "parent theory" at the very high scale $\mu=M_{PV}$ (the Pauli-Villars scale). For this, we start at some scale $\mu_o=(\rm several)\lm$ from \eqref{(8.1)} with $m\ra 0,\, \mx\ra 0$. As it is, such theory is the UV free unbroken ${\cal N}=2$ theory with the massless one-loop NSVZ $\beta$-function and $2N_c-N_F>0$. It is well defined at $\mu>\mu_o$. Therefore, when evolved to the very large scale $\mu=M_{PV}$ its Lagrangian $L_{PV}$ looks as it was at $\mu=\mu_o$ with the only replacement of $g^2(\mu_o)=O(1)$ by $g_{PV}^2=g^2(\mu=M_{PV})\ll 1$. And the scale factor $\Lambda_{PV}$ of the gauge coupling $g_{PV}(\mu=M_{PV}/\Lambda_{PV})$ is $\Lambda_{PV}=\lm$ in this simple case. It is worth to remind that the RG evolution in the range $\mu_o<\mu<M_{PV}$ in this theory with $m\ra 0,\,\mx\ra 0$ is independent of $m$ and $\mx$, so that $g_{PV}^2=g^2(\mu=M_{PV}/\Lambda_{PV})$ {\it is also independent} of $m$ and $\mx$. (Clearly, $L_{PV}$ will evolve back to \eqref{(8.1)} when diminishing the scale from $\mu=M_{PV}$ down to $\mu_o$).\\
2) The second step is to continue in the matter superpotential of $L_{PV}$ the mass parameters $\mx$ and $m$ from $\mx\ra 0,\, m\ra 0$ to those values that we wish, e.g. $0<m\ll\lm,\, \lm\ll\mx\ll M_{PV}$.

We note that the adjoint field $X^{\rm adj}_{SU(N_c)}$, with higgsed $\langle X^{\rm adj}_{SU(N_c)}\rangle\sim\lm$ at $\mx\ll\lm$, becomes unhiggsed {\it in all vacua} at $\mx>(\rm several)\,\lm$. The reason is that it becomes too heavy and too short ranged, while its physical (i.e. path dependent) $SU(N_c)$ phases induced by interactions with effectively massless gluons fluctuate then freely at all scales from the high energy down to at least $\mu^{\rm low}_{\rm cut}=\mx^{\rm pole}/({\rm several})={\it g}^2\mx/({\rm several})$. This results in the zero value of its mean value integrated from high energies down to $\mu^{\rm low}_{\rm cut}:\, \langle X^{\rm adj}_{SU(N_c)}\rangle_{\mu^{\rm low}_{\rm cut}}=0$. Therefore, e.g. at $\mx\gg\la$, {\it  after it decouples as heavy} at $\mu<\mx^{\rm pole}/(\rm several),\, \lm\ll\la\sim [\lm^{2N_c-N_F}\mx^{N_c}]^{1/(3N_c-N_F)}\ll\mx^{\rm pole}$, {\it it gives no contributions to particle masses} of the ${\cal N}=1$ SQCD Lagrangian written at the scale $\mu=\mu^{\rm low}_{\rm cut}=\mx^{\rm pole}/({\rm several})$, and does not affect the lower energy dynamics. As a result, all those fields which received masses $\sim\lm$ from higgsed $\langle X^{\rm adj}_{SU(N_c)}\rangle\sim\lm$ at $\mx\ll\lm$ have lost now these contributions to their masses at $\mx>(\rm several)\,\lm$. Their real masses are determined now by the dynamics of the genuine ${\cal N}=1$ SQCD at scales $\mu<\mu^{\rm low}_{\rm cut}$ (see e.g. section 5 in \cite{ch7}, the field $X^{adj}_{SU(N_c)}$ can be replaced by $\Phi^i_j$ at $\mx\gg\la$).~
\footnote{\,
At the same time, e.g. the total $SU(N_c)$ invariant mean value $\langle\,{\rm Tr}\,(\sqrt{2}\,X^{\rm adj}_{SU(N_c)})^2\rangle\approx m\,\nt m_1\ll\lm^2$ in \eqref{(8.1.1)}, integrated {\it by definition} down to $\mu^{\rm lowest}_{\rm cut}=0$, is holomorphic in $\mx$ and stays intact at $\mx\lessgtr\lm$. But, unlike e.g. $\langle\,{\rm Tr}\,(\sqrt{2}\,X^{\rm adj}_{SU(N_c)})^2\rangle\sim\lm^2$ in L or Lt-vacua with the spontaneously broken $Z_{2N_c-N_F}$ symmetry, as a result of the unbroken $Z_{2N_c-N_F}$ symmetry in br2 and S-vacua, the small total mean values $\ll\lm^2$ in these vacua do not reflect the scale $\sim\lm$ even at $\mx\ll\lm$, when $\langle X^{\rm adj}_{SU(N_c)}\rangle\sim\lm$ is higgsed and gives contributions $\sim\lm$ to corresponding particle masses.
}
\\
3) At the last step, starting from $L_{PV}$ with $\lm\ll\mx\ll M_{PV}$, we diminish now the scale from $\mu=M_{PV}$ down to $\la\ll\mu<\mx^{\rm pole}={\it g}^2 (\mu=\mx^{\rm pole})\mx\ll M_{PV}$, and the heavy field $X^{\rm adj}_{SU(N_c)}$ decouples. The scale factor $\la$ of the lower energy ${\cal N}=1$ theory is determined from the matching at $\mu=\mx^{\rm pole}$ of couplings $g^{(+)}(\mu/\Lambda_{PV})$ at $\mu>\mx^{\rm pole}$ and $g^{(-)}(\mu/\la)$ at $\mu<\mx^{\rm pole}$. In the considered simple case this results (in the leading-$\log$s approximation, see section 11 in \cite{ch4} or section 12 in \cite{ch5} for more details) in\,:  $\la^{3N_c-N_F}\sim\Lambda_{PV}^{2N_c-N_F}\mx^{N_c}=\lm^{2N_c-N_F}\mx^{N_c}$. It is clear that, before continuing from $\mx\ll\lm$ to $\lm\ll\mx\ll M_{PV}$, the definite value of $\la$ can not be determined without specifying first completely the parent Lagrangian $L_{PV}$ in the UV region $\mu=M_{PV}$ (to be able to calculate after the value of its coupling $g^{(+)}(\mx^{\rm pole}/\Lambda_{PV})\,$).

But trying to apply this procedure to the theory \eqref{(8.1.3)} {\it by itself}, it is seen that even the first step is impossible to perform without the appropriate UV completion at $\mu>\lm$. The reason is that already at scales $\mx\ll\mu<\lm$ the theory \eqref{(8.1.3)} is the effectively massless unbroken ${\cal N}=2$ SQCD theory with $\langle \textsf{x}\rangle\lesssim m\ll\mx\ll\lm$ and with the one-loop NSVZ $\beta$-function. But $2\nd-N_F<0$ now, so that its coupling $g^2(\mu>\lm)$ becomes negative, this is unphysical. \\

The original idea of \cite{APS} to derive Seiberg's duality in ${\cal N}=1$ SQCD, with in particular $N_c+1<N_F<3N_c/2$, was as follows. To take first the electric theory \eqref{(8.1)} with $m\ra 0$ (i.e. considering $m$ as an inessential parameter) and $\mx\ll\lm$. To consider its low energy form, to find its vacua and light particles therein and to choose the appropriate vacua (the baryonic branch of \eqref{(8.1)} with the low energy $SU(\nd)$ was chosen in \cite{APS}, i.e. the br1 and SQCD vacua of \eqref{(8.1.3)} with all quark and gluino condensates flowing to the origin and coalesce at $m\ra 0$, see \eqref{(8.1.8)} and \eqref{(8.2.2)}, so that all reasonings below concern equally well the $SU(\nd)$ br1 and SQCD vacua in \eqref{(8.1.3)}\,).
\footnote{\,
Really, these reasonings concern also all other non-baryonic vacua of \eqref{(8.1)} with $n_1<N_F/2$ as their lower energy IR free theories with $SU(n_1)$ colors and $N_F>2n_1$ flavors also require, {\it by itself}, the UV completions at $\mu>\lm$.
}

And second, continuing then to $\mx\gg\la$ with fixed $\la\,\, (\la^{3N_c-N_F}=\lm^{2N_c-N_F}
\mx^{N_c},\,\, \lm\ll\la\ll\mx$), to obtain simultaneously the genuine ${\cal N}=1$ SQCD theory with $N_c$ colors from the total Lagrangian \eqref{(8.1)}, and its Seiberg's dual variant with $\nd$ colors from those in \eqref{(8.1.3)} with $m\ra 0$, {\it by itself}. This idea looks strange in any case as it ignores not only the known existence of many particles with masses $\sim\lm$ in the parent theory \eqref{(8.1)} with $\mx\ll\lm$ from which \eqref{(8.1.3)} originated at lower energies, but even the existence of the lighter $U^{(1)}\times U(1)^{2N_c-N_F}$ Abelian sector with its mass scale $\sim\sqrt{\mx\lm}\ll\lm$ (and which in its turn also needs by itself the UV completion at $\mu>\lm$ before continuing from $\mx\ll\lm$ to $\mx\gg\lm$).

Even ignoring as in \cite{APS} all other heavier particles and taking \eqref{(8.1.3)} as it is, it was also ignored in \cite{APS} that the chosen $SU(\nd)$ theory needs then the appropriate UV-completion at scales $\mu>\lm$ before continuing from $\mx\ll\lm$ to $\mx\gg\lm$. And the explicit form of this UV completion is crucial for most properties of the appropriately continued theory \eqref{(8.1.3)} at scales $\mu<\la\ll\mx$. The UV completion has to satisfy a number of stringent conditions to have desired properties of Seiberg's dual ${\cal N}=1$ SQCD in the theory \eqref{(8.1.3)} with $m\ra 0$ continued to large $\mx$\,: a) the numbers of colors $\nd$ and flavors $N_F$ of \eqref{(8.1.3)} have to be preserved,\,\, b) there should appear $N_F^2$ additional light mesons $M^i_j$ with quantum numbers of $({\ov Q}_j Q^i)$ and the additional term $\sim {\rm Tr}\,({\otQ}{\rm M} \tQ)$ in the superpotential (with ${\otQ}, \tQ$ considered as magnetic quarks in \cite{APS}), and do not appear other unwanted light particles with masses $m_i<\la$. And, as was described above, the form of Lagrangian should be as in \eqref{(8.3.3)}. Besides, after the desired UV completion is found, the scale factor $\Lambda^\prime$ of this continued ${\cal N}=1$ "dual" theory with $\nd$ colors has to be determined and should be $\Lambda^\prime\sim\Lambda_q\sim\la$ (see the beginning of this section). At small $m\neq 0$ the value of $\Lambda^\prime$ is important for the mass spectrum of this continued theory and so for comparison with the mass spectrum of the genuine ${\cal N}=1\,\, SU(N_c)$ SQCD with quarks ${\ov Q}, Q$ and with the scale factor $\la$ of the gauge coupling, $\la^{3N_c-N_F}\sim\lm^{2N_c-N_F}\mx^{N_c}\,$.

By the way, as was shown in \cite{ch4,ch5}, if we are interested in the ${\cal N}=1$ SQCD vacua with the unbroken $U(N_F)$ flavor symmetry in the theory \eqref{(8.1)} at large $\mx\gg\la$ and fixed $\la$, these can be reached not simply at $\mx=(\rm several)\la$, but only at $\mx>\mu_{\rm x,o}=\la(\la/m)^{(2N_c-N_F)/N_c}
\gg\la$. But at $m\ra 0$ this region with $\mx>\mu_{\rm x,o}$ is unreachable at all, i.e. the limits $\mx\ra\infty$ and $m\ra 0$ do not commute in \eqref{(8.1)}.

Besides, there are problems with the flavor symmetry and connected with it multiplicities of vacua at small $m\neq 0$. To have unbroken $U(N_F)$ at large $\mx$ in SQCD vacua of this "dual" theory \eqref{(8.1.3)} and naturally supposing that the desired UV completion preserves $U(N_F)$, one has to proceed then as follows. To start at $m\ll\mx\ll\lm$ and small $m\neq 0$ in the theory \eqref{(8.1.3)} from its own SQCD vacua with unbroken $U(N_F)$ (see section 8.2), and to continue then $\mx\gg\la$.  But one of the problems with these SQCD vacua of \eqref{(8.1.3)} is that they have the multiplicity $\nd=N_F-N_c$ inappropriate for the vacua with the multiplicity $N_c$ of the genuine ${\cal N}=1$ SQCD with $N_c$ colors (see section 8.2 for more details).

And finally, supposing that this highly non-trivial appropriate for the Seiberg's duality UV completion at scales $\mu>\lm$ for the theory \eqref{(8.1.3)} {\it by itself} can be found (this is not evident at all because the requirements on it are very stringent), the physical question will remain about the origin of additional fields which enter this UV completion and turn on at $\mu>\lm$. Besides, if this theory \eqref{(8.1.3)} by itself allows also other UV completions, the question will arise then why they can not be used.\\

Now, in short about the recent paper \cite{SY2} of M.Shifman and A.Yung. To "improve" their previous results in \cite{SY1}, the authors proceeded in this their subsequent paper as follows.

{\bf a)}\, For vacua with the unbroken non-trivial $Z_{2N_c-N_F\geq 2}$ symmetry, replaced by hand in their $U(\nd)$ analog of \eqref{(8.1.3)} the light electric quarks $\tQ^i$ in their previous paper \cite{SY1} by light dyons, $\tQ^i \ra D_i$. And replaced all other light electric fields in \eqref{(8.1.3)} by dyonic ones with nonzero magnetic charges, implying this time that all charged pure electric $SU(\nd)$ fields in \eqref{(8.1)} acquired really large masses $\sim\lm$ and disappeared from the lower energy Lagrangian \eqref{(8.1.3)}, while the whole light {\it dyonic} theory \eqref{(8.1.3)},\eqref{(8.1.5)} appeared instead. Besides, these dyons in \eqref{(8.1.5)} with $\tQ^i \ra D_i$ where claimed arbitrarily to belong the antifundamental representation "${\ov N}_F$" of $SU(N_F)_L$ (unlike $Q^i$ belonging to the fundamental representation $N_F$).

This replacement of the electric $SU(\nd)$ by the dyonic one has been made following assumed analogy with their previous result in \cite{SY3} for the $U(N_c=3), N_F=5, \no=2$ theory. The properties of this example were freely extrapolated then in \cite{SY2} to all very special and br2 vacua (these are respectively $r=N_c$ and zero vacua in \cite{SY2}). Unfortunately, it was overlooked that the example from \cite{SY3} is not typical but exceptional because $2N_c-N_F=1$ in this example, so that the discrete $Z_{2N_c-N_F=1}$ symmetry becomes trivial in this case and gives no restrictions on the form of $\langle X^{\rm adj}_{SU(N_c)}\rangle$. While in all cases with the non-trivial unbroken $Z_{2N_c-N_F\geq 2}$ symmetry it forbids the appearance of the quark mass terms like $\sim\lm ({\ov Q} Q)$ in the $SU(\nd)$ sector of the superpotential at the scale $\mu\sim\lm\,$. And it does not matter whether such terms originated from higgsed $\langle X^{\rm adj}_{SU(N_c)}\rangle\sim\lm$ or from unrecognized "outside" (see Introduction and section 6 in \cite{ch7.3} for additional critical remarks and discussions of this point).

This dyonic theory \eqref{(8.1.5)} was "continued" then in \cite{SY2} to $\mx\gg\la\gg\lm$ {\it ignoring the need for the UV completion}, and this is a principled drawback.

Moreover, the expression \eqref{(8.1.4)}, $\lt\sim\lm (\lm/\mx)^{\frac{N_F-N_c}{3N_c-2N_F}}$, for the scale factor $\lt$ of the $SU(\nd)$ gauge coupling valid {\it only} at scales $m\ll\mu<\mx\ll\lm$ where $\lt\gg\lm$ was used also at $\mx\gg\la$ where $\lt\ll\lm$, with all consequences following from this erroneous continuation. As was explained above in the text, such a naive continuation is not right because the properties of the theory \eqref{(8.1.3)} with $\mx\ll\lm$ continued to $\mx\gg\la$, and in particular the concrete value of the gauge coupling scale factor $\Lambda^\prime$, can not be determined without the explicit form of the necessary UV completion. For instance, with the theory \eqref{(8.1)} used for the UV completion, the scale factor of the gauge coupling at $\mx\gg\lm$ will be $\la\sim (\lm^{2N_c-N_F}\mx^{N_c})^{1/(3N_c-N_F)}\gg\lm\gg\lt$.\\

{\bf b)}\, Simply replaced {\it by hand} at $\mx\gg\la$ the (dyonic) Lagrangian \eqref{(8.1.5)}, valid as it is at $m\ll\mu<\mx\ll\lm$ only, by their \eqref{(8.3.2)} (with chosen arbitrarily $\kappa^2\sim\mx m$) introducing in such a way desirable $N_F^2$ light Seiberg's mesons $M^i_j$. It was claimed that these mesons have to originate from the Abelian $U^{\bb}(1)$ sector continued to $\mx\gg\la$ because they are needed in Seiberg's dual ${\cal N}=1\,\,\,SU(\nd)$ theory. This form \eqref{(8.3.2)} was criticized above in this section, but in any case, the approach with using the known properties of Seiberg's dual Lagrangian in attempts, similarly to \cite{APS}, to derive its field content and its form at $\mx\gg\la$ from \eqref{(8.1.5)} (even with added Abelian sector), looks very strange.\\

On the whole, our results in this section 8 on properties of direct theories and their Seiberg's dual differ essentially from both those in \cite{SY1} and \cite{SY2} and we presented above our criticism concerning these two papers (see also \cite{ch7.3} for a number of additional details and critical remarks).

\section{Conclusions}

\hspace*{5mm} This paper continues \cite{ch5} by studying  the direct ${\cal N}=1\,\, SU(N_c)$ SQCD-like theories and their Seiberg's dual, with various numbers $N_F$ of quark flavors and with $N_F^2$ additional colorless but flavored fields $\Phi^j_i$. Unlike \cite{ch5}, where only the region $1<N_F<2N_c$ was considered, we calculated here the mass spectra of the direct and dual theories in the region $2N_c<N_F<3N_c$\,.

The calculations in this article were performed within the dynamical scenario introduced in \cite{ch3}. This scenario assumes that, when such ${\cal N}=1$ SQCD-like theories with $m_Q\neq 0$ are in the strong coupling regime $a(\mu)\gtrsim 1$, the quarks can be in the two {\it standard} phases only. - These are either the HQ (heavy quark) phase where they are confined or the Higgs phase where they are condensed with (some components of) $\langle Q\rangle=\langle \ov Q\rangle\neq 0$. The word {\it standard} also implies here that, unlike e.g. the very special ${\cal N}=2$ SQCD theories with colored adjoint scalar fields, no "unexpected" {\it additional} (i.e. in addition to the standard Nambu-Goldstone particles due to the spontaneous breaking of the global flavor symmetry) parametrically lighter particles like magnetic monopoles or dyons appear in these ${\cal N}=1$ SQCD-like theories without colored adjoint scalars (see also the footnote \ref{(f2)}).

The calculations of the mass spectra are based on finding first the quark and gluino condensates in various vacua. This was done for $N_F>2N_c$ in section 2, where the multiplicities of various vacua with the unbroken or spontaneously broken flavor symmetry were also found. It is worth noting that the explicit expressions for the total number and multiplicities of various vacua in the three regions: $1<N_F<N_c,\, N_c<N_F<2N_c$, and $N_F>2N_c$ are different and are not analytic continuations of each other, see \cite{ch5} and section 2. And the hierarchies among the quark condensates are also different.

The mass spectra of the direct theories and their Seiberg's dual at $2N_c<N_F<3N_c\,,\,\,\mph\gg\la$, were calculated in various vacua and compared with each other in sections 3-6. These mass spectra are parametrically different, in general, in direct and dual theories.

We calculated also in section 7 the mass spectra of the weakly coupled dual theory with $\nd=N_F-N_c$ colors and $N_c<N_F<3N_c/2$ flavors. Besides of interest by itself, the results were useful for the next section 8.\\

In this last section 8, ${\cal N}=2\,\, SU(N_c)$ SQCD \eqref{(8.1)} with $N_c$ colors and $N_c+1<N_F<3N_c/2$ flavors of original direct (electric) quarks  $Q^i$ was considered. The scale factor of the $SU(N_c)$ gauge coupling  is $\lm$ and the quark mass term in the superpotential is $m{\rm Tr}({\ov Q}Q)$. ${\cal N}=2$ is broken down to ${\cal N}=1$ by the mass term $\mx {\rm Tr}(X^2)$ of the adjoint scalar field $X$, $\,m\ll\mx\ll\lm\,$.

The spectra of masses smaller than $\mx$ were calculated {\it in vacua with the unbroken non-trivial $Z_{2N_c-N_F\geq 2}$ discrete R-symmetry} and with the unbroken or spontaneously broken $U(N_F)\ra U(n_1)\times U(n_2)$ flavor symmetry. This theory was considered previously in \cite{APS} (for the case $m=0$) and \cite{CKM}, and later by M.Shifman and A.Yung in the large series of papers, see e.g. recent papers \cite{SY1,SY2} and references therein.

We calculated in section 8.1, in vacua with the unbroken non-trivial discrete symmetry $Z_{\bb\geq 2}$ and spontaneously broken flavor symmetry $U(N_F)\ra U(n_1)\times U(n_2)$, the spectra of masses smaller than $\mu_{\rm x}$ in the lower energy $SU(\nd)$ theory \eqref{(8.1.3)} with $\nd=N_F-N_c$ colors and $N_F$ flavors of light quarks $\tQ^i,\, \otQ_j$. These $SU(\nd)$ quarks are a subset of original direct (electric) $SU(N_c)$ quarks $Q^i,\, \ov Q_j$ which have not received large masses $\sim\lm$ due to higgsing $\langle X^{\rm adj}_{SU(N_c)}\rangle\sim\lm$ of the adjoint scalar $X^{\rm adj}_{SU(N_c)}$, with $SU(N_c)\ra SU(\nd)\times U(1)^{2N_c-N_F}$, and remained light, i.e. with masses $\ll\lm$ (as well as the $SU(\nd)$ gluons and adjoint scalars $\textsf{x}^{\rm adj}_{SU(\nd)}$), while all other charged electric particles acquired large masses $\sim\lm$.

The mass spectra in these vacua were calculated also in two variants of Seiberg's dual theories. The first variant was the literal Seiberg's dual to \eqref{(8.1.3)},\eqref{(8.1.5)} strongly coupled theory \eqref{(8.1.10)},\eqref{(8.1.14)} with $N_c=N_F-\nd$ colors and $N_F$ flavors of quarks $\tq_i$ (dual to $\tQ^i$). It was shown that the spectrum of masses smaller than $\mu_{\rm x}$  in the direct theory \eqref{(8.1.3)} with the quarks $\tQ$ and in the dual one \eqref{(8.1.10)} with the quarks $\tq$ are parametrically different (see also \cite{ch7} for more details).

The second variant was the strongly coupled direct theory \eqref{(8.1.18)} with $N_c$ colors and $N_F$ flavors of original direct quarks $Q$ and its weakly coupled Seiberg's dual theory \eqref{(8.1.19)},
\eqref{(8.1.20)} with $\nd$ colors and $N_F$ flavors of quarks $q$ (dual to $Q$), and with especially chosen values of parameters $m_Q, \mph$ and $\la$ \eqref{(8.1.24)}. It was shown that, with this choice \eqref{(8.1.24)}, the spectra of masses smaller than $\mu_{\rm x}$ are the same: i) in the pair of weakly coupled theories with $\nd$ colors, i.e. in the  theory \eqref{(8.1.3)} with direct quarks $\tQ,\, \otQ$ and in \eqref{(8.1.20)} with dual quarks $q,\, \ov q$\,;\,\, ii) in the pair of strongly coupled theories with $N_c$ colors, i.e. in the  theory \eqref{(8.1.18)} with direct quarks $Q,\, \ov Q$ and in \eqref{(8.1.10)} with dual quarks $\tq,\, \otq$ (but the mass spectra in the strongly coupled pair of theories are different from those in the weakly coupled pair\,). The same equalities of mass spectra of these pairs of theories were shown in section 8.2 for vacua with the unbroken $U(N_F)$ and $Z_{2N_c-N_F\geq 2}$ symmetries (see \cite{ch7} for more details).

We emphasized also that, in spite of equalities of mass spectra in each pair of theories, their quarks are not the same particles, i.e. $\tQ\neq q$ and $Q\neq \tq$, as they behave differently under $SU(N_F)_L$ flavor transformations, resulting in different values of the $SU^3(N_F)_L$ 't Hooft triangles. As a result, with the choice of parameters as in \eqref{(8.1.24)}:\\
1) the pair of weakly coupled theories with $\nd$ colors, e.g. \eqref{(8.1.3)},\eqref{(8.1.20)}, have the same spectra of mass smaller than $\mx$ but different $SU^{\,3}(N_F)_L$ triangles (and the same for the pair \eqref{(8.1.10)},\eqref{(8.1.18)} of strongly coupled theories with $N_c$ colors)\,;\\
2) the dual pair with different number of colors, e.g. \eqref{(8.1.5)} with $\nd$ colors and \eqref{(8.1.10)} with $N_c$ colors, have the same $SU^{\,3}(N_F)_L$ triangles (but only in the specially chosen intervals of scales, see section \eqref{(8.1)}\,), but different spectra of mass smaller than $\mx$ (and the same for the dual pair \eqref{(8.1.18)},\eqref{(8.1.23)}\,)\,.\\

On the whole, our results in the last section 8 of this paper  on properties of direct theories and their Seiberg's dual differ essentially from those obtained by M.Shifman and A.Yung in \cite{SY1,SY2}. In particular, we presented in section 8.3 a number of critical remarks about the form of Seiberg's dual theory used in \cite{SY1,SY2} and about the approach used in \cite{SY2} to describe the transition of the IR free $SU(N_F-N_c)$ theory with $N_c+1<N_F<3N_c/2$ light quark flavors from $\mx\ll\lm$ to $\mx\gg\lm$. (See also \cite{ch7.3} for additional critical remarks and detailed explanations ).\\

\appendix
\numberwithin{equation}{section}
\section{Anomalous dimensions in the strong coupling regime}

The purpose of this appendix is to find the anomalous dimensions of quarks and colorless but flavored fields $\Phi^j_i$ in the theory \eqref{(8.1.18)} (or $M^i_j$ in \eqref{(8.1.10)} \,) in the strong coupling regime in the range of scales where they are dynamically relevant, i.e. their running masses at the scale $\mu$ are smaller than $\mu$. To be specific, we consider below the theory \eqref{(8.1.18)} with parameters \eqref{(8.1.24)}, $m_Q\ll\lt\ll\mph$, but the results for anomalous dimensions are general and valid also for \eqref{(8.1.10)}.\\

Note first that there is the first generation of all $N_F^2$ fions $\Phi_i^j$ with masses $\mu_1^{\rm pole}(\Phi)\sim\mph\gg\lt$. As shown in \cite{ch4,ch5}, the anomalous dimension $\gamma^{(+)}_{\Phi}$ of fions in the range of scales $\mos<\mu<\lt$ where they are still heavy and dynamically irrelevant is:\, $\gamma^{(+)}_{\Phi}=\,-\,2\gamma^{(+)}_Q$.

Besides, there is also the second generation of all $N_F^2$ fions with masses, see \eqref{(8.1.4)},\eqref{(8.1.24)},
\bq
{\hspace*{-0.5cm}}\mu_2^{\rm pole}(\Phi)\sim\mos\sim\lt\Bigl (\frac{\lt}{\mph}\Bigr )^{\frac{1}{2\gamma^{(+)}_Q-1}}=\lt\Bigl (\frac{\lt}{\mph}\Bigr )^{\frac{\nd}{5N_c-3N_F}}\sim\lm\Bigl (\frac{\lm}{\mx}\Bigr )^{\frac{\nd}{5N_c-3N_F}},\,\, \lm\ll\mos\ll\lt,\,\, \mph=-\mx\,,\label{(A.1)}
\eq
and all $N_F^2$ "heavy" fions become effectively massless and dynamically relevant in the direct theory \eqref{(8.1.18)} at $\mu<\mos$.\\

{\bf a}) Now, as for the anomalous dimensions of strongly coupled quarks in the direct theory
\eqref{(8.1.18)}, i.e. $\gamma_Q^{(+)}$ in the range $\mos<\mu<\lt$ where the fions are still irrelevant and ${\wt\gamma}_Q^{\,(+)}$ at $m_{Q,1}^{\rm pole}<\mu<\mos$ where the fions became relevant. We use first the approach \cite{ch1} (see section 7). For this, we have to introduce the dual theory which has the same 't Hooft triangles (in a given range of scales) as the direct theory \eqref{(8.1.18)}.

In the range $\mos\ll\mu\ll\lt$ this dual theory can be taken as \eqref{(8.1.20)}. The effectively massless particles in this range of scales in the direct theory \eqref{(8.1.18)} are all quarks and gluons, while in the dual one \eqref{(8.1.20)} these are the dual quarks and gluons and mions $M^i_j$. As shown in \cite{S2}, all 't Hooft triangles will be the same in these two theories.

Now, in the approach \cite{ch1}, we equate two NSVZ $\,{\widehat\beta}_{ext}$ - functions of the external baryon and $SU(N_F)_{L}$ - flavor vector fields in the direct and dual theories,
\bq
\frac{d}{d\,\ln \mu}\,\frac{2\pi}{\alpha_{ext}}={\widehat\beta}_{ext}= -\frac{2\pi}{\alpha^2_{ext}}\,\beta_{ext}= \sum_i T_i\,\bigl (1+\gamma_i\bigr )\,,\label{(A.2)}
\eq
where the sum runs over all fields which are effectively massless at scales considered, the unity in the brackets is due to one-loop contributions while the anomalous dimensions $\gamma_i$  of fields represent all higher-loop effects, $T_i$ are the coefficients. It is worth noting that these general NSVZ forms \eqref{(A.2)} of the external "flavored" $\widehat\beta$-functions are independent of the kind of massless perturbative regime of the internal gauge theory, i.e. whether it is conformal, or the strong coupling or the IR free one.

For the baryon charge \eqref{(A.2)} looks as
\bq
N_F N_c\,\Bigl ( B_Q=1 \Bigr )^2\,(1+\gamma^{(+)}_Q)=N_F \nd \,\Bigl ( B_q=\frac{N_c}{\nd}
\Bigr )^2\,(1+\gamma_q)\,.\label{(A.3)}
\eq

In \eqref{(A.3)}: the left-hand side is from the direct theory while the right-hand side is from the dual one. The dual theory \eqref{(8.1.23)} is in the IR free logarithmic regime at $\mu\ll\lt$ with the dual coupling ${\ov a}(\mu)\ll 1$ and $\gamma_q\sim {\ov a}\ra 0$. Then from \eqref{(A.3)}, see section 7 in \cite{ch1}
\bq
\gamma^{(+)}_Q=\frac{2N_c-N_F}{N_F-N_c},\quad\frac{d\,a(\mu\ll\lt)}{d\log\mu}=\beta_{NSVZ}(a)=
\frac{a^2(\mu)}{a(\mu)-1}\,\frac{\bo-N_F\gamma_Q^{(+)}}{N_c}\,\,\,
\xrightarrow{a(\mu)\gg 1}\,\,\, -\,\nu^{(+)}\,a(\mu),\label{(A.4)}
\eq
\bbq
 a(\mos\ll\mu\ll\lt)\sim (\lt/\mu)^{\nu^{(+)}}\gg 1,
\,\,\,\,\,\, \nu^{(+)}=\frac{3N_c-2N_F}{N_F-N_c}\,>\,0,\,\,\,  \bo=3N_c-N_F\,.
\eeq

For the flavor charge \eqref{(A.2)} looks as
\bq
N_c(1+\gamma^{(+)}_Q)=\nd(1+\gamma_q)+N_F(1+\gamma_{M})\,.\label{(A.5)}
\eq
Both $\gamma_q$ and $\gamma_{M}$ are logarithmically small at $\mu\ll\lt$ and \eqref{(A.5)} is incompatible with \eqref{(A.4)} (and with the NSVZ $\beta$-function, see section 7 in \cite{ch1}). Therefore, we will not use \eqref{(A.2)} for the flavor charge (here and below).\\

{\bf b}) Now, for the range $m_{Q,1}^{\rm pole}\ll\mu\ll\mos$ where the fions in \eqref{(8.1.18)} became relevant in the strong coupling regime $a(\mu)\gg 1$, the dual theory with the same 't Hooft triangles as in \eqref{(8.1.18)} (now with effectively massless fions $\Phi$) can be taken as \eqref{(8.1.20)} but without the mion fields $ M^i_j$ (i.e. only massless dual quarks ${\ov q}^{\,j}, q_i$ and $SU(\nd)$ dual gluons).  But because the baryon charge of fions $\Phi^j_i$ in the theory \eqref{(8.1.18)} is zero, \eqref{(A.3)} and \eqref{(A.4)} remain the same, i.e. the anomalous dimension ${\wt\gamma}^{\,(+)}_Q$ of quarks ${\ov Q}, Q$ remains the same after fions became relevant, ${\wt\gamma}^{\,(+)}_Q=\gamma^{(+)}_Q=(2N_c-N_F)/\nd$.\\

{\bf c}) Because \eqref{(A.2)} for the flavor charge is not fulfilled, the above method does not allow to find the anomalous dimension ${\wt\gamma}^{\,(+)}_{\Phi}$ at $\mu<\mos$ where the fions become effectively massless and dynamically relevant. Therefore, on the example of the strongly coupled theory \eqref{(8.1.18)}, we present now independent way to connect the anomalous dimensions of quarks and $\Phi$, $\,{\wt\gamma}^{\,(+)}_Q$ and ${\wt\gamma}^{\,(+)}_{\Phi}$, at those scales where $\Phi$ became relevant, i.e. at $\mu<\mos$. It follows from the internal self consistency when the quarks ${\ov Q}_1, Q^{1}$ decouple as heavy ones at scales $\mu<m_{Q,1}^{\rm pole}$ in the theory \eqref{(8.1.18)}, while all $N_F^2$ fions $\Phi^j_i$ still remain effectively massless in the large range of scales $\mu^{\rm pole}_{3}
(\Phi^1_1)<\mu<\mos,\,\, \mu^{\rm pole}_{3}(\Phi^1_1)\ll m^{\rm pole}_{Q,1}\ll\mos$ (the spectrum of masses smaller than $\mx$ is the same in \eqref{(8.1.18)} and in \eqref{(8.1.10)}, see \cite{ch7}\,).

The Lagrangian of the direct theory \eqref{(8.1.18)}, which is in the HQ (heavy quark) phase in br2 vacua, with $m_{Q,1}^{\rm pole}\gg m_{Q,2}^{\rm pole}$) has the form at the scale $\mu=m_{Q,1}^{\rm pole}\ll\mx\ll\lm$
\bbq
K=K_{\Phi}+K_{Q}=z^{(+)}_{\Phi}{\rm Tr}\,(\Phi^\dagger\Phi)+z^{(+)}_Q{\rm Tr}\,(Q^\dagger Q+Q\ra {\ov Q})=
{\rm Tr}\,(\delta{\Phi_R}^{\dagger}\delta{\Phi_R})+{\rm Tr}\,({Q_R}^\dagger{Q_R}+Q_R\ra {\ov Q}_R),
\eeq
\bbq
z^{(+)}_Q=z^{(+)}_Q(\lt,m_{Q,1}^{\rm pole})=z^{(+)}_Q(\lt,\mos)\,{\wt z}^{\,\,(+)}_Q(\mos,m_{Q,1}^{\rm pole})=\Bigl (\frac{\mos}{\lt}\Bigr )^{\gamma^{(+)}_Q}\Bigl (\frac{m_{Q,1}^{\rm pole}}{\mos}\Bigr )^{{\wt\gamma}^{\,(+)}_Q}\,,
\eeq
\bq
\quad z^{(+)}_{\Phi}=z^{(+)}_{\Phi}(\lt,m_{Q,1}^{\rm pole})=z^{(+)}_{\Phi}(\lt,\mos)\,{\wt z}^{\,\,(+)}_
{\Phi}(\mos,m_{Q,1}^{\rm pole})=\Bigl (\frac{\lt}{\mos}\Bigr )^{2\gamma^{(+)}_Q}\Bigl (\frac{m_{Q,1}^{\rm pole}}{\mos}\Bigr )^{{\wt\gamma}^{\,(+)}_{\Phi}}\,,\label{(A.6)}
\eq
\bq
m_{Q,1}^{\rm pole}=\frac{\langle m_{Q,1}^{\rm tot}\rangle}{z^{(+)}_Q(\lt,m_{Q,1}^{\rm pole})}\,,
\,\,\langle m_{Q,1}^{\rm tot}\rangle=\frac{\langle\Qt\rangle}{\mph}\sim m_Q\,,\,\, m_Q^{\rm tot}=
m_Q-\Phi=\langle m_Q^{\rm tot}\rangle-\delta\Phi\,,\,\, \langle\delta\Phi\rangle=0\,,\label{(A.7)}
\eq
where $\Phi_R$ and $Q_R$ are the canonically normalized fields and $\delta\Phi_R$ is a pure quantum fluctuation,
\bq
{\cal W}_{\rm matter}={\cal W}_{\Phi}+{\rm Tr}\,({\ov Q}\,m_Q^{\rm tot} Q)\,,\label{(A.8)}
\eq
${\cal W}_{\Phi}$ is given in \eqref{(8.1.18)}, see \cite{ch7} for details.

At $\mu<m_{Q,1}^{\rm pole}$ all quarks ${\ov Q}_1, Q^1$ decouple and can be integrated out as heavy ones. As a result, there will appear a series of higher dimension operators in D-terms of $\Phi$, e.g.
\bq
O_{\rm n}\sim a^{(1)}_f\delta{\Phi_R}^{\dagger}\delta{\Phi_R}\Biggl [a^{(1)}_f\frac{\delta{\Phi_R}^{\dagger}
\delta{\Phi_R}}{\Bigl (m_{Q,1}^{\rm pole}\Bigr )^2}\,\Biggr ]^{\rm n}\,,\quad a^{(1)}_f=\frac{a_f(\mu=\lt)= 1}{z^{(+)}_{\Phi}\,(z^{(+)}_Q)^2}\,\,,\label{(A.9)}
\eq
where $a^{(1)}_f=a_f(\mu=m_{Q,1}^{\rm pole})$ is the Yukawa coupling at the scale $\mu=m_{Q,1}^{\rm pole}$. These terms originate from the expansion in powers of $\delta{\Phi_R}$ of the heavy quark loop integrated over the non-parametric interval of momenta $p\sim m_{Q,1}^{\rm pole}$,
\footnote{\,
Because quarks ${\ov Q}_1, Q^1$ are weakly confined, they form a set of colorless hadrons with the characteristic scale of hadron masses ${\ov M}_{\rm hadr}\sim m_{Q,1}^{\rm pole}$. Therefore, gluons with momenta $\lesssim m_{Q,1}^{\rm pole}$ effectively decouple from these quarks, and for this reason gluons are absent in \eqref{(A.10)}. On the other hand, there are no reasons to decouple for the colorless fions $\Phi^j_i$.
}
\bq
\Delta_{\Phi}\sim\int d^4 p\,[\, p^2+M^\dagger M\, ]^{-1}\,,\quad M=m_{Q,1}^{\rm pole}\Biggl (1-\frac{\delta{\Phi_R}}{z^{(+)}_Q {\sqrt{z^{(+)}_{\Phi}}\,m_{Q,1}^{\rm pole}}}\Biggr ),\quad \langle M\rangle=m_{Q,1}^{\rm pole}.\label{(A.10)}
\eq

When propagating in the loop with the momentum "k", the canonical fields $\delta{\Phi_R}\sim k$. Therefore, in order that heavy quarks ${\ov Q}_1, Q^1$ really completely decouple as it should be, the contributions from
$O_{\rm n}$ in \eqref{(A.9)} have to be small at momenta $k < m_{Q,1}^{\rm pole}$ in comparison with the canonical Kahler term $\delta{\Phi_R}^{\dagger}\delta{\Phi_R}$. Then, because the expansion \eqref{(A.9)} of \eqref{(A.10)} breaks down at larger momenta $> m_{Q,1}^{\rm pole}$ (in any case, the running quark mass diminishes very quickly with increasing momentum), all terms $O_{\rm n}/[\delta{\Phi_R}^{\dagger}
\delta{\Phi_R}]$ will be $O(1)$ at the momenta $\sim m_{Q,1}^{\rm pole}$. For this, it should be parametrically $a^{(1)}_f\sim 1/[z^{(+)}_{\Phi}(z^{(+)}_Q)^2]\sim 1$, see \eqref{(A.6)},\eqref{(A.7)},\eqref{(A.9)},
\bq
a^{(1)}_f=O(1)\,\,\ra\,\, z^{(+)}_{\Phi}\,\Bigl ( z^{(+)}_Q\,\Bigr )^2=
z^{(+)}_{\Phi}(\mos,m_{Q,1}^{\rm pole})\,\Bigl ( z^{(+)}_Q(\mos,m_{Q,1}^{\rm pole})\,\Bigr )^2\sim
1\,\,\ra\,\, {\wt\gamma}^{\,(+)}_{\Phi}=-2\,{\wt\gamma}^{\,(+)}_Q\,.\label{(A.11)}
\eq
\eqref{(A.11)} ensures then that corrections from $O_{\rm n}$ will be small at momenta $k<m_{Q,1}^{\rm pole}$, so that quarks ${\ov Q}_1, Q^1$ really decouple completely.

Therefore, $\gamma_{\Phi}=-2\gamma_Q$ not only at $\mos<\mu<\lt$ where the fions $\Phi^j_i$ were irrelevant, but also ${\wt\gamma}_{\Phi}=-2\,{\wt\gamma}_Q$ at lower scales $\mu<\mos$ where they became relevant.\\

{\bf d}) And finally, we present also independent reasonings leading to the same results. First, we point out that the gauge coupling $a(\mu\ll\lt)\gg 1$ entered already into a strong coupling regime, while $\gamma^{(+)}_{\Phi}=-2\gamma^{(+)}_Q$ at $\mos<\mu<\lt$ where the fions are still irrelevant \cite{ch1}. Therefore, the gauge and Yukawa couplings look at $\mu=\mos\ll\lt$ as, see \eqref{(A.4)},
\bq
a(\mu=\mos)=(\lt/\mos)^{\nu^{(+)}\,>\,0}\gg 1,\quad a_f(\mu=\mos)=\frac{a_f(\mu=\lt)= 1}
{z^{(+)}_{\Phi}(\lt,\mos)[\,z^{(+)}_Q(\lt,\mos)\,]^2}\sim 1\,.\label{(A.12)}
\eq

Consider now the Feynman diagrams in the strongly coupled direct theory \eqref{(8.1.18)} contributing to the renormalization factors $z_{\Phi}(\mu)$ and $z_{Q}(\mu)$ at $m^{\rm pole}_{Q,1}\ll\mu\ll\mos$ where both the quarks and fions $\Phi$ are effectively massless. Order by order in the perturbation theory the extra loop with the exchange of the field $\Phi$ is $a_f(\mos)/a(\mos)\sim (\mos/\lt)^{\nu^{(+)}\,>\,0}\ll 1$ times smaller than the extra loop but with the exchange of gluon.

In all cases with a resummation of perturbative series, a standard assumption is that the leading contribution to the sum originates from summation of leading terms at each order. With this assumption, we can neglect all exchanges of $\Phi$ in comparison with those of gluons, order by order. Therefore, the fact that the fields $\Phi^j_i$ became relevant at $\mu<\mos$ is really of no importance for the RG-evolution, so that both ${\wt\gamma}^{\,\rm (+)}_{Q}$ and ${\wt\gamma}^{\,(+)}_{\Phi}$ {\it remain the same} at $m^{\rm pole}_{Q,1}\ll\mu\ll\mos$ as they were at $\mos\ll\mu\ll\lt$ when fions were irrelevant:\, ${\wt\gamma}^{\,\rm(+)}_{Q}=\gamma^{\rm(+)}_{Q},\,\,{\wt\gamma}^{\,(+)}_{\Phi}=\gamma^{(+)}_{\Phi}=
\,-2\,\gamma^{\rm (+)}_{Q}$  (i.e. the Yukawa coupling $a_f(\mu)$ still stays intact at $a_f(\mu)\sim 1$ also at $m^{\rm pole}_{Q,1}\ll\mu\ll\mos$, as it was at $\mos\ll\mu\ll\lt$). The equality ${\wt\gamma}^{\,\rm (+)}_{Q}=\gamma^{\rm (+)}_{Q}$ agrees with "${\bf b}$" above, while ${\wt\gamma}^{\,(+)}_{\Phi}=\,-2\,{\wt\gamma}^{\,\rm (+)}_{Q}$  agrees with "${\bf c}$" above.\\

It is also worth to remind that, once the fions $\Phi^j_i$ become effectively massless and dynamically relevant with respect to internal interactions, they simultaneously begin to contribute to the 't Hooft triangles.

\end{document}